\documentclass[prd,aps,twocolumn,nofootinbib,superscriptaddress,preprintnumbers,balancelastpage,longbibliography,floatfix]{revtex4-1}

\usepackage{amsmath}	
\usepackage{mathtools}
\usepackage{fontawesome} 
\usepackage[dvipsnames]{xcolor}
\usepackage{hyperref}
\usepackage{xspace}
\usepackage{booktabs}

\newcommand{\nbicon}{{\color{linkcolor}\faFileCodeO}\xspace}
\newcommand{\nblink}[1]{\href{https://github.com/smsharma/astrometry-lensing-correlations/blob/arXiv-v1/notebooks/#1.ipynb}{\nbicon}}
\newcommand{\githubmaster}{\href{https://github.com/smsharma/astrometry-lensing-correlations/}{\faGithub}\xspace}

\definecolor{linkcolor}{rgb}{0.7752941176470588, 0.22078431372549023, 0.2262745098039215}

\newcommand{\Gaia}{\emph{Gaia}\xspace}
\newcommand{\vect}[1]{\boldsymbol{\mathbf{#1}}}
\newcommand{\dd}{\mathrm{d}}
\newcommand{\sun}{\odot}

\newcommand{\vesc}{v_\text{esc}}
\newcommand{\ellm}{{\ell m}}
\newcommand{\GN}{G_\mathrm{N}}
\newcommand{\Cinv}{(C^{-1})}

\def\deltabar{{\mathchar '26\mkern -10mu\delta}}

\newcommand{\ddbar}{{\rm d}\hspace*{-0.15em}\bar{}\hspace*{0.1em}}
\newcommand{\deltabarthree}{\deltabar^{(3)}}
\newcommand{\appropto}{\mathrel{\vcenter{
  \offinterlineskip\halign{\hfil$##$\cr
    \propto\cr\noalign{\kern2pt}\sim\cr\noalign{\kern-2pt}}}}}

\definecolor{deepblue}{rgb}{0.2,0.4,0.8}

\definecolor{deepred}{rgb}{0.8,0.2,0.2}

\definecolor{deepgreen}{rgb}{0.2,0.8,0.2}

\hypersetup{colorlinks=true,
linkcolor=linkcolor,
citecolor=linkcolor,
urlcolor=linkcolor,
linktocpage=true,
pdfproducer=medialab}

\begin{document}

\title{The Power of Ha$\boldmath\ell$ometry}
\author{Siddharth Mishra-Sharma}
\email{sm8383@nyu.edu}
\thanks{ORCID: \href{https://orcid.org/0000-0001-9088-7845}{0000-0001-9088-7845}}
\affiliation{Center for Cosmology and Particle Physics, Department of Physics, New York University, New York, NY 10003, USA}
\author{Ken Van Tilburg}
\email{kvt@kitp.ucsb.edu}
\thanks{ORCID: \href{https://orcid.org/0000-0001-7085-6128}{0000-0001-7085-6128}}
\affiliation{Kavli Institute for Theoretical Physics, University of California, Santa Barbara, California 93106, USA}
\affiliation{Center for Cosmology and Particle Physics, Department of Physics, New York University, New York, NY 10003, USA}
\affiliation{School of Natural Sciences, Institute for Advanced Study, Princeton, NJ 08540, USA}
\author{Neal Weiner}
\email{neal.weiner@nyu.edu}
\thanks{ORCID: \href{https://orcid.org/0000-0003-2122-6511}{0000-0003-2122-6511}}
\affiliation{Center for Cosmology and Particle Physics, Department of Physics, New York University, New York, NY 10003, USA}

\date{\today}

\begin{abstract}
Astrometric weak gravitational lensing is a powerful probe of the distribution of matter on sub-Galactic scales, which harbor important information about the fundamental nature of dark matter. We propose a novel method that utilizes angular power spectra to search for the correlated pattern of apparent motions of celestial objects induced from time-dependent lensing by a population of Galactic subhalos. Application of this method to upcoming astrometric datasets will allow for the direct measurement of the properties of Galactic substructure, with implications for the underlying particle physics. We show that, with near-future astrometric observations, it may be possible to statistically detect populations of cold dark matter subhalos, compact objects, as well as density fluctuations sourced by scalar field dark matter. Currently-unconstrained parameter space will already be accessible using upcoming data from the ongoing {\Gaia}~mission. \githubmaster
\end{abstract}

\maketitle

\tableofcontents

\section{Introduction}
The Standard Model of particle physics has been remarkably successful in explaining every observed laboratory phenomenon on Earth. Its failure to successfully describe almost anything about the growth of cosmological structure is therefore quite striking. Evidence has grown for decades now that the addition of a simple pressureless component of matter---dark matter (DM)---along with a cosmological constant $\Lambda$ can allow a complete description of nearly all known cosmological data through the $\Lambda$ Cold Dark Matter ($\Lambda$CDM) paradigm. 

The evidence for cold dark matter is now quite robust: precision studies of the Cosmic Microwave Background and large-scale structure provide strong evidence of the gravitational influence of DM in the early Universe~\cite{Aghanim:2018eyx}. Gravitational lensing and the measurements of galactic rotations have provided local evidence for dark matter and its distribution. Consistently, on large scales, the $\Lambda$CDM model has provided an excellent account of observations.

On smaller scales, the evidence for CDM is less clear. On length scales below that of the Milky Way, perturbations are nonlinear, necessitating expensive numerical simulations which require an understanding of the role of baryonic matter in the process. Moreover, phenomenona such as reionization, galaxy quenching, and interaction with a host galaxy lead to suppressed star formation in structures below $\lesssim10^9 \, \mathrm{M_\odot}$~\cite{Efstathiou:1992zz,Fitts:2016usl,Read:2017lvq}, implying that they are unlikely to be associated with any luminous matter. As a consequence, we must rely on purely gravitational techniques to study them.

Observing and understanding DM fluctuations on these smaller scales is critical, as it is the most likely place where the simple pressureless fluid model will break down. A whole host of phenomenona can suppress or enhance power at small (sub-Galactic) scales. These range from simple changes in the phase space in warm dark matter scenarios~\cite{Bond:1983hb,Bode:2000gq,Dalcanton:2000hn,Boyanovsky:2008he,Boyanovsky:2010pw}, to attractive self-interactions that enhance small-scale structure in the early Universe~\cite{Arvanitaki:2019rax}, to dissipative processes that allow compact structures to form~\cite{Agrawal:2017pnb,Agrawal:2017rvu,Buckley:2017ttd,Fan:2013yva,Vogelsberger:2015gpr,Chang:2018bgx,Essig:2018pzq}, to kpc-sized de Broglie wavelengths of ultralight scalar DM that prevent small structures from forming at all~\cite{Hu:2000ke,Hui:2016ltb,Mocz:2017wlg,Du:2016zcv,Bar-Or:2018pxz,Mocz:2017wlg}. This is especially pressing in light of the absence of a robust signal in direct, indirect, and collider searches for dark matter to date~\cite{Aprile:2018dbl,Aaboud:2019yqu,Sirunyan:2016iap,Fermi-LAT:2016uux}. 

As a consequence, there have been a wide range of recent efforts to detect these smaller structures. Some of them rely on understanding the effects that small dark matter halos would have on visible structures. The phase space of stars in the Milky Way may show signs of passing halos~\cite{Buschmann:2017ams}. Perturbations of stellar streams are promising avenues~\cite{2018ApJ...867..101B,Johnston:1998bd,Carlberg:2011xj}, with recent claims of both a detection of the statistical imprint of dark matter substructure~\cite{Banik:2019cza,Banik:2019smi}, and of a collision with a single dense subhalo~\cite{Bonaca:2020psc,Bonaca:2018fek}.

Complementary to these methods, Ref.~\cite{VanTilburg:2018ykj} (hereby V18) proposed using measurements of time-varying astrometric perturbations induced by Galactic subhalos on distant sources as a way to probe substructure in our Galaxy. In particular, correlated induced motions of a large number of background sources due to Galactic subhalos with typical velocity dispersion were shown to be a promising way to look for a population of extended subhalos within our Galaxy, within reach of current and future astrometric surveys. Broadly, the proposed searches took two different forms: {\em local template} analyses, which could point to a specific pattern of motions as signaling the existence of a dark object, and techniques based on {\em global correlations}, where no single source is clearly identified but imprints of a population of dark objects could be discerned. The first constraints on the abundance of compact dark objects using template methods and data from \Gaia's second data release were recently presented in Ref.~\cite{Mondino:2020rkn}. Prospects for detecting the astrometric effects of individual, dense dark matter objects were additionally studied in Refs.~\cite{Erickcek:2010fc,Li:2012qha,Zackrisson:2009rc}.

In this paper, we propose a new technique to characterize the population properties of Galactic substructure through its collective lensing effect on distant sources. This method extends the global correlation observables presented in V18 and recasts them in the language of angular power spectra, ubiquitous in cosmology. We present a general framework for calculating the power spectrum decomposition of induced velocities and accelerations due to a population of Galactic subhalo lenses characterized by arbitrary population (\emph{e.g.,} mass spectrum) as well as internal (\emph{e.g.,} density profile) properties. We apply this formalism to a few motivated scenarios---cold dark matter, compact object populations, ultralight scalar dark matter, and enhanced primordial fluctuations---to assess the sensitivity of future astrometric surveys to these cases. We emphasize that unconstrained parameter space can already be probed using future data releases of the ongoing \Gaia mission. We point out several handles that can be used to distinguish a lensing signal induced due to substructure from that of astrophysical or systematic origin. In particular, we shall show that properties of the induced motions allow us to separate out a pure noise channel, providing a cross check of an orthogonal channel containing both signal and noise. Additionally, we describe how the preferred motion of the Sun in our Galaxy would lead to a directionality in the substructure-induced lensing signal which would imprint itself as an asymmetry over the azimuthal correlation modes.

\begin{figure*}[tbp]
\centering
\includegraphics[width=0.95\textwidth]{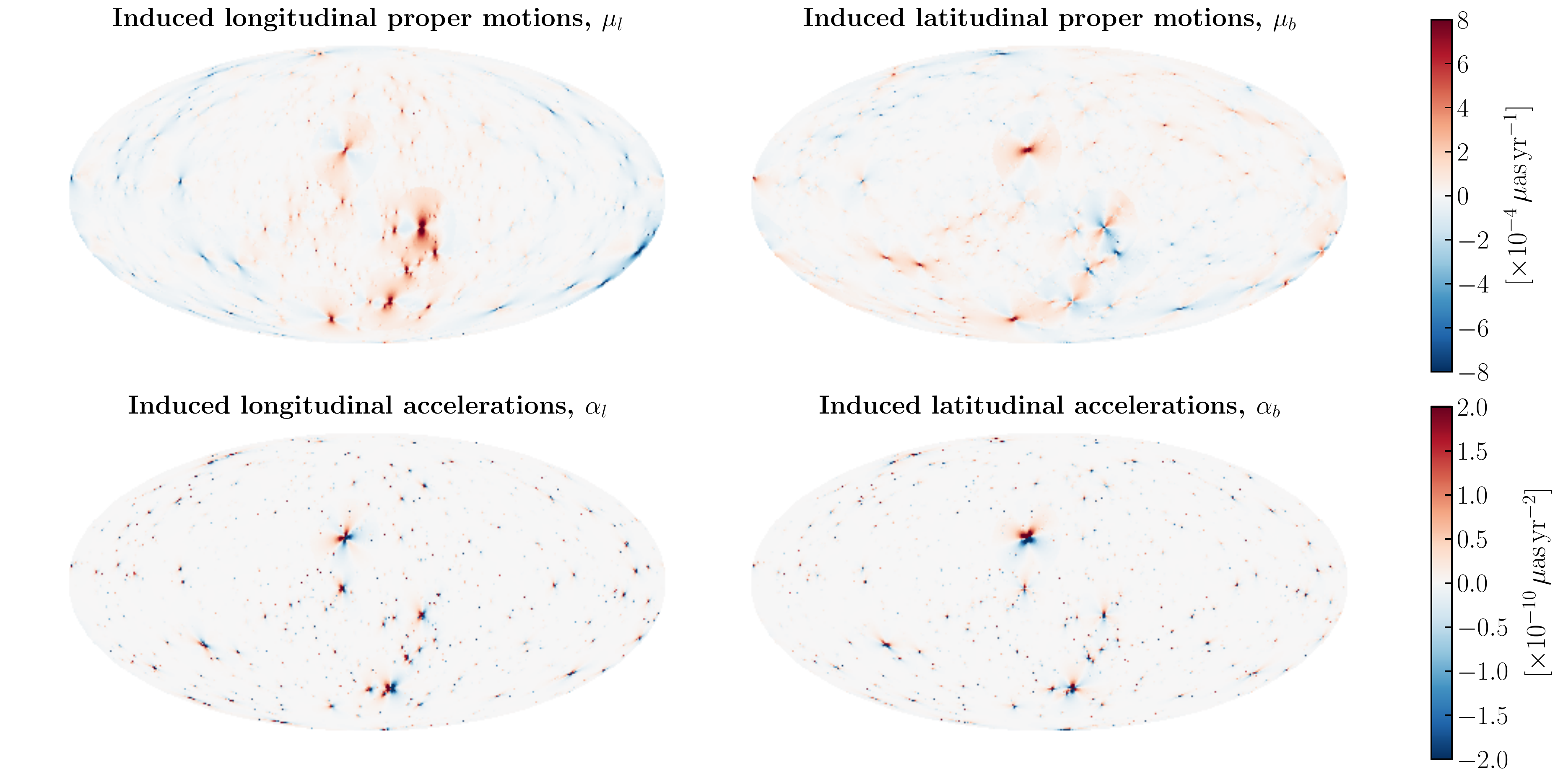}
\caption{In Galactic coordinates, the expected proper motion {(top row)} and proper acceleration {(bottom row)} induced on a population of background celestial sources due to a simulated realization of subhalos using the CDM-inspired fiducial configuration described in Sec.~\ref{sec:cdmpop}. Galactic longitude {(left column)} and Galactic latitude {(right column)} components are shown on a \texttt{HEALPIX} grid with resolution \texttt{nside}=128. \nblink{01_simulations}}
\label{fig:population_maps}
\end{figure*}

An example of the induced proper motion and proper acceleration components for a realization of cold dark matter subhalos (described in Sec.~\ref{sec:cdmpop}) is shown in the top and bottom rows of Fig.~\ref{fig:population_maps}, respectively, in Galactic coordinates. These are split into the motion component in the longitudinal direction (left columns) and those in the latitudinal direction (right columns). Searching for global evidence of such a pattern of apparent motions and using this to infer the properties of the underlying subhalo distribution will be the main subject of this work.  

This paper is organized as follows. In Sec.~\ref{sec:singlesh}, we introduce the lens-induced proper motion and proper acceleration power spectrum formalism, illustrating the signal characteristics with toy examples. Section~\ref{sec:noise} describes extragalactic as well as Galactic background source populations that could be used to measure this collective signal, and presents anticipated noise levels on the astrometric properties of these sources that could be deliverable by ongoing (\emph{e.g.}, \Gaia) and future (\emph{e.g.}, SKA and WFIRST) surveys. Section~\ref{sec:forecasts} presents forecasts on a few motivated benchmark scenarios achievable using future measurements of correlated induced lensing effects. In Sec.~\ref{sec:handles}, we discuss several handles that can be used to distinguish a putative signal from unmodeled sources of noise in a power spectrum measurement. In Sec.~\ref{sec:gaia-quasars} we demonstrate the feasibility of the techniques introduced by constructing a vector power spectrum estimator and applying it to the proper motions of quasars in \Gaia's second data release. We conclude in Sec.~\ref{sec:conclusions}.

We use units with $\hbar = c  = 1$ and the \textit{Planck} 2018 cosmology~\cite{Aghanim:2018eyx} throughout this work. In the spirit of reproducibility, the code used to obtain the results of this study is available on GitHub \githubmaster and a link below each figure (\nbicon) provides the Python code with which it was generated.

\section{Formalism}
\label{sec:singlesh}
Our ultimate goal is to develop a technique to detect the presence of lenses in front of a set of sources statistically, without necessarily resolving the presence of any individual lens. There are a wide range of possible methods to do this. One of the most developed techniques to deal with such correlations involves understanding two-point functions within a dataset. 
We aim to detect the presence of dark matter substructure (\emph{i.e.}, a population of Galactic subhalos) by measuring the two-point correlation function of lens-induced velocities and accelerations. V18 proposed using a ``correlation'' observable to this effect. Here we offer a substantially improved framework for studying the two-point function of lens-induced motions of background sources, namely one based on a vector power spectrum decomposition. As we shall see, this framework allows for additional handles and discriminants that can reject systematics and other spurious effects, and allows for a more detailed interpretation of any positive signal.

We provide a general overview of this formalism in Sec.~\ref{sec:formalism}, leaving details of the derivations to App.~\ref{app:derivations}. We will apply this formalism to a few simple subhalo lens case studies in Sec.~\ref{sec:examples} in order to build intuition for how the signal characteristics are affected by the properties of the underlying substructure population. The reader may refer to App.~\ref{app:significance} for details about the statistical tools and scaling of the expected signal significance with various parameters characterizing the signal and noise properties.

\subsection{General framework}
\label{sec:formalism}

\subsubsection*{Lens-induced proper motions and accelerations}

In the thin-lens regime, the angular deflection $\Delta \boldsymbol{\theta}$ of a source at angular diameter distance $D_s$ due to a lens at angular diameter distance $D_l$ is given by~(see \emph{e.g.}, V18)
\begin{equation}
\Delta \boldsymbol{\theta}(\mathbf b)=-\left(1-\frac{D_{l}}{D_{s}}\right) \frac{4 G_\mathrm{N} M(b)}{b} \hat{\mathbf{b}}
\end{equation}
where $\vect{b}$ is the physical impact parameter between the source and lens, and $M(b)=2 \pi \int_{-\infty}^{+\infty} \mathrm{d} x \int_{0}^{b} \mathrm{d} b^{\prime} b^{\prime} \rho\left(\sqrt{x^{2}+b^{\prime 2}}\right)$ is the enclosed mass function of the spherically-symmetric lens within a cylinder of radius $b$. As discussed in V18, the induced deflections are typically too small to be disentangled from naturally-occurring and systematic variations in the angular number density of sources, either individually or collectively.

Effects in the time domain offer more promise. Since dark matter substructure has a characteristic velocity dispersion, an effective lens velocity $\vect{v}_l \equiv {\dd \vect{b}}/{\dd t}$ induces an apparent velocity on the luminous sources. This angular velocity correction $\vect{\mu}\equiv\Delta\boldsymbol{\dot \theta}$ can be written as 
\begin{align}
\vect{\mu}(\vect{b}) = 4 \GN \left\lbrace \frac{M(b)}{b^2} \left[ 2 \hat{\vect{b}} (\hat{\vect{b}} \cdot \vect{v}_l) -  \vect{v}_l \right] -\frac{M'(b)}{b} \hat{\vect{b}} (\hat{\vect{b}} \cdot \vect{v}_l) \right\rbrace
\label{eq:mureal}
\end{align}
where we denote the angular separation $\vect{\beta} \equiv \vect{b}/D_l = \vect{\theta}_l -  \vect{\theta}_s$  between celestial positions of the lens and source, $\vect{\theta}_l$ and $\vect{\theta}_s$, respectively. 
In Eq.~\eqref{eq:mureal} above, we have ignored the $(1-{D_l}/{D_s})$ geometric factor in the limit of large source distance $D_s$ relative to the line-of-sight distance $D_l$ to the lens. Equation~\eqref{eq:mureal} represents a dipole-like pattern centered at the lens position. This can also be seen in the top row of Fig.~\ref{fig:population_maps}, which shows a map of induced velocities across the full sky induced by a realization of simulated CDM subhalos (described in Sec.~\ref{sec:cdmpop}).

The induced acceleration can be calculated similarly by taking an additional derivative of Eq.~\eqref{eq:mureal} (see V18 for details). This results in a quadrupole-like pattern centered on the lens position, which can be seen in the bottom row of Fig.~\ref{fig:population_maps} for the same lens population as in the top panel. Induced accelerations are suppressed by characteristic factors of $\sim v_l/b$ compared to induced velocities. A key feature of induced accelerations compared to velocities can be seen in Fig.~\ref{fig:population_maps}---while the velocity signal is dominated by the heaviest objects, the acceleration signal is democratically sensitive to structure at all scales, including populations of dense, low-mass subhalos contributing at smaller angular scales. This will be explored in detail in Sec.~\ref{sec:examples} and App.~\ref{app:significance} below.

\subsubsection*{Vector spherical harmonic decomposition}

The power spectrum decomposition of vector fields on a sphere, in our case the measured proper motions and proper accelerations of celestial objects, relies on the vector spherical harmonic (VSH) decomposition. This is an extension of the scalar spherical harmonic decomposition ubiquitous in astrophysics and cosmology, and within astronomy has previously been applied to astrometric datasets for studying systematics biases in and calibration of celestial reference frames~\cite{Mignard:2012xm,Undefined:2018amf,2018A&A...609A..19L}. We briefly outline the basic formalism here; for further details see, \emph{e.g.}, Refs.~\cite{arfken2013mathematical,Kostelec:2000:CHA:351065.351076}.

Simply, the VSH decomposition amounts to decomposing a vector field into a \emph{curl-free} component (also known as poloidal), and a \emph{divergence-free} component  (also known as toroidal). More precisely, a given vector field $\vect \mu = \vect \mu(\vect \theta)$ on a sphere admits a multipole expansion
\begin{align}
\vect{\mu} = \sum_{\ell m} \mu^{(1)}_{\ell m} \vect{\Psi}_{\ell m} + \mu^{(2)}_{\ell m} \vect{\Phi}_{\ell m},
\label{eq:vsh_decomposition}
\end{align}
with poloidal ($\vect{\Psi}$) and toroidal ($\vect{\Phi}$) amplitudes 
\begin{align}
\mu^{(1)}_{\ell m} =  \int \dd \Omega \, \vect{\mu} \cdot \vect{\Psi}^*_{\ell m};\quad
\mu^{(2)}_{\ell m} =  \int \dd \Omega \, \vect{\mu} \cdot \vect{\Phi}^*_{\ell m},
\label{eq:harmdecomposition}
\end{align}
where the vector spherical harmonics are defined in terms of the spherical harmonics $Y_{\ell m}$ as
\begin{align}
\vect{\Psi}_{\ell m} = \frac{\vect{\nabla}_{\vect{\theta}} Y_{\ell m}}{\sqrt{l(l+1)}}; \quad \vect{\Phi}_{\ell m} = \hat{\vect{r}} \times 
\vect{\Psi}_{\ell m} \label{eq:PsiPhidef}
\end{align}
in spherical coordinates $\lbrace r, \theta, \phi \rbrace$ with $\theta$ and $\phi$ Galactic colatitude and longitude respectively (sometimes combined in the 2D vector $\vect{\theta} = \lbrace \theta, \phi \rbrace$ with corresponding angular gradient $\vect{\nabla}_{\vect{\theta}}$), and $\vect{r}$ the radial line-of-sight vector. The above VSH are normalized such that they are orthonormal $\int \dd \Omega \, \vect{V}_{\ell m} \cdot \vect{V}^*_{\ell' m'}  = \delta_{\vect{V}'\vect{V}} \delta_{\ell' \ell} \delta_{m' m}$ with $\vect{V} = \{\vect{\Psi},\vect{\Phi}\}$, and form a complete basis for a vector field on the celestial sphere. The power per mode for each component can then be obtained as usual by averaging over the azimuthal modes: 
\begin{equation}
C_{\ell}^{\mu} \equiv \frac{1}{2\ell + 1} \sum_{m = -\ell}^{\ell} \left| \mu_{\ell m} \right|^2.
\end{equation}
As we stated above, Eq.~\eqref{eq:vsh_decomposition} corresponds physically to decomposing a vector field into a curl-free component (poloidal), which can be written as the gradient of a sourcing scalar potential, and a divergence-free component (toroidal), which can be written as the curl of a sourcing vector potential. 

Application of the vector spherical harmonic decomposition formalism to lens-induced observables is straightforward. Since the lensing deflection can be written as the gradient of an effective projected (scalar) lensing potential $\psi$, $\Delta\theta \sim \vect{\nabla}_{\vect{\theta}} \psi$, it follows that the angular deflection field and, in fact, \emph{all} lensing observables, only have overlap with poloidal power spectrum modes. This can be seen explicitly for the lens-induced angular velocity correction of Eq.~\eqref{eq:mureal}, which can be written as
\begin{align}
\vect{\mu} =  \frac{\dd}{\dd t} \vect{\nabla}_{\vect{\theta}} \psi = - \frac{1}{D_l} \vect{\nabla}_{\vect{\theta}} \left(\vect{v} \cdot \vect{\nabla}_{\vect{\theta}} \psi \right) \label{eq:deflectionpotential}.
\end{align} 
We therefore immediately have that $\mu_{\ell m}^{(2)}$, and the corresponding power per mode $C_{\ell}^{\mu (2)}$, are identically zero after integrating by parts in Eq.~\eqref{eq:harmdecomposition} and noting that $\vect{\nabla}_{\vect{\theta}} \cdot \vect{\Phi}_{\ell m} = 0$.

The proper motion power per lens can be calculated once the lens properties (effective transverse velocity, enclosed mass function, and angular diameter distance) are specified (see App.~\ref{app:derivations} for derivations and details). The per-lens expected power is given by
\begin{align}
C_{\ell}^{\mu (1)} &\equiv \frac{1}{2\ell + 1} \sum_{m = -\ell}^{\ell} \left| \mu_{\ell m}^{(1)} \right|^2   \label{eq:pspec_mu}\\
&\simeq \sum_l \left(\frac{4 \GN v_l}{D_l^2}\right)^2 \frac{\pi}{2} \ell^2 \left[\int_0^\infty \dd \beta M(\beta D_l) J_1(\ell \beta) \right]^2, \nonumber
\end{align}
where the sum over $l$ is a sum over different lenses, appropriately weighed, which can in general have different $D_l$, $v_l$, and enclosed mass functions $M(b)$.

The proper acceleration power spectrum can be computed similarly and is compactly expressed in terms of the proper motion power spectrum (see App.~\ref{app:derivations} for derivation): 
\begin{align}
C_{\ell}^{\alpha (1)} &\equiv \frac{1}{2\ell + 1} \sum_{m = -\ell}^{\ell} \left| \alpha_{\ell m}^{(1)} \right|^2 = \frac{3}{4} \sum_l \frac{\ell^2 v_l^2}{D_l^2} C_{\ell,l}^{\mu (1)},
\label{eq:pspec_alpha}
\end{align}
where $C_{\ell,l}^{\mu (1)}$ is the proper motion power spectrum per lens, given by the term within the $l$-summation in Eq.~\eqref{eq:pspec_mu}. Equation~\eqref{eq:pspec_alpha} represents the contribution from an isotropically-distributed population of lenses. Just like for the proper motion power spectrum, the total acceleration power spectrum can be obtained as the sum over all the individual lens contributions, appropriately weighted.

In practice, the astrometric vector field of interest will be nonuniformly sampled with variable noise over a subset of the celestial sphere, and a straightforward application of Eq.~\eqref{eq:harmdecomposition} will introduce biases. An unbiased estimator is required (see Refs.~\cite{2019arXiv190800041L,Dahlen:2007sv,Kostelec:2000:CHA:351065.351076} for examples), and we describe a simple quadratic maximum-likelihood estimator for computing vector spherical harmonic coefficients in App.~\ref{app:estimator}. In Sec.~\ref{sec:gaia-quasars}, we use this estimator to compute the VSH decomposition of quasar proper motions from \emph{Gaia}'s second data release (DR2)~\cite{Prusti:2016bjo,Brown:2018dum} as a proof-of-principle application.

\begin{figure*}[htbp]
\centering
\includegraphics[width=0.45\textwidth]{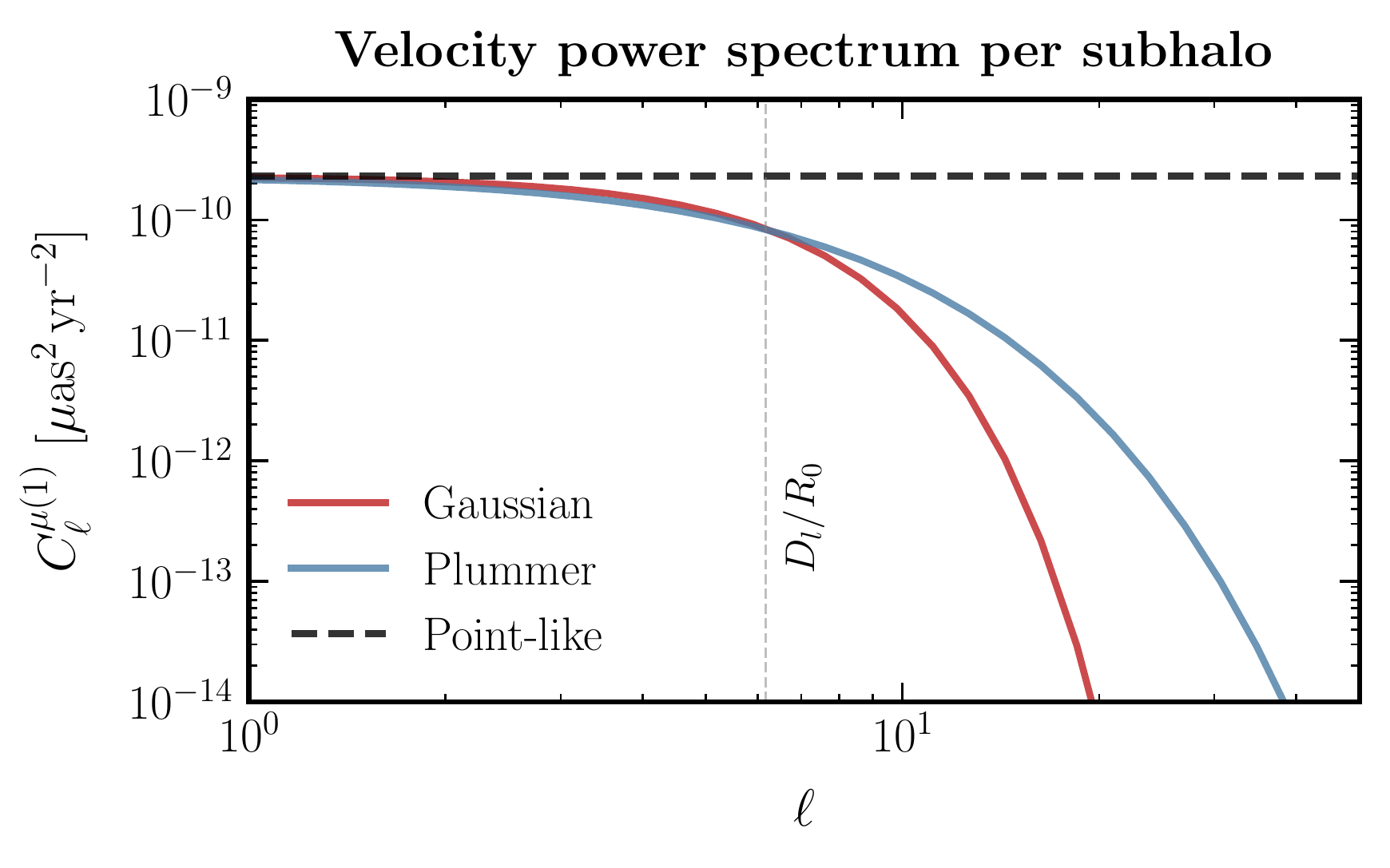}
\includegraphics[width=0.45\textwidth]{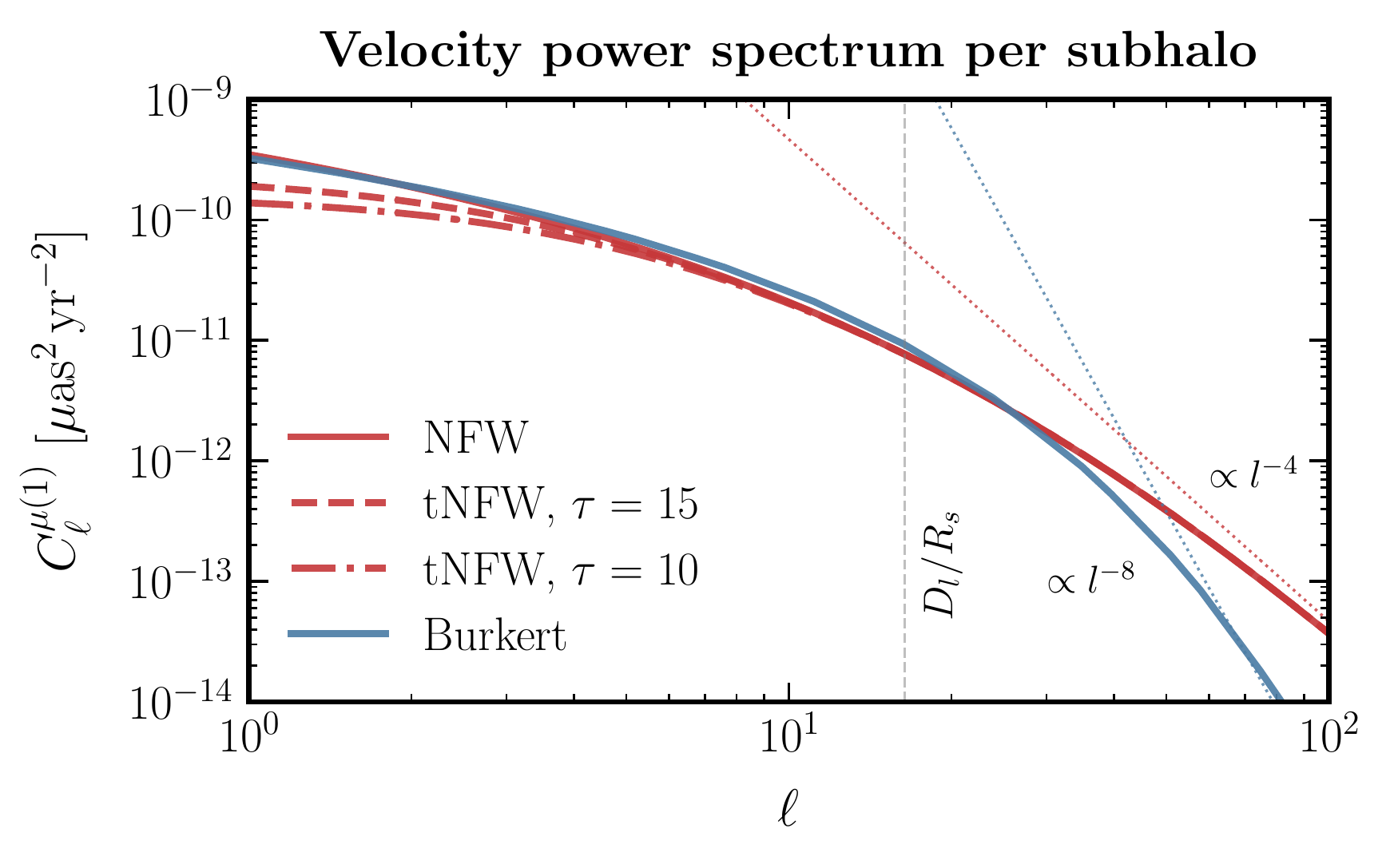}
\includegraphics[width=0.45\textwidth]{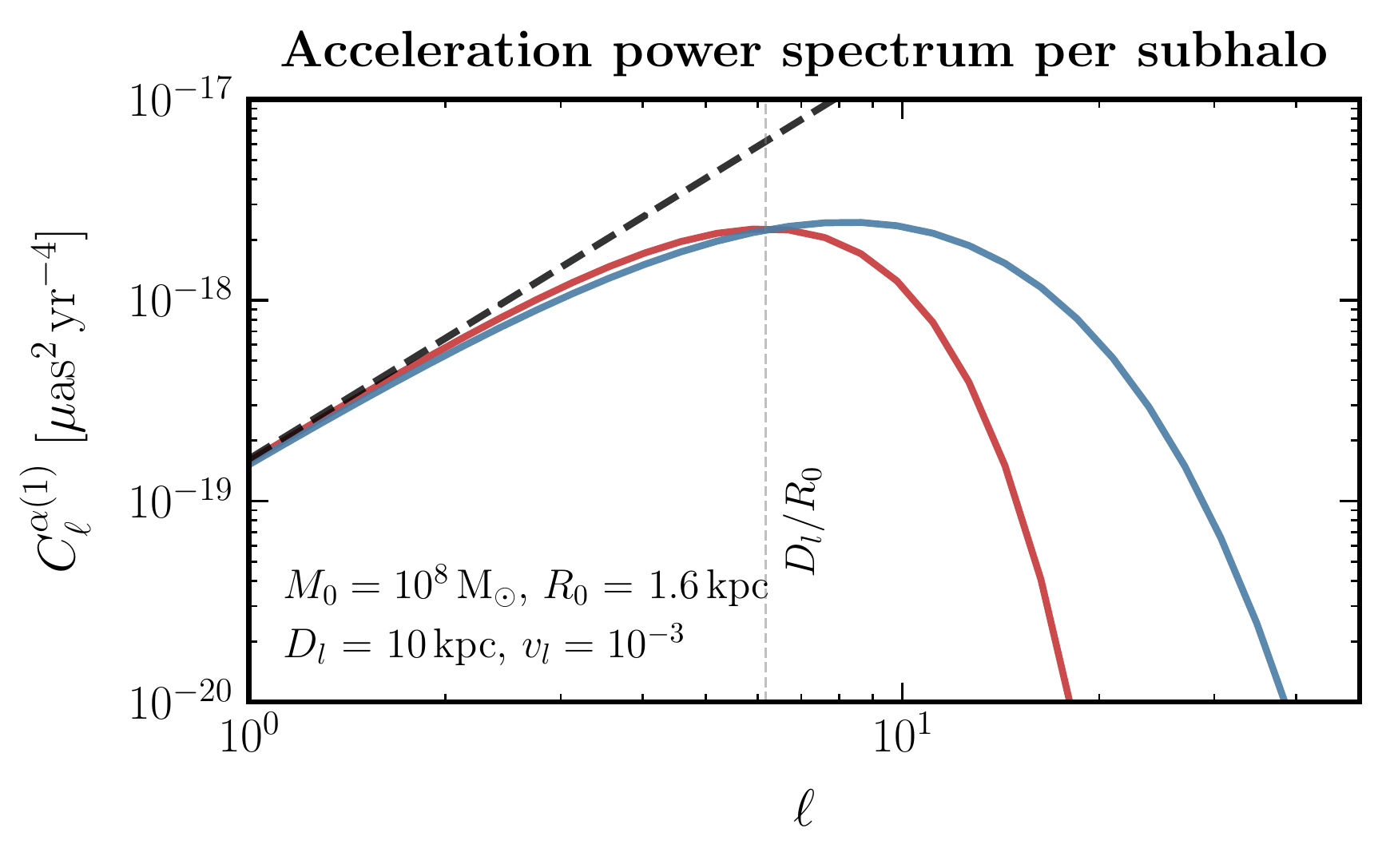}
\includegraphics[width=0.45\textwidth]{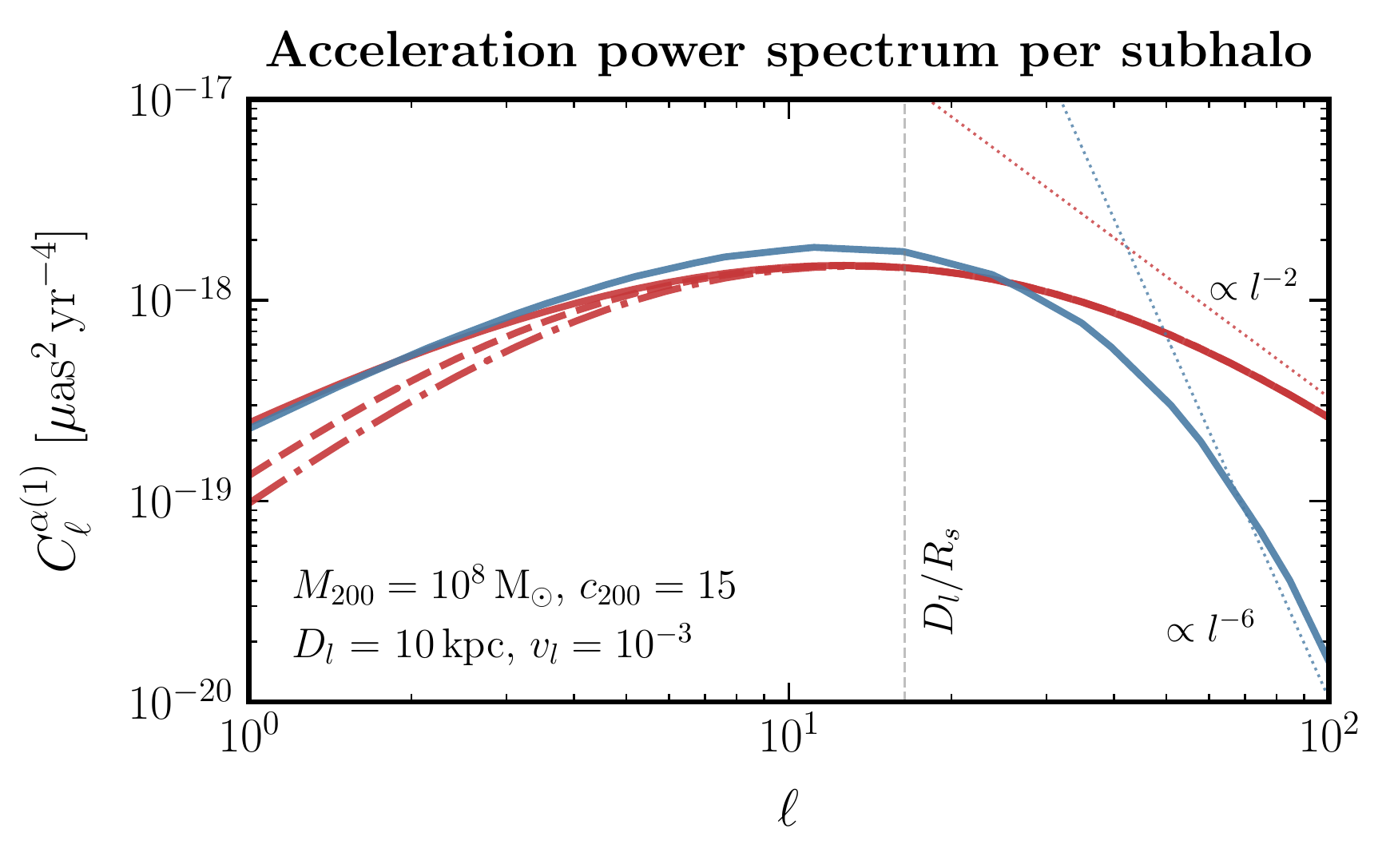}
\caption{Expected lens-induced proper motion (top row) and proper acceleration (bottom row) power spectra per subhalo for a homogeneous subhalo population. Shown for lenses with Gaussian and Plummer profiles (left column, red and blue lines respectively) of mass $M_0 = 10^{8}$\,M$_\odot$, size $R_0=1.6$\,kpc, transverse velocity  $v_l=10^{-3}$ at a distance $D_l=10$\,kpc from us. Also shown for NFW, two different truncated NFW ($\tau \equiv r_\mathrm{t}/r_\mathrm{s} = 10, 15$), and Burkert profile lenses (right column, solid, dashed, dot-dashed red and solid blue lines, respectively) with virial mass $M_{200}=10^{8}$\,M$_\odot$, concentration $c_{200}=15$, transverse velocity $v_l=10^{-3}$ at a distance $D_l=10$\,kpc from us. Asymptotic behavior at high $\ell$ for the NFW and Burkert profiles is illustrated with the thin dotted lines in the right column. \nblink{02_power_per_subhalo}} 
\label{fig:single_sub}
\end{figure*}

\subsection{Specific examples}
\label{sec:examples}

Once the lens properties have been specified, the lens-induced proper motion and proper acceleration signal power spectra can readily be calculated from Eqs.~\eqref{eq:pspec_mu} and~\eqref{eq:pspec_alpha}. We illustrate this for a few specific cases, starting from simple scenarios and going on to progressively more realistic ones in order to gain some intuition for what the signal looks like in angular space. We defer a more detailed discussion of how the signal and its significance depend on properties of the subhalo population, as well as which angular scales contribute to the signal in various cases, to App.~\ref{app:significance}. Unless otherwise specified, all spectra refer to the respective poloidal components $C_{\ell}^{\mu/\alpha (1)}$, with the toroidal signal components $C_{\ell}^{\mu/\alpha (2)}$ identically vanishing (\emph{cf.} Eq.~\eqref{eq:deflectionpotential}).

In contrast to the signal power spectra $C_{\ell}^{\mu/\alpha}$, the noise power spectra $N_{\ell}^{\mu/\alpha}$ are approximately scale-invariant and are given by
\begin{equation}
N_{\ell}^{\mu/\alpha} \simeq \frac{\sigma_{\mu/\alpha}^2}{\Sigma_q} \label{eq:noise_powspec}
\end{equation}
where $\sigma_{\mu/\alpha}$ is the typical measurement error on the proper motions or accelerations and $\Sigma_q \equiv N_q / 4\pi $ is the angular density of background celestial objects ($N_q$ is the all-sky equivalent number of sources), assumed here to be uniformly distributed over the sky.

\subsubsection{Population of point lenses}

We start by considering a population of points lenses of mass $M_0$ located at a constant distance $D_l$ from us, moving with random transverse velocity with magnitude $v_l$. In this case, the azimuthally-averaged power per $\ell$ mode per lens is given by (Eq.~\eqref{eq:pspec_mu})
\begin{equation}
C_{\ell}^{\mu (1)} \simeq \left(\frac{4 \GN M_0 v_l}{D_l^2}\right)^2 \frac{\pi}{2}.
\end{equation}
This scale-invariant spectrum is shown on the top left plot of Fig.~\ref{fig:single_sub} as the dotted black line, for the lens population properties specified in the inset.

For a population of point lenses uniformly distributed between $D_l^{\mathrm{min}}$ and $D_l^{\mathrm{max}}$ and making up a fraction $f_\mathrm{DM}$ of the local dark matter density $\rho_\mathrm{DM}$, we instead have
\begin{equation}
C_{\ell}^{\mu (1)} \simeq \frac{32 \pi^{2} G_{\mathrm{N}}^{2} M_{0} v_l^{2} \rho_{\mathrm{DM}} f_\mathrm{DM} \left(D_{l}^{\mathrm{max}}-D_{l}^{\mathrm{min}}\right)}{D_{l}^{\mathrm{max}} D_{l}^{\mathrm{min}}}.
\label{eq:mu_pspop}
\end{equation}
For a point-lens population, the proper motion power per $\ell$ mode is scale-invariant, with higher modes having greater mode multiplicity. 

The acceleration power spectrum can be calculated similarly using Eq.~\eqref{eq:pspec_alpha} and shown in the bottom left plot of Fig.~\ref{fig:single_sub} as the dotted black line for the lens properties in the inset text. Unlike the velocity power spectrum, it is not scale invariant, and grows as $C_\ell^\alpha \appropto \ell^2$. 

\subsubsection{Population of extended lenses}
\label{sec:extended_pop}

In order to motivate the study of spatially extended subhalos, we consider a population of Gaussian lenses with density profile 
\begin{equation}
\rho(r) =\frac{M_0}{2\sqrt{2}\pi^{3/2}R_0^3} e^{-r^2/2R_0^2} 
\label{eq:Gaussianrho}
\end{equation} 
where $M_0$ is the total lens mass and $R_0$ its characteristic size. The integral in Eq.~\eqref{eq:pspec_mu} can be carried out analytically, yielding
\begin{equation}
C_{\ell}^{\mu (1)} \simeq \left(\frac{4 \GN M_0 v_l}{D_l^2}\right)^2 \frac{\pi}{2} e^{-\ell^2\beta_0^2}
\label{eq:mu_ext}
\end{equation}
where $D_l$ is the distance to the lens and $\beta_0 \equiv R_0/D_l$ its characteristic angular scale. The power spectrum per lens is shown in the top left plot in Fig.~\ref{fig:single_sub} as the red line and has an exponential suppression at characteristic scale $\ell \sim D_l/R_0$.

This gives us some intuition for the more realistic case of a population of such lenses distributed between $D_l^{\mathrm{min}}$ and $D_l^{\mathrm{max}}$. In this case, we have the azimuthally-averaged power per $\ell$ mode 
\begin{align}
C_{\ell}^{\mu (1)} \simeq 16 \pi ^{5/2} &G_\mathrm{N}^2 M_0 v_l^2 \rho_\mathrm{DM} f_\mathrm{DM}\times\nonumber\\ &\left[\frac{{\mathrm{erf}}\left(\frac{\ell
{R_0}}{D_l^{\mathrm{min}}}\right)-{\mathrm{erf}}\left(\frac{\ell {R_0}}{{D_l^{\mathrm{max}}}}\right)}{\ell {R_0}}\right].
\label{eq:mu_extpop}
\end{align}
We see that the power decreases as $\propto 1/\ell$ up to $\ell \sim D_l^\mathrm{max}/R_0$ and falls off steeply beyond that point. Combined with the higher multiplicity of modes at high $\ell$, this implies that each \emph{logarithmic bin} in $\ell$ (\emph{e.g.}, each $e$-fold or each decade) up to $\ell \sim D_{l}^\mathrm{max} / R_0$ contributes equally to the overall signal significance, as we shall see in App.~\ref{app:significance}.

A similar story holds for proper accelerations. The power spectrum per lens can be calculated from Eq.~\eqref{eq:pspec_alpha} and is shown as the red line in the bottom left of Fig.~\ref{fig:single_sub}. As in the case of point lenses, the per-lens signal grows with multipole as $C_\ell^\alpha \appropto \ell^2$, but now with an exponential suppression at $\ell \sim D_l/R_0$. The total power power mode in accelerations for a uniformly distributed population of lenses is given by (setting $D_{l}^{\mathrm{min}} = 0$ for simplicity)
\begin{align}
C_{\ell}^{\alpha (1)} \simeq & \, 6 \pi ^2  G_\mathrm{N}^2  M_0 v_l^4 \rho_\mathrm{DM} f_\mathrm{DM} \frac{1}{\ell R_0^3} \times\nonumber  \\ & \left[\frac{2 \ell R_0}{D_l^{\mathrm{max}}} e^{-({\ell
R_0}/{D_{l}^{\mathrm{max}}})^2}+\sqrt{\pi } \text{erfc}\left(\frac{\ell R_0}{D_{l}^{\mathrm{max}}}\right)\right].
\label{eq:alpha_extpop}
\end{align}
Just as for proper motions, the acceleration power spectrum decreases as $\propto 1/\ell$ up to $\ell \sim D_l^\mathrm{max}/R_0$, beyond which it is exponentially suppressed. This again implies that each logarithmic bin in $\ell$, or equivalently each logarithmic bin in line-of-sight distance $D_l$, contributes equally to the signal significance (\emph{cf.}, App.~\ref{app:significance}).

For illustration, we also show in the left column of Fig.~\ref{fig:single_sub} the velocity and acceleration power spectra per lens for lenses described as Plummer spheres~\cite{Plummer:1911zza} (blue lines), commonly used in the literature as an analytically tractable subhalo profile closer to a realistic subhalo descriptions than the Gaussian lens. Here, the density is given $\rho(r)= 3M_0/(4\pi R_0^3)(1 + r^2/R_0^2)^{-5/2}$ and proper motion power per $\ell$ mode $C_{\ell}^{\mu (1)} \simeq \left({4 \GN M_0 v_l/D_l^2}\right)^2 \frac{\pi}{2} \ell^2\beta_0^2 k_1(\ell \beta_0)^2$, where $k_1$ is the first-order modified Bessel function of the second kind. Properties and overall scalings similar to those of Gaussian lenses are observed in this case.

\subsubsection{Realistic subhalo profiles}

Realistic subhalo density profiles are modeled with input from $N$-body simulations. We consider two different profiles: a (truncated) Navarro-Frenk-White (NFW) profile as expected for standard CDM halos~\cite{Navarro:1995iw,2008gady.book.....B}, and a cored Burkert profile favored, \emph{e.g.}, in the case of self-interacting dark matter (SIDM) halos~\cite{Burkert:1995yz}. The truncated NFW profile is parameterized as~\cite{Baltz:2007vq}
\begin{equation}
\rho_\text{tNFW}(r)=\frac{M_\mathrm{s}}{4\pi r(r + r_\mathrm{s})^2}\left(\frac{r_\mathrm{t}^2}{r^2 + r_\mathrm{t}^2}\right)\,,
\end{equation}
where $M_\mathrm{s} = 4\pi \mathrm{s}^3\rho_\mathrm{s}$ is the NFW scale mass and $r_\mathrm{t}$ is the truncation radius accounting for the stripping away of the outer halo mass due to tidal forces towards the Galactic center. The Burkert profile is parameterized as
\begin{equation}
\rho_{\rm Burkert}(r) = \frac{M_{\mathrm B}}{4\pi(r+r_{\mathrm B})(r^2 + r_{\mathrm B}^2)}\,,
\end{equation}
where $M_{\mathrm B} = 4\pi r_{\mathrm B}^3\rho_{\mathrm B}$ is the Burkert scale mass and  the Burkert scale radius $r_{\mathrm B}$ can be related to the NFW scale radius as $r_{\mathrm B} \simeq 0.7 r_\mathrm{s}$~\cite{Bartels:2015uba,Lisanti:2017qoz}.

We show induced power spectra for (truncated) NFW and Burkert profiles in the right column of Fig.~\ref{fig:single_sub}, for proper motions (top) and proper accelerations (bottom). The NFW truncation radius is parameterized through $\tau\equiv r_\mathrm{t}/r_\mathrm{s}$ and the concentration is taken to be $c_{200}=15$. The cases $\tau=10$ (dashed red) and $\tau=15$ (dot-dashed red) are shown for illustration as typical truncation scales. It can be seen that truncation effects generally manifest at larger scales, as expected. On the other hand, the presence of a core in the Burkert profile leads to a suppression of power at smaller scales compared to NFW subhalos (blue line). 
Asymptotic high-$\ell$ behavior is indicated, with the induced velocity power per $\ell$ mode scaling as $\propto \ell^{-4}$ and $\propto \ell^{-8}$ for the NFW and Burkert profiles respectively, and as $\propto \ell^{-2}$ and $\propto \ell^{-6}$ in the case of induced acceleration power. Note that these scalings are the same as those obtained in Ref.~\cite{Rivero:2017mao} for the case of substructure convergence power spectra in strong lensing systems. 

Armed with the proper motion velocity and acceleration power spectra for a homogeneous subhalo population, we are in a position to calculate the expected signal due to a Galactic substructure population. Since power adds stochastically, the total power spectra $C_\ell^{\mu/\alpha(1/2)}$ for a population of lenses with properties drawn from some distribution (\emph{e.g.,} those characterizing the mass function and spatial distribution of lenses) can be obtained as appropriately-weighed sums. Hence, to obtain the expected aggregate signal we can convolve the per-lens subhalo power spectrum, evaluated from the Earth's location, with the subhalo distribution in our Galaxy. In particular, for a Galactic subhalo population with Earth-frame dark matter velocity distribution $f_\oplus(\vect{v}_l, t)$ and number density $n_l(M, \mathbf r)$ we have 
\begin{equation}
C_\ell^{\mathrm{tot}}=\int \dd \vect{v}_l \,\dd \mathbf r \,\dd M_l \,f_\oplus(\vect{v}_l, t)\,n_l(M, \mathbf r)\,\,C_\ell(M_l, \mathbf{v}_l, D_l(\mathbf{r}))
\label{eqn:popcell}
\end{equation}
with line-of-sight distance $D_l^2 = |\vect{r}|^2 + R_\sun^2 +2|\vect{r}|R_\sun\cos\theta_\mathrm{Gal}$, where $\theta_\mathrm{Gal}$ is the angle between the subhalo and Galactic plane from the Galactic center. In our case, $n_l = \dd^2N/(\dd M_l \dd \mathbf r)$ depends on the assumed spatial distribution and mass function of subhalos.

\section{Source Targets and noise levels}
\label{sec:noise}

So far, we have been agnostic about the population of luminous background sources onto which correlations due to Galactic subhalos may imprint themselves. We describe here two such celestial populations for which precise astrometry will be available in the near future and where substructure lens-induced astrometric effects could be observed over intrinsic noise. We summarize our assumed noise configurations in Tab.~\ref{tab:noise_specs}.

\begin{table}[h]
\begin{center}
\begin{tabular}{cccccc} 
\hline\hline
Observation & $\ell$-range & $f_{\rm sky}$ & $\Sigma_q$\,[\,sr$^{-1}$] & $\sigma_\mathrm{eff}$ \\ 
\hline
SKA-like $\mu$ & $[10,\,5000]$ & $1.0$ & $10^7$ & $1\,\mu$as\,yr$^{-1}$\\
WFIRST-like $\alpha$ & $[50,\,5\times10^5]$ &  $0.05$ &  $10^{11}$  & $0.1\,\mu$as\,yr$^{-2}$ \\
\Gaia~$\alpha$ & $[50,\,5\times10^4]$ &  $0.05$ &  $5\times 10^{9}$  & $2\,\mu$as\,yr$^{-2}$\\
\hline
\end{tabular}
\end{center}
\caption{Assumed specifications---multipole range, sky fraction, full-sky equivalent source number, and effective astrometric precision---for future experiments providing measurements of extragalactic (quasar) proper motions, denoted $\mu$, and Galactic (stellar) proper accelerations, denoted $\alpha$. }
\label{tab:noise_specs}
\end{table}

\subsection{Extragalactic proper motions}

Galaxies outside of our own are numerous, and their motions are expected to be measured with unprecedented precision with future surveys. Ideal candidates for our purpose are quasi-stellar objects (QSOs), also known as quasars which, owing to their large distances from us, are expected to have small intrinsic proper motions. Known systematic effects (\emph{e.g.}, correlation induced by the drift of the Solar System's barycenter towards the Galactic center~\cite{Titov:2010zn,Titov:2013qk}) can be modeled and subtracted. It is expected that future very long baseline interferometry surveys (VLBI), in particular the Square Kilometer Array (SKA)~\cite{Fomalont:2004hr,Jarvis:2015tqa} will be able to measure the motions of $\sim 10^8$ quasars across the full sky with proper motion precision of $\sigma_\mu\sim1\,\mu$as\,yr$^{-1}$ (see V18 for details). We assumed these characteristics for an ``SKA-like'' survey for our sensitivity projections.

With this source density, multipoles up to $\ell_\mathrm{max}\sim10^4$ should be accessible, with measurements on larger scales dominated by shot noise. We assume the multipole range $\ell\in[10,\,5000]$, discarding smaller multipoles due to sample variance and larger multipoles due to potential small-scale systematic effects.

\subsection{Galactic proper accelerations}

Compared to quasars, the stellar population within the Milky Way is characterized by a much higher source density. The prospect of using proper motion correlations for our purposes is limited by the large intrinsic velocity dispersion of the stars. The use of acceleration correlations, however, shows promise. Current (\emph{e.g.}, \Gaia~\cite{Prusti:2016bjo}) and future (\emph{e.g.}, WFIRST~\cite{2017arXiv171205420S}) optical surveys are expected to map out the motions of a sizeable fraction of all stars in the Milky Way, corresponding to angular densities of sources listed in Tab.~\ref{tab:noise_specs} over a limited region of the sky $f_\mathrm{sky} = 0.05$ described by the Galactic disk and bulge, with unprecedented precision. Averaged over the stellar population, proper acceleration precision of $\sigma_\alpha=2(0.1)\,\mu$as\,yr$^{-2}$ could be achievable over an observation time of 10 years by \emph{Gaia}(a WFIRST-like survey)  (see V18 for further details). Large-scale correlations due to the Galactic gravitational potential can be modeled and subtracted, and the effect of small-scale systematics (\emph{e.g.}, correlated motions of unresolved binaries) can be modeled and marginalized over.

With the assumed source densities, smaller scales up to $\ell_\mathrm{max}\sim10^6$ should be accessible to a WFIRST-like survey, and scales up to $\ell_\mathrm{max}\sim10^5$ to \Gaia by its end of mission. In our fiducial set-up, we consider the multipole range $\ell\in[50,\,5\times10^5]$ for a WFIRST-like survey and $\ell\in[50,\,5\times10^4]$ for \Gaia. To account for the fact that only stars behind the Galactic lenses can be considered, we consider the subhalo population only within 1\,kpc of the Solar position, assuming that stars beyond this radius can be considered as background sources. 
It can be shown that acceleration observables are preferentially sensitive to the most nearby lenses compared to velocity power spectra (\emph{cf.} the additional $D_l^{-2}$ factor for acceleration power spectra in Eq.~\eqref{eq:pspec_alpha}), this is expected to capture a dominant portion of the signal while being conservative.
Increasing this cut to 2\,kpc has a negligible impact on our results.

\section{Sensitivity Forecasts}
\label{sec:forecasts}

We assess the sensitivity of global astrometric correlations in the context of several illustrative benchmark scenarios. In addition to the standard cold dark matter paradigm which predicts a broad mass spectrum of subhalos evolved from a nearly scale-invariant primordial spectrum of fluctuations, we also consider a scenario representative of enhanced power on small scales parameterized by a kink in the primordial power spectrum. Enhancement to the power spectrum over a limited range of scales may cause dense dark matter clumps of a characteristic mass and size to constitute a significant fraction of the overall dark matter abundance~\cite{Berezinsky:2013fxa}, which we explore in the context of compact dark objects. Finally, we consider the detectability of dynamical fluctuations due to interference effects in the case of ultralight scalar field dark matter. We note that these scenarios are motivated examples but do not exhaust the applicability of methods presented here.

\subsection{Cold dark matter}
\label{sec:cdmpop}

The cold dark matter paradigm has been extremely successful in explaining the distribution of structure at large scales, with theory and simulations additionally predicting a broad spectrum of subhalo masses down to sub-Galactic scales~\cite{Madau:2008fr,Springel:2008cc}. We study the sensitivity of our methods to a population of CDM subhalos, and start by describing the main ingredients in our CDM-inspired models.

\paragraph*{Subhalo mass function:} Numerical simulations show that the (sub)halo mass function in $\Lambda$CDM can be well described by a power-law distribution of the form $\dd N/\dd M\propto M^{-\gamma}$ with $\gamma\approx1.9$--$2$~\cite{Moline:2016pbm} over a large range of masses. We set $\gamma=1.9$ in our fiducial configuration, consistent with simulations of Milky Way-sized halos~\cite{Madau:2008fr,Springel:2008cc}. We also investigate a steeper mass function with $\gamma=2$, leading to a larger relative abundance of lower-mass subhalos. To calibrate the amplitude of the subhalo mass function, we require $150$ subhalos in expectation between $10^8$--$10^{10}$\,M$_\odot$~\cite{Hutten:2016jko}, consistent with results from recent hydrodynamical simulations~\cite{Mollitor:2014ara,Sawala:2015cdf}. The threshold minimum and maximum allowed subhalo masses are fixed at $10^{-6}$\,M$_\odot$ and $0.05\,M_\mathrm{MW}$~\cite{Hiroshima:2018kfv} respectively. This configuration leads to $\sim20$\% of the total Milky Way mass bound in substructure. We also investigate a more subhalo-rich configuration, with $300$ subhalos in the $10^8$--$10^{10}$\,M$_\odot$ mass range, consistent with the results of DM-only simulations~\cite{Roos:2012cc}. Note that in all case, we do not take into account subsubstructure (\emph{i.e.}, subsubhalos within subhalos).

\paragraph*{Spatial distribution of subhalos:}  While the unevolved (infall) subhalo spatial distribution is expected to follow the smooth Milky Way halo profile, tidal disruption due to the gradient of the Galactic potential towards the Galactic center is expected to deplete the fraction of mass bound in substructures in this region. We account for this by describing the spatial distribution of subhalos using an Einasto profile with a fit to the results of the Aquarius simulation~\cite{Roos:2012cc,Hutten:2016jko}, 

\begin{equation}
\rho(r) = \exp\left\lbrace-\frac{2}{\gamma_\mathrm{E}}\left[\left(\frac{r}{r_\mathrm{E}}\right)^{\gamma_\mathrm{E}} - 1\right]\right\rbrace
\end{equation}
with $r_\mathrm{E}=199$\,kpc and $\gamma_\mathrm{E}=0.678$.

There are indications that some portion of the subhalo tidal disruption effects observed in simulations could be numerical in origin~\cite{vandenBosch:2017ynq,vandenBosch:2018tyt}. We account for this possibility by investigating an alternative scenario where the evolved distribution of subhalos traces the smooth Galactic dark matter profile.

\paragraph*{Subhalo profile:} We model the subhalos with an NFW profile, using the Galactocentric distance-dependent concentration-mass relation from Ref.~\cite{Moline:2016pbm} in our fiducial set-up which takes into account the larger concentration of subhalos as compared to field halos closer to the Galactic center due to tidal disruption effects. We explore the dependence of the power spectrum signal on the concentration-mass parameterization by using the alternate model from Ref.~\cite{Correa:2015dva}, which does not take into account Galactocentric distance-dependent tidal effects.

\paragraph*{Dark matter velocity distribution:} In the Galactic frame and asymptotically far away from the Sun's gravitational potential, we take the velocity distribution of dark matter $f_\infty (\vect{v})$ to be given by the Standard Halo Model (SHM),
\begin{equation}{
f_{\infty} (\vect{v}) = \left\{ \begin{array}{ll}
N \left( {\frac{1}{\pi v_0^2}} \right)^{3/2} e^{- \vect{v}^2 / v_0^2 } \qquad &|\vect{v}| < \vesc \\
0 \, \qquad &\text{otherwise} \,,
\end{array}
\right.
} \label{eq:fvinfty}
\end{equation}
where $N$ is a normalization factor, and we take $v_{0}=220$~km\,s$^{-1}$~\cite{Kerr:1986hz} and the escape velocity $\vesc=550$~km\,s$^{-1}$~\cite{Piffl:2013mla}.

To a first approximation, the velocity distribution at the Earth's location may be found simply by applying a Galilean transformation to $f_\infty (\vect{v})$ from the Galactic frame to the lab frame, so that    
\begin{equation}
f_\oplus (\vect{v}) \approx f_{\infty} \left( \vect{v} + \vect{v}_\sun \right) \,, \label{eq:fvoplus}
\end{equation}
where $\vect{v_\sun} = (11, 232, 7)$ km\,s$^{-1}$~\cite{Schoenrich:2009bx} is the velocity of the Sun in Galactic coordinates. Note that the Earth-frame velocity acquires a time-dependence $f_\oplus (\vect{v},t)$ due to the motion of the Earth around the Sun and leads to a fractional annual modulation in the signal. We conservatively ignore this effect here and postpone its study to future work. We use this velocity distribution in the rest of the scenarios presented in this paper. Additionally, in practice we impose a lower cutoff on the line-of-sight integration in Eq.~\eqref{eqn:popcell} corresponding to the distance within which a single subhalo is expected, for a given set-up, in order to mitigate the effect of Poisson noise in the limit of a small number of lenses.

\begin{figure}[htbp]
\centering
\includegraphics[width=0.45\textwidth]{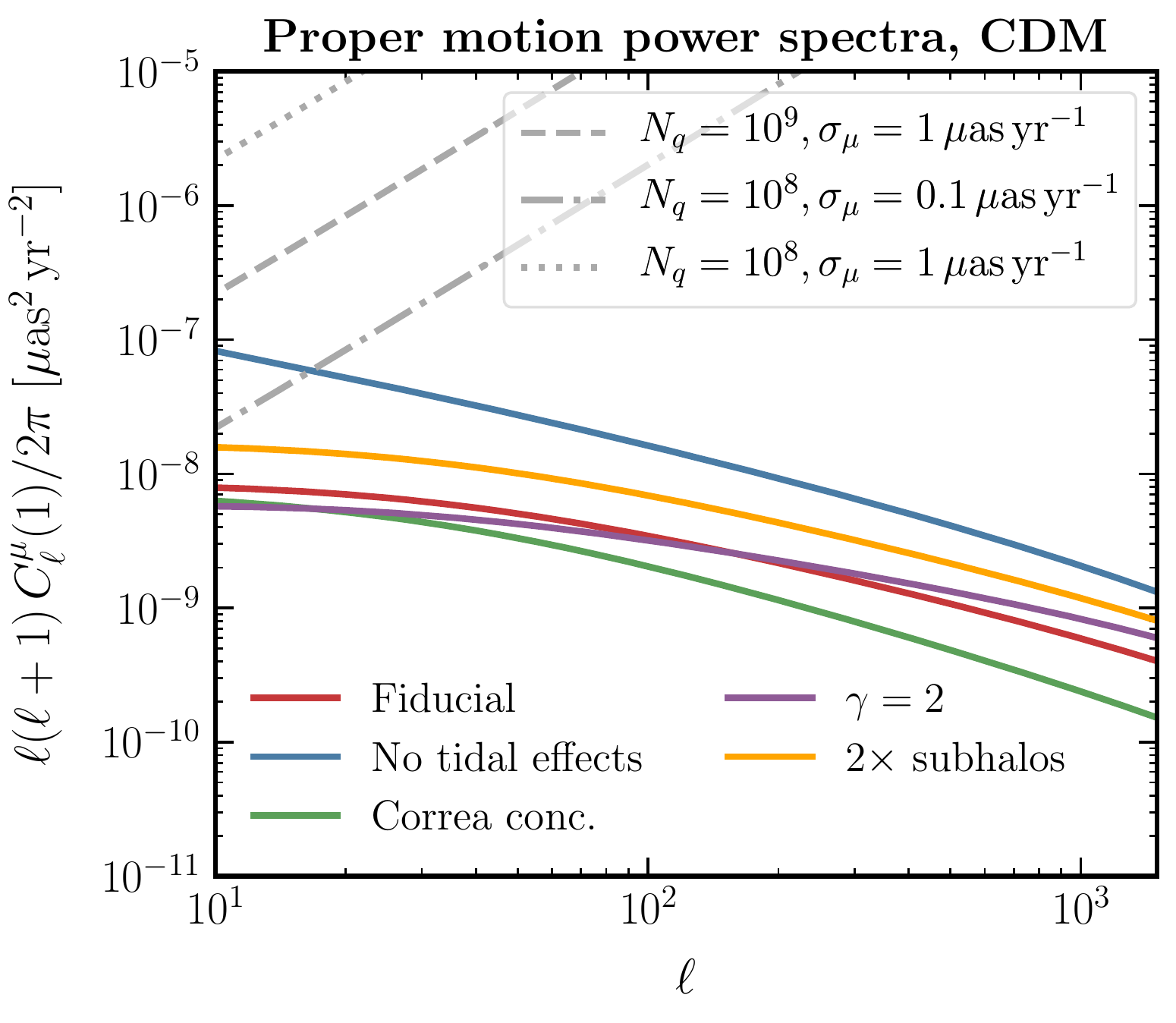}
\caption{The total signal power spectrum expected for various CDM-inspired subhalo configuration described in Sec.~\ref{sec:cdmpop}. The fiducial configuration is shown in red. A signal without accounting for effects of tidal disruption is shown in blue. The effect of using an alternative concentration model from Ref.~\cite{Correa:2015dva} which does not accounting for Galactocentric distance-dependent effects is shown in green. A steeper subhalos mass function (slope $\gamma = 2$ instead of $1.9$) is shown in purple. A  more subhalo-rich configuration with twice the number of subhalos compared to the fiducial configuration is shown in orange. The grey dashed, dot-dashed, and dotted lines correspond to different noise spectra with number of background sources $N_q$ and effect astrometric precision $\sigma_\mu$, \{$N_q , \sigma_\mu$\} =  \{$10^9$, $1\,\mu$as\,yr$^{-1}$\}, \{$10^8$, $0.1\,\mu$as\,yr$^{-1}$\}, and \{$10^8$, $1\,\mu$as\,yr$^{-1}$\} respectively. \nblink{05_cdm}} 
\label{fig:lcdm_theory}
\end{figure}

The total poloidal induced proper motion power spectrum signal expected for the fiducial configuration, as well as the alternate modeled scenarios, is shown in Fig.~\ref{fig:lcdm_theory}. The fiducial CDM model is shown as the red line. Not accounting for tidal disruption~\cite{vandenBosch:2017ynq,vandenBosch:2018tyt} preferentially brings subhalos closer to the Galactic center and boosts the signal by about an order of magnitude at all scales (blue line). A steeper subhalo mass function with $\gamma=2$ (purple line) results in a larger number of low-mass subhalos, slightly boosting the signal on small scales and depressing it on larger scales. A more subhalo-rich configuration with twice the number of lower-mass subhalos as compared to our fiducial model (but still consistent with observations on large scales) is shown as the orange line, with the signal boosted by a factor of two. Using the alternate concentration model from Ref.~\cite{Correa:2015dva} depresses the overall signal (green line) as it doesn't account for the increased concentration of subhalos closer to the Solar position due to tidal stripping effects.

\begin{figure*}[!htbp]
\centering
\includegraphics[width=0.45\textwidth]{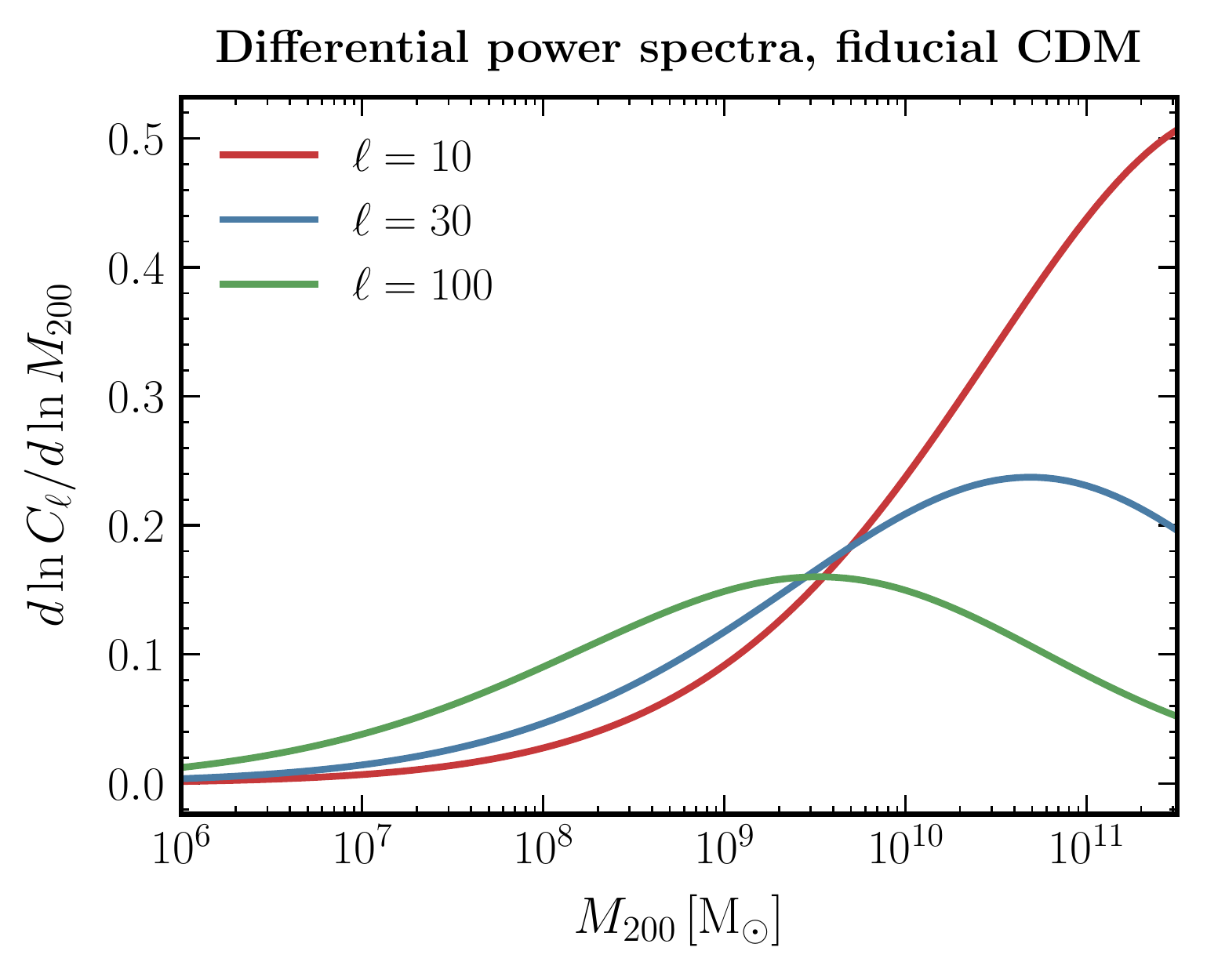}
\includegraphics[width=0.45\textwidth]{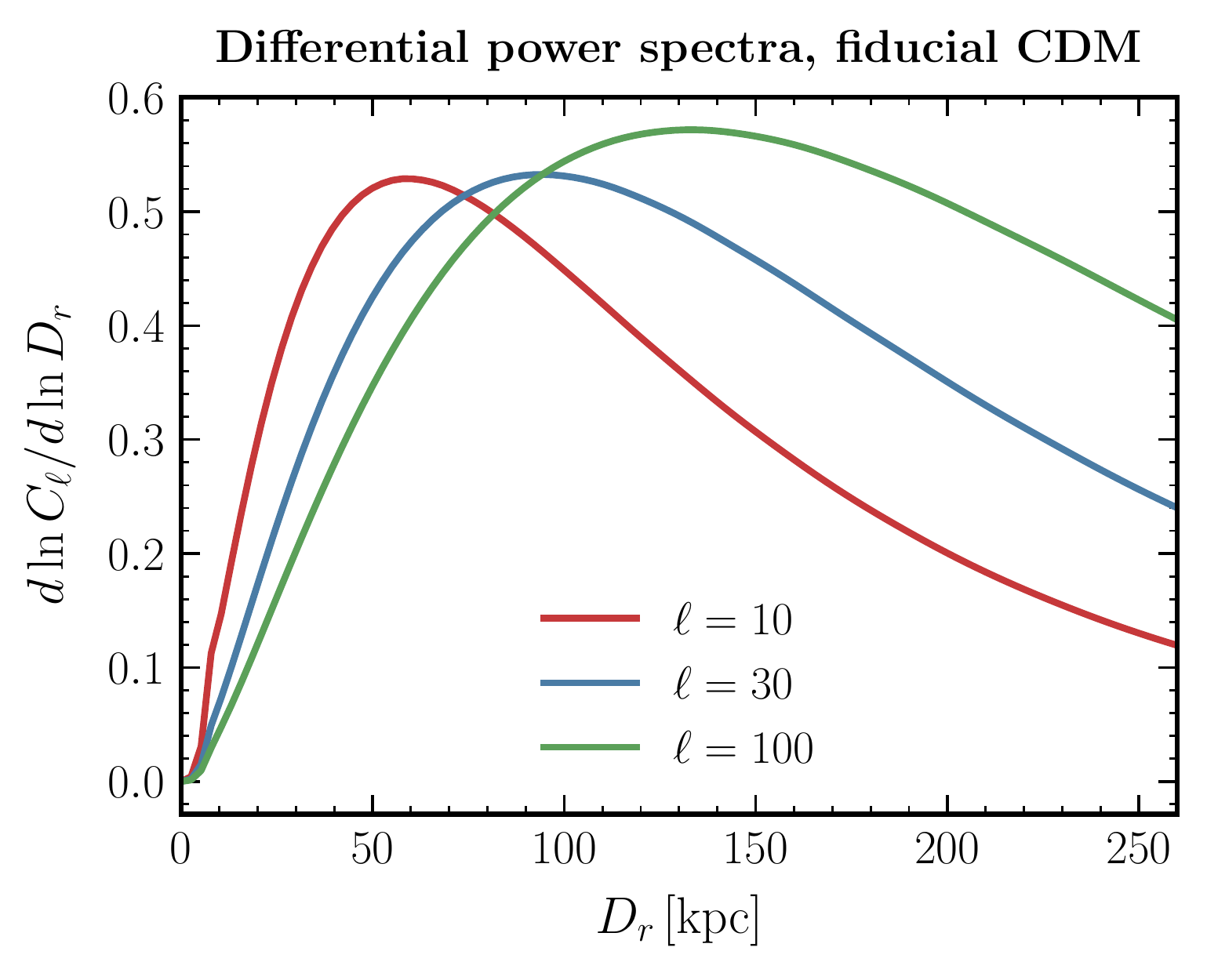}
\caption{\emph{(Left)} The differential proper motion power spectrum for the fiducial CDM configuration as a function of virial subhalo mass at multipoles $\ell=10$ (red), $\ell=30$ (blue), and $\ell=100$ (green). Larger multipoles (smaller scales) are preferentially sensitive to less massive subhalos; overall, the signal is still dominated by more massive CDM subhalos. \emph{(Right)} The differential proper motion power spectrum for the fiducial CDM configuration as a function of Galactocentric distance of the subhalo  at multipoles $\ell=10$ (red), $\ell=30$ (blue), and $\ell=100$ (green) Fractionally, most of the sensitivity comes from subhalos in the bulk of the Milky Way rather than from those close to the Solar position. \nblink{05_cdm}} \label{fig:pspec_differential}
\end{figure*}

It is instructive to ask which regions of the subhalo mass and spatial distribution phase space contribute to the total power spectrum signal. The differential spectra $\dd\ln C_\ell/\dd\ln M_{200}$ and $\dd\ln C_\ell/\dd\ln R$ are shown in Fig.~\ref{fig:pspec_differential} for multipoles $\ell=10, 30, 100$. It can be seen that larger scales receive preferential contribution from subhalos that are massive and/or closer in Galactocentric radius, as  expected. It can also been seen that, at accessible scales, the dominant contribution comes from the population of subhalos at intermediate Galactocentric radii ($R\sim50$--$150$\,kpc). This underscores the fact that the lensing signal is derived in aggregate from a \emph{population} of subhalos, and thus that the power spectrum measurement is probing the substructure population in the bulk Galactic halo rather than being sensitive to individual, nearest subhalos.

\begin{figure*}[!htbp]
\centering
\includegraphics[width=0.45\textwidth]{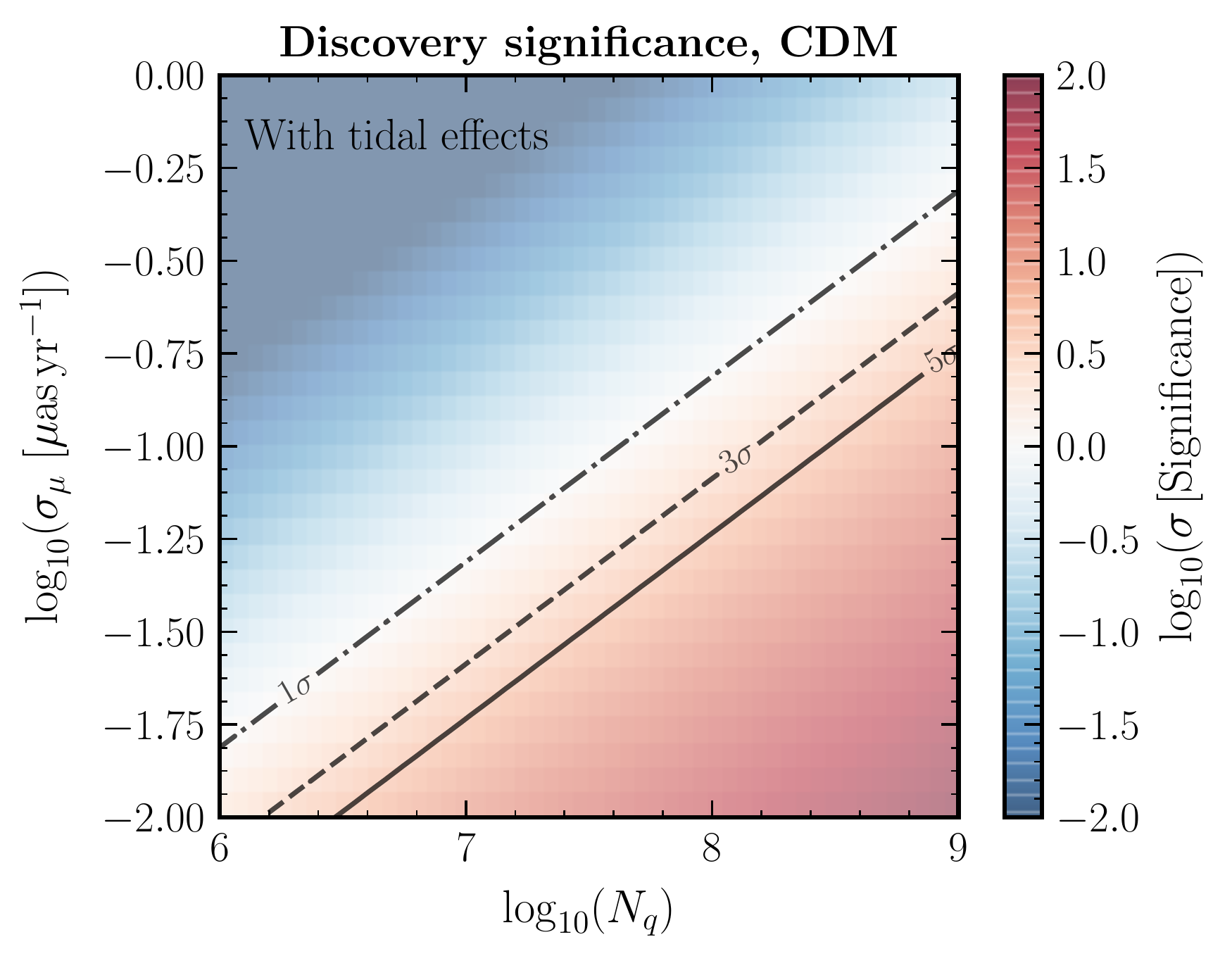}
\includegraphics[width=0.45\textwidth]{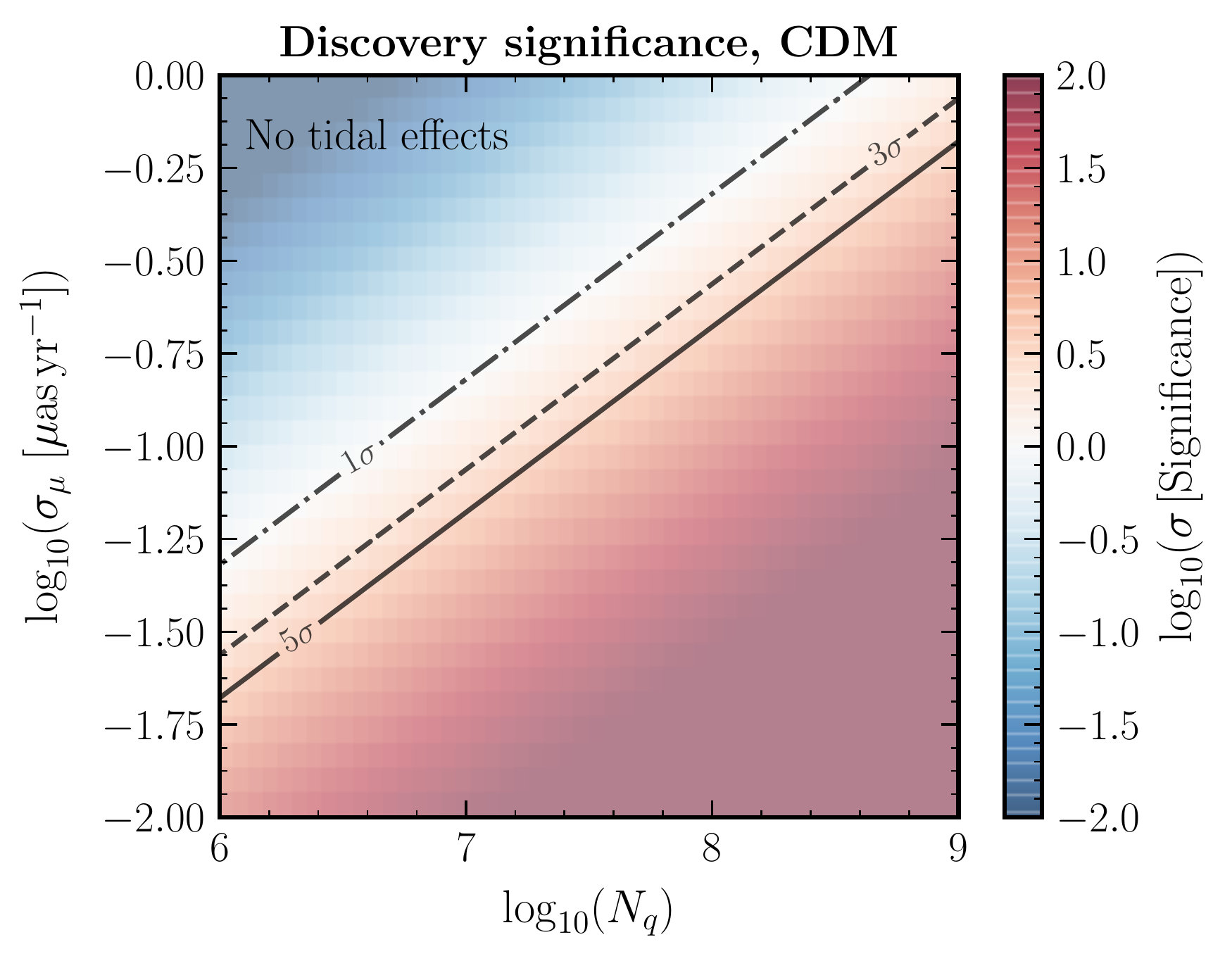}
\caption{Discovery significance of the fiducial CDM configuration (left) and an optimistic CDM configuration without accounting for tidal disruption effects (right) achievable using proper motion measurements, shows as a heatmap for different values of the measured proper motion noise $\sigma_\mu$ and number of observed background sources $N_q$. Also shown are the 1-, 2- and 3-$\sigma$ detection contours as dot-dashed, dashed, and solid black lines. \nblink{05_cdm}} \label{fig:lcdm_disc}
\end{figure*}

We may finally obtain the forecasted sensitivity of a given set of observations to a given signal configuration. Figure~\ref{fig:lcdm_disc} shows the discovery significance for the fiducial CDM configuration (left panel) and the optimistic configuration without tidal stripping (right panel), using quasar proper motion power spectra, for different values of the proper motion noise $\sigma_\mu$ and number of observed quasars $N_q$. We see that the optimistic scenario may be within reach of the next generation of interferometric telescopes, assuming noise levels $\sigma_\mu\approx 1\,\mu$as\,yr$^{-1}$ and $N_q\approx10^8$. Prospects assuming the fiducial scenario accounting for tidal disruption are less promising, and will require astrometric precision beyond that expected from next-generation surveys or methods beyond those based on two-point correlations presented in this work.

\subsection{Compact objects}
\label{sec:compact}

\begin{figure*}[htbp]
\centering
\includegraphics[width=0.45\textwidth]{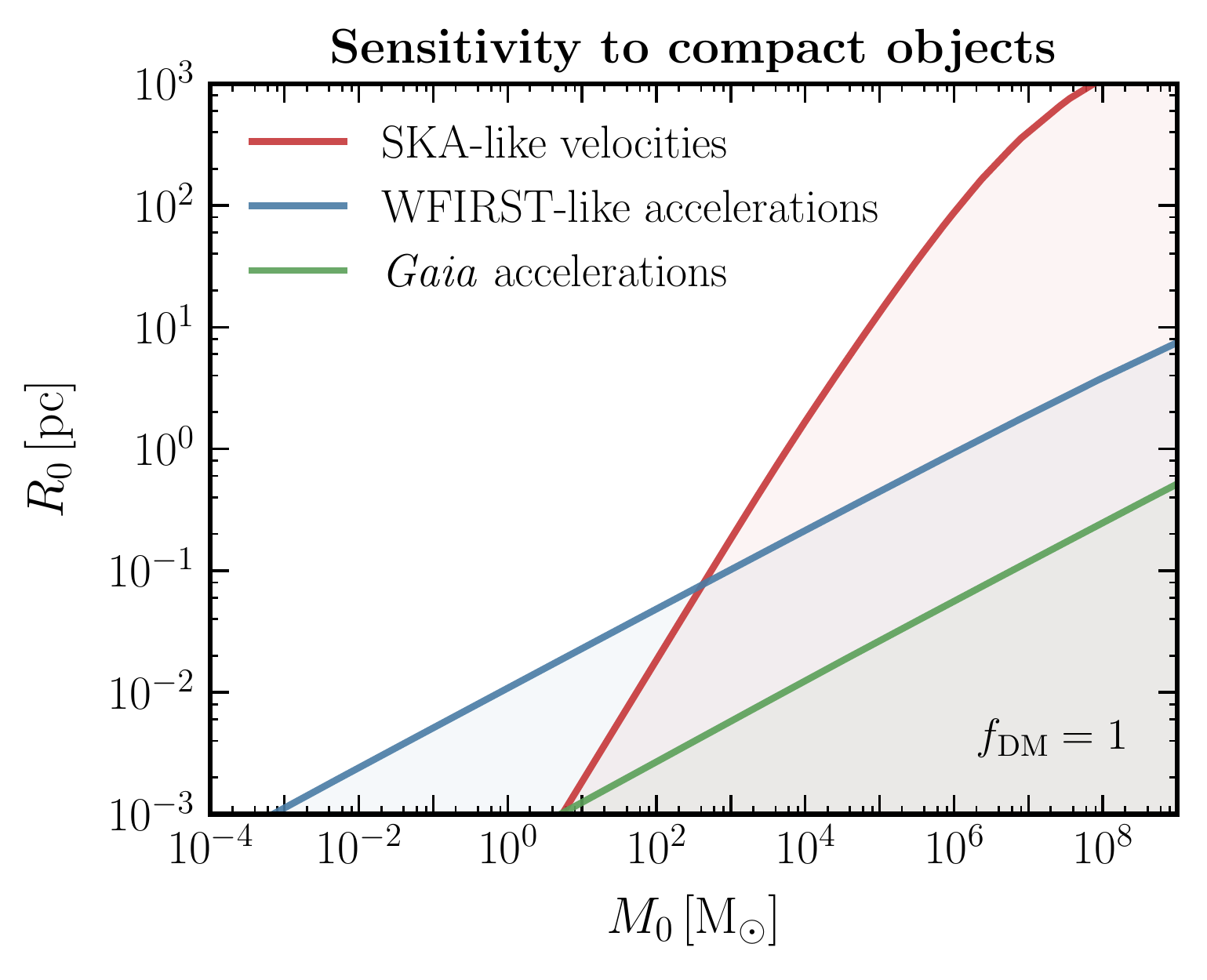}
\includegraphics[width=0.48\textwidth]{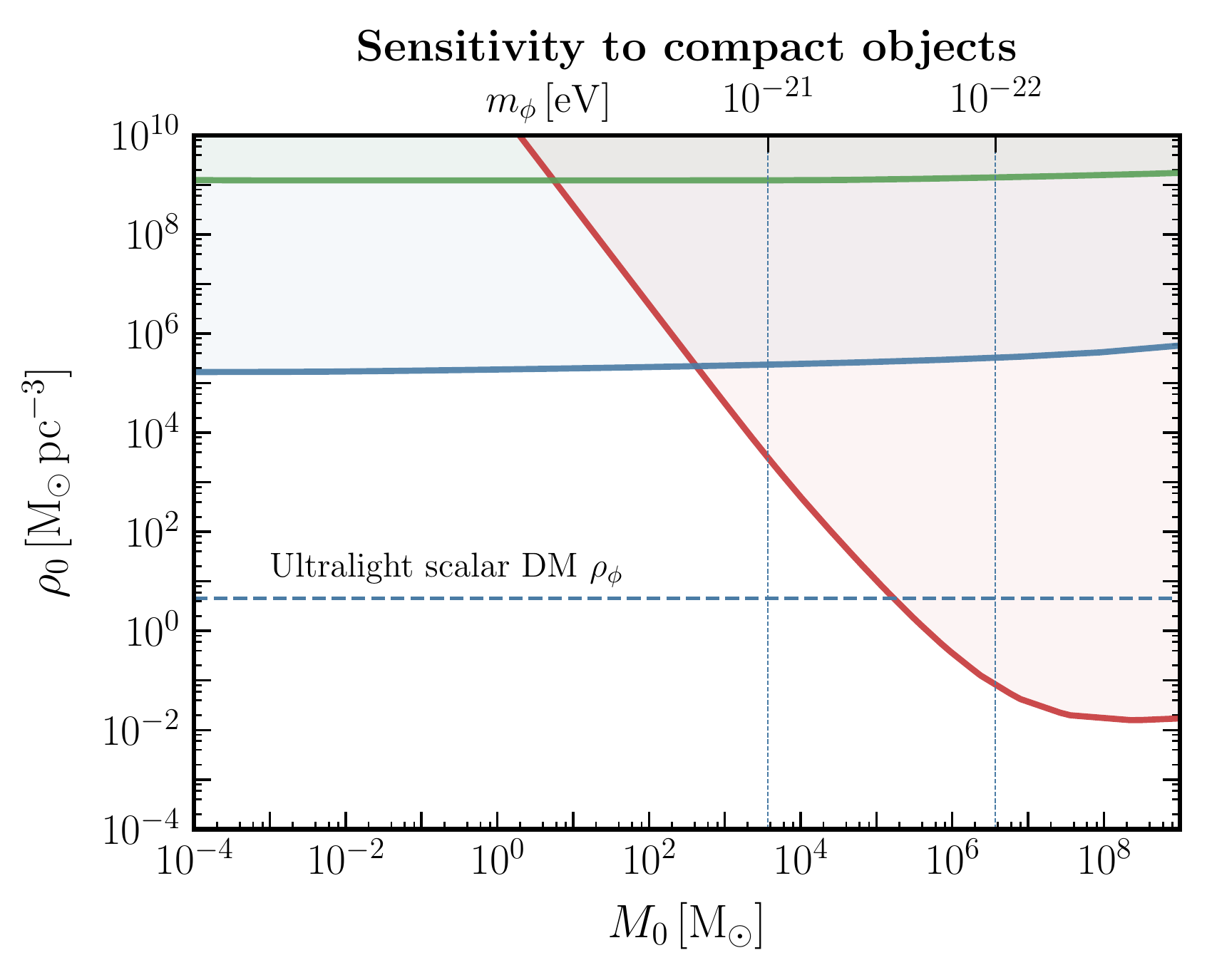}
\caption{\emph{(Left)} Maximum subhalo size $R_0$ that can be constrained at 95\% confidence as a function of subhalo mass $M_0$, and \emph{(Right)} Maximum subhalo density $\rho_0$ that can be constrained at 95\% confidence as a function of subhalo mass $M_0$, assuming dark matter fraction $f_\mathrm{DM}=1$. In each case, achievable constraints using WFIRST-like Galactic stellar proper accelerations (blue line) and using SKA-like (red line) extragalactic proper motions are shown. The dashed lines on the right plot represent the density (horizontal) and masses (vertical) of unbound fluctuations in the case of scalar field dark matter with benchmark masses $m_\phi = 10^{-22}$ and $10^{-21}$\,eV. \nblink{04_compact_objects}} 
\label{fig:compact_sens}
\end{figure*}

While primordial black holes (PBHs) have been studied as canonical examples of compact dark objects that would form due to large primordial overdensities and may constitute a fraction of the dark matter~\cite{Carr:1975qj, Niemeyer:1999ak}, dense compact objects of a finite size such as ultracompact minihalos (UCMHs)~\cite{Delos:2017thv,Delos:2018ueo} or supermassive dark matter clumps (SDMCs)~\cite{Berezinsky:2013fxa} may form in regions of intermediate overdensities~\cite{Ricotti:2009bs,Bringmann:2011ut} and are predicted in a wide range of inflationary models (see, \emph{e.g.}, Refs.~\cite{Chluba:2012we,Aslanyan:2015hmi,Delos:2017thv} and references therein) and non-standard early-Universe evolution~\cite{Schmid:1998mx,Erickcek:2011us}. Here, we study the sensitivity of global astrometric correlations to a general population of compact objects parameterized by a size $R_0$ and mass $M_0$. Their profile is modeled as Gaussian following to Eq.~\eqref{eq:Gaussianrho} and subhalos are assumed to be uniformly distributed within the Milky Way's smooth dark matter halo, whose density distribution is taken to be NFW with scale radius $r_\mathrm{s} = 18$\,kpc.

The sensitivities achievable with measurements of extragalactic proper motions (assuming SKA-like specifications, red line) and Galactic proper accelerations (assuming WFIRST-like and end-of-mission \Gaia specifications, blue and green lines respectively) are shown in Fig.~\ref{fig:compact_sens} in the mass-radius (left panel) and mass-density (right panel) parameter planes, assuming the lenses make up the totality of the Galactic dark matter. Currently unconstrained parameter space can already be probed using near-future \Gaia~astrometry. In App.~\ref{app:aristotle}, we show projected sensitivities using a simplified scenario of lenses uniformly distributed in an Aristotelian ball, directly described by Eqs.~\eqref{eq:mu_extpop} and~\eqref{eq:alpha_extpop}, showing excellent agreement with the results in Fig.~\ref{fig:compact_sens}.

Unlike traditional Galactic substructure searches based on photometric microlensing where magnification effects are strongly suppressed for lens radii larger than the characteristic Einstein radius~\cite{Paczynski:1985jf,Croon:2020wpr}, globally correlated astrometric effects are directly sensitive to a subhalo population of much larger radii. On the other hand, we note that the methods presented here are not ideally suited to searches for pointlike or very compact objects such as primordial black holes, where photometric lensing observations~\cite{Croon:2020wpr} and techniques based on detecting local astrometric lensing effects and transients (see V18 for examples and details) are more appropriate.

\subsection{Enhanced primordial power}\label{sec:enhanced}

We investigate a scenario in which the spectrum of primordial perturbations at small scales is enhanced compared to the standard $\Lambda$CDM expectation. This leads to a relative overabundance of low-mass halos which, having collapsed at earlier times, would also be significantly denser compared to those in the standard cosmological evolution. We parametrize this enhancement phenomenologically by introducing a kink in the dimensionless power spectrum of Gaussian curvature perturbations $\Phi$ parametrized by a break at $k_\mathrm{B}$ and high-$k$ slope $n_\mathrm{B}$,
\begin{align}
\mathcal{P}_{\Phi}(k) = \begin{cases} 
A_s \left( \frac{k}{k_*} \right)^{n_s -1} & k < k_\mathrm{B} \\ 
A_s \left( \frac{k_\mathrm{B}}{k_*} \right)^{n_s -1}\left( \frac{k}{k_\mathrm{B}} \right)^{n_\mathrm{B} -1} & k \ge k_\mathrm{B} 
\end{cases}
\end{align}
where we take $A_s = 2.105\times10^{-9}$, $n_s=0.9665$, and $k_*\equiv0.05$\,Mpc$^{-1}$~\cite{Aghanim:2018eyx}.

\begin{figure*}[!htbp]
\centering
\includegraphics[width=0.32\textwidth]{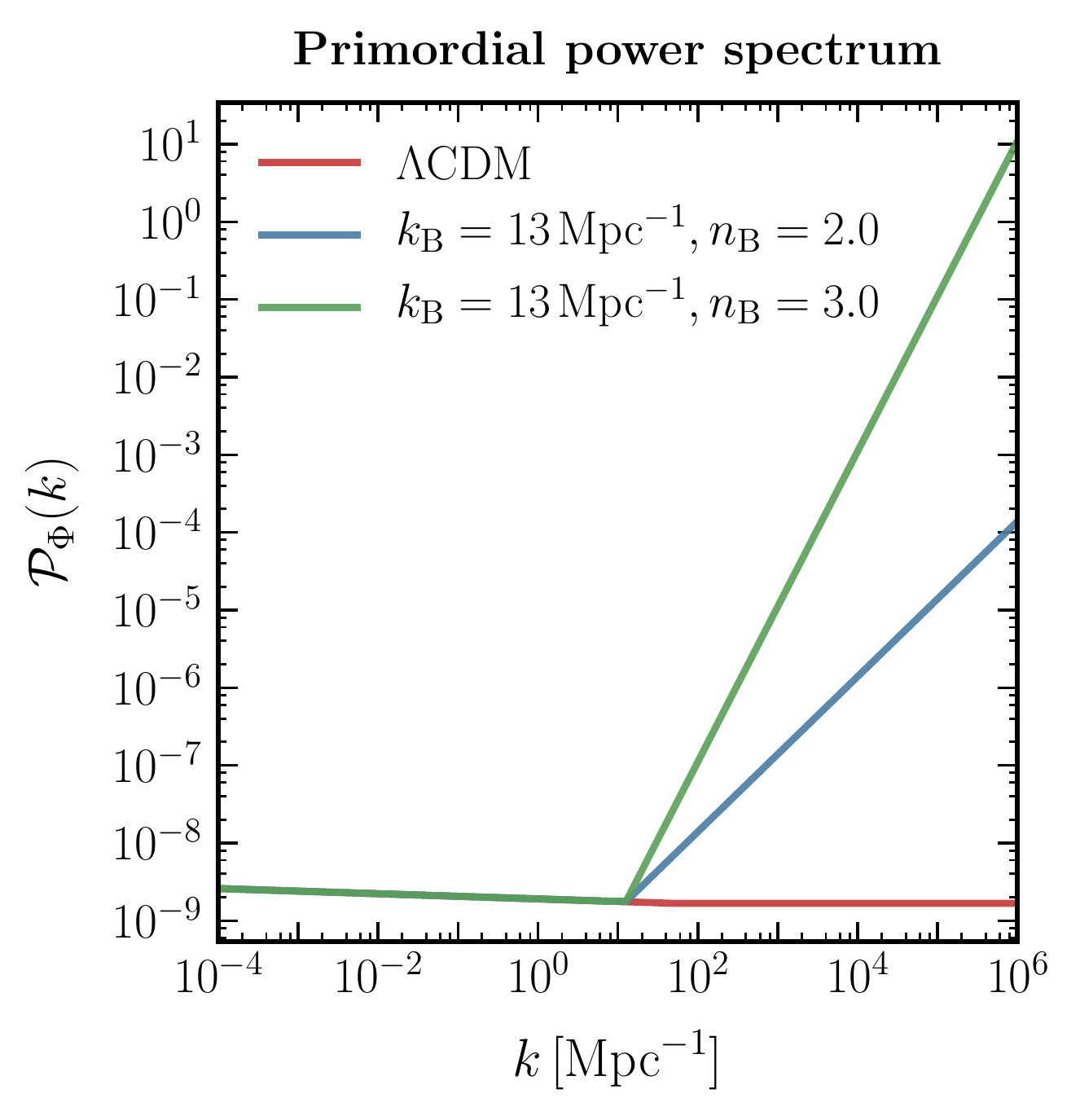}
\includegraphics[width=0.32\textwidth]{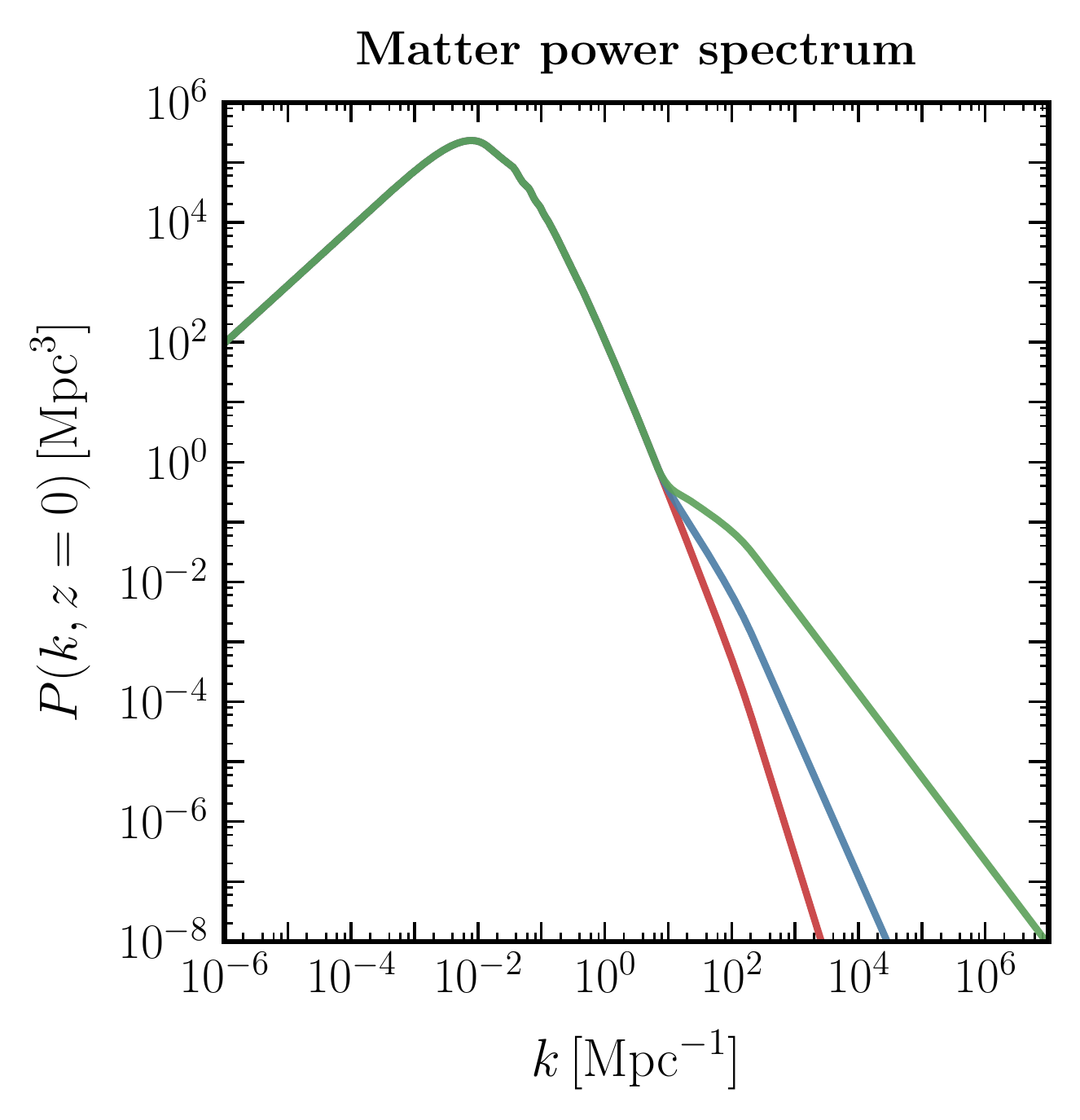}
\includegraphics[width=0.32\textwidth]{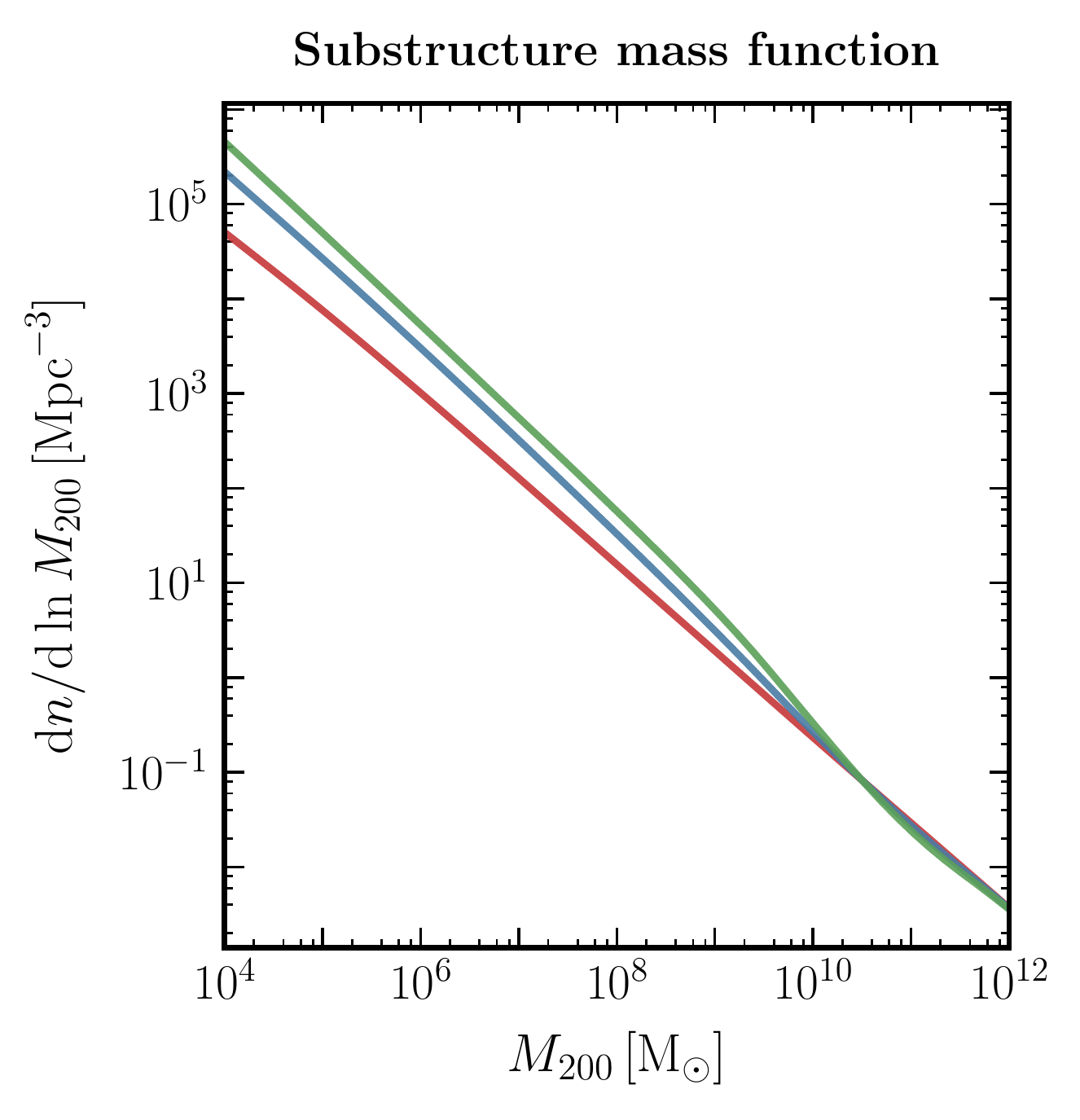}
\caption{Primordial power spectra with small-scale enhancement through a kink (left panel), derived present-day matter power spectra (middle panel), and present-day mass functions (right panel) shown for the representative set of kink parameters $k_\mathrm{B}=13$\,Mpc$^{-1}$, $n_\mathrm{B}=2(3)$ in blue(green). The standard $\Lambda$CDM prediction is shown as the red line in each case. A kink in the primordial power spectrum results in an overabundance of dense, low-mass subhalos. \nblink{06_primordial_kink}}
\label{fig:kinked_specs}
\end{figure*}

The dimensionless matter power spectrum at a given wavenumber and redshift can be obtained through the matter transfer function $D(k,z)$ as
\begin{align}
\mathcal{P}_{\delta}(k,z) = \left|D\left(k,z\right)\right|^2 \mathcal{P}_{\Phi}(k) 
\end{align}
and we use \texttt{CLASS}~\cite{Blas:2011rf} to compute the transfer function. Given the present-day matter power spectrum, the mass variance (encoding the amplitude of fluctuations within a sphere of radius $R$) can be computed as 
\begin{align}
\sigma^2(R) = \int \dd (\ln k) \, \mathcal{P}_{\Phi}(k) \left|D\left(k,z=0\right)\right|^2  \left| W(k,R) \right|^2
\end{align}
where  the window function $W(k,R) = 3 (kR)^{-3}[\sin(kR)-kR\cos(kR)]$ is the Fourier transform of a top-hat smoothing function with smoothing scale $R = (M / (4/3\pi\overline{\rho}_\mathrm{m}))^{1/3}$ with $\overline{\rho}_\mathrm{m}$ the mean density of the Universe.
With the mass variance in hand, the modified mass spectrum of subhalos in this scenarios is computed using the Tinker mass function~\cite{Tinker:2008ff} implemented in the \texttt{COLOSSUS}~\cite{Diemer:2017bwl} code and given by
\begin{equation}
\frac{\dd n}{\dd M}=f(\sigma) \frac{\overline{\rho}_\mathrm{m}}{M} \frac{\dd \ln \sigma^{-1}}{\dd M}
\end{equation}
where $f(\sigma)$ is parameterized as
\begin{equation}
f(\sigma)=A\left[\left(\frac{\sigma}{b}\right)^{-a}+1\right] e^{-c / \sigma^{2}}
\end{equation}
with the constants $A, a, b$ and $c$ calibrated to ($\Lambda$CDM) simulations (see Refs.~\cite{Tinker:2008ff,Diemer:2017bwl} for further details). We calibrate the overall number of subhalos such that an unkinked power spectrum yields the same number of subhalos as in the $\Lambda$CDM case we have considered in Sec.~\ref{sec:cdmpop} (150 between $10^8$--$10^{10}$\,M$_\odot$~\cite{Hutten:2016jko}) with this pipeline. The left panel of Fig.~\ref{fig:kinked_specs} shows representative examples of kinked primordial power spectra, with the derived present-day matter power spectra and mass functions shown in the middle and right panels respectively.

The present-day density (or equivalently, concentration) of subhalos is calculated following the procedure outlined in Ref.~\cite{Ludlow:2016ifl}. Specifically, we assume that the mean density $\langle\rho_\mathrm{s}\rangle$ of subhalos (modeled as NFW) within the scale radius $r_\mathrm{s}$ is proportional to the critical density of the Universe at collapse redshift $z_\mathrm{coll}$,
\begin{equation}
\frac{\langle\rho_\mathrm{s}\rangle}{\rho_0} = C\frac{\rho_c(z_\mathrm{coll})}{\rho_0} = C\left[\frac{H(z_\mathrm{coll})}{H_0}\right]^2
\label{eq:rho_s}
\end{equation}
where $C$ is a constant to be determined. The collapse redshift corresponds to the time at which the current characteristic mass $M_{200}$ was contained in progenitors more massive than a fraction $f$ of this current mass.

Extended Press-Schechter theory can be invoked to relate the current characteristic mass $M_{200}$ to the scale mass~\cite{Lacey:1994su},
\begin{equation}
\frac{{M}_\mathrm{s}}{{M}_{200}} \equiv \operatorname{erfc}\left(\frac{\delta_{\mathrm{sc}}\left(z_\mathrm{coll}\right)-\delta_{\mathrm{sc}}\left(z=0\right)}{\sqrt{2\left(\sigma^{2}\left(f {M}_{200}\right)-\sigma^{2}\left({M}_{200}\right)\right)}}\right)
\label{eq:M_s}
\end{equation}
where $\delta_{\mathrm{sc}}(z) \approx \delta_\mathrm{c} / D(z)$, with $\delta_\mathrm{c} = 1.686$, is the density threshold for collapse of a spherical top-hat perturbation and $D(z)$ the linear growth factor. The left sides of Eqs.~\eqref{eq:rho_s} and \eqref{eq:M_s} depend on the halo profiles through the concentration $c_{200} \equiv r_{200}/r_\mathrm{s}$. Given a present-day characteristic mass $M_{200}$, they can be simultaneously and iteratively solved to yield consistent solutions for the concentration $c_{200}$ and collapse redshift $z_\mathrm{coll}$. The constant $C$ is calibrated to yield concentrations for cluster-mass ($M_{200}\sim10^{13}\,\mathrm{M}_\odot$) halos consistent with observations for $f = 0.01$~\cite{Ludlow:2016ifl}. We choose to go down to a minimum subhalo mass of $10\,\mathrm M_\odot$ in this expository scenario to avoid extrapolating the derived concentration-mass relations and mass functions to even smaller values.

\begin{figure}[!htbp]
\centering
\includegraphics[width=0.45\textwidth]{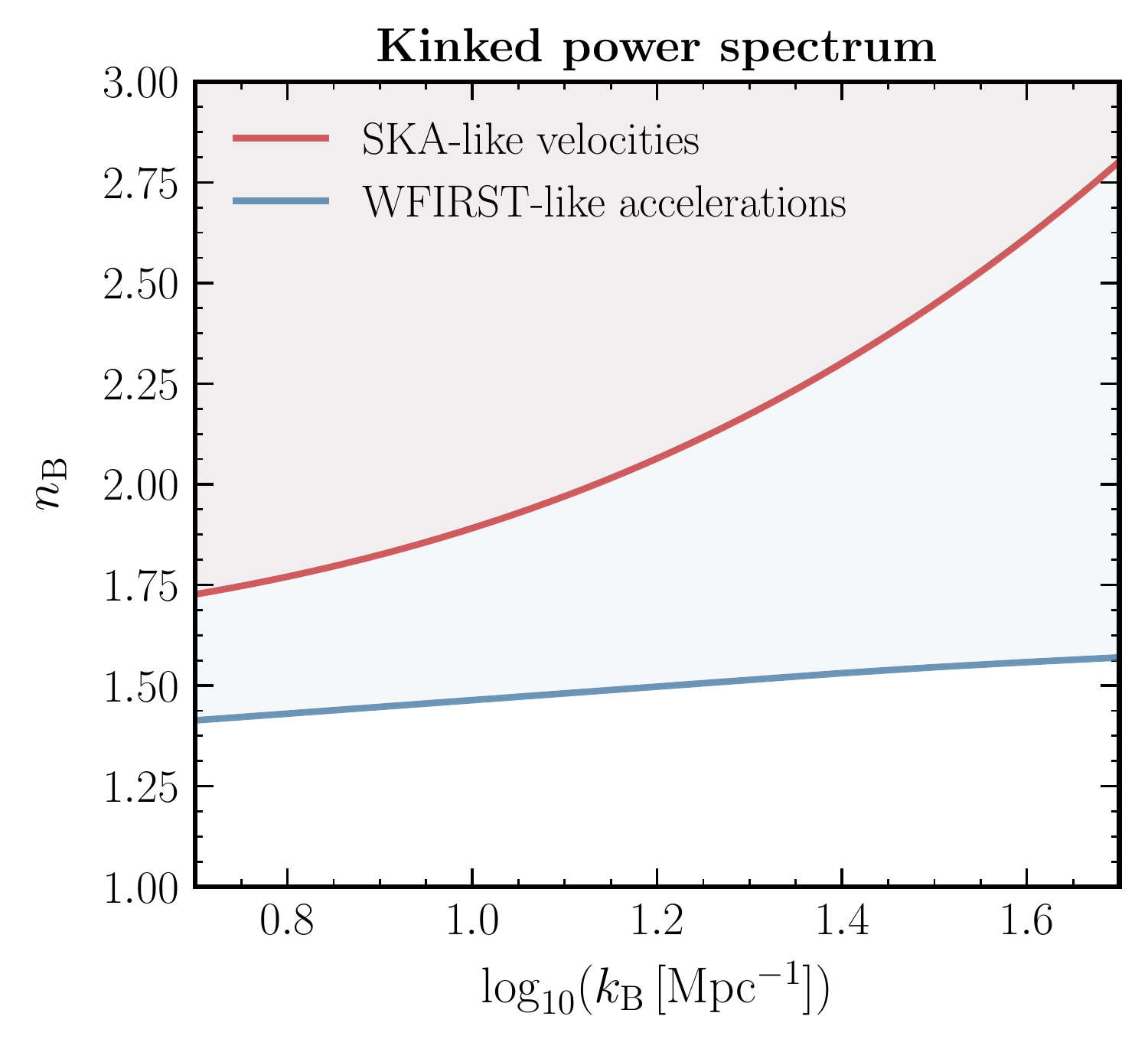}
\caption{95\% confidence interval sensitivity forecasts on scenarios with a kink in the power spectrum parameterized by the break location $k_\mathrm{B}$ and kink slope $n_\mathrm{B}$, and assuming a minimum bound substructure mass of $10\,\mathrm M_\odot$. Shown are sensitivities achievable using quasar proper motion measurements with an SKA-like survey (red) as well as observations of Galactic stellar proper accelerations by a WFIRST-like survey (blue). \nblink{06_primordial_kink}}
\label{fig:kink_ps}
\end{figure}

Sensitivity forecasts on the break $k_\mathrm{B}$ and high-$k$ slope $n_\mathrm{B}$ of an enhanced primordial power spectrum are displayed in Fig.~\ref{fig:kink_ps}. Shown are constraints achievable at the 95\% confidence interval using quasar proper motion measurements with an SKA-like survey, as well as observations of Galactic stellar proper accelerations by a WFIRST-like survey.

We caution that our treatment is simplistic in several ways---it is anchored to CDM simulations at higher masses, while necessitating extrapolation of the modified power spectrum down to small scales and of the subhalo mass function and concentration-mass relation down to small subhalo masses. Although a more accurate treatment would necessarily involve \emph{N}-body simulations consistent with the modified primordial spectra, our simple semi-analytic prescription captures the essential physics while making the point that enhancement of structure on small scales can be effectively probed with near-future astrometric observations.

\subsection{Scalar dark matter}\label{sec:scalar}

Dark matter may constitute of ultralight scalar fields, sometimes denoted ``fuzzy'' dark matter, with masses potentially as low as $10^{-22}\,\mathrm{eV}$~\cite{Hu:2000ke,Hui:2016ltb}. Scalar field dark matter automatically exhibits \emph{unbound} substructure due to interference effects, sourcing $\mathcal{O}(1)$ fractional density fluctuations that can cause a stochastic weak gravitational lensing signal~\cite{Bar-Or:2018pxz,Hui:2016ltb}. 
The contribution to the power spectra described below (and calculated in App.~\ref{app:scalar}) is irreducible because it originates from the unavoidable density fluctuations of a free scalar field at the scale of the typical de Broglie wavelength in a thermal ensemble. 

Assuming the velocity spectrum and density distribution of the halo is known, the velocity and acceleration power has only \emph{one free parameter}---the scalar field's mass $m$. The density fluctuations of real scalar dark matter can be attributed to random interference fringes, which have a typical mass $M_0$ and radius $R_0$ equal to
\begin{align}
M_0 &= C \rho_0 \left(\frac{\pi}{\sigma_k}\right)^3 \approx  5 \times 10^5 M_\odot \, C \left(\frac{10^{-22}\,\mathrm{eV}}{m}\right)^3,\label{eq:scalarM0}\\
R_0 &= \frac{1}{2\sigma_k} \approx 58\,\mathrm{pc} \left(\frac{10^{-22}\,\mathrm{eV}}{m}\right); \label{eq:scalarR0}
\end{align}
where $C$ is an $\mathcal{O}(1)$ constant, $\rho_0$ is the local mean DM energy density, and $\sigma_k = m \sigma_v$ is set by the scalar mass $m$ and the known velocity dispersion in the Milky Way.

These density fluctuations unavoidably constitute a substructure fraction of 100{\%}. We relegate the detailed calculation of the velocity and acceleration power spectra to App.~\ref{app:scalar}. The results of that calculation support the interpretation of the scalar's density fluctuations as a 100\% substructure fraction of dark matter, with mass and size given by Eqs.~\eqref{eq:scalarM0} and \eqref{eq:scalarR0}. Indeed, with some simplifying assumptions (spatially constant $\rho_0$, infinite source distance, and no velocity asymmetry $v_\odot = 0$), the velocity and acceleration power spectra for scalar dark matter are \emph{identical} to those of a population of Gaussian lenses with masses $M_0$ and radii $R_0$ that make up all of the dark matter, provided we take $C = 4/(3\pi^{3/2})$ for velocities and $C = 32/(15\pi^{3/2})$ for accelerations in Eq.~\eqref{eq:scalarM0}. Without those simplifying assumptions, the formula for the velocity power spectrum is given by Eq.~\eqref{eq:mulmscalar} with a completely analogous formula for the acceleration power spectrum, also using the formulae of Eqs.~\eqref{eq:Prhodot} and \eqref{eq:Prhoddot} for the power spectra of the time derivatives of the density fluctuations.

Because of the correspondence to the Gaussian-lens power spectrum, we can indicate on the right panel of Fig.~\ref{fig:compact_sens} the mass-independent halo density relation implied by Eqs.~\eqref{eq:scalarM0} and \eqref{eq:scalarR0} by the horizontal blue dashed line. The effective ``halo'' mass of these density fluctuations is indicated by the vertical solid lines for $m = 10^{-21}\,\mathrm{eV}$ and $m = 10^{-22}\,\mathrm{eV}$. Our future projections imply that fuzzy dark matter at very low masses should be detectable with the assumed survey parameters. The proper acceleration power spectrum signal is approximately scale independent and thus could be a potential probe at higher scalar field masses, although the magnitude of the signal is still out of reach of near-future surveys using the methods presented here.

\begin{figure*}[!htbp]
\centering
\includegraphics[width=0.49\textwidth]{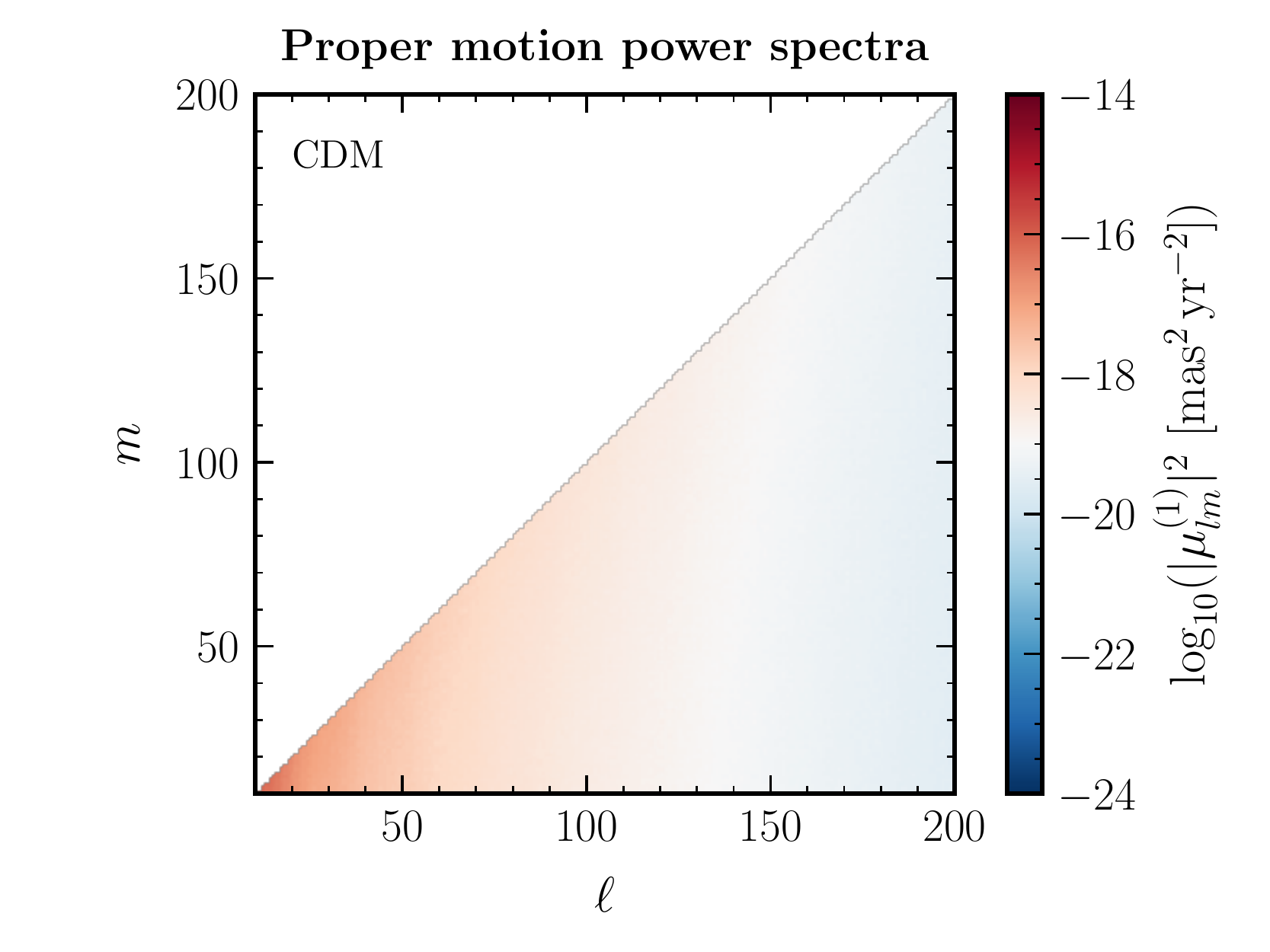}
\includegraphics[width=0.49\textwidth]{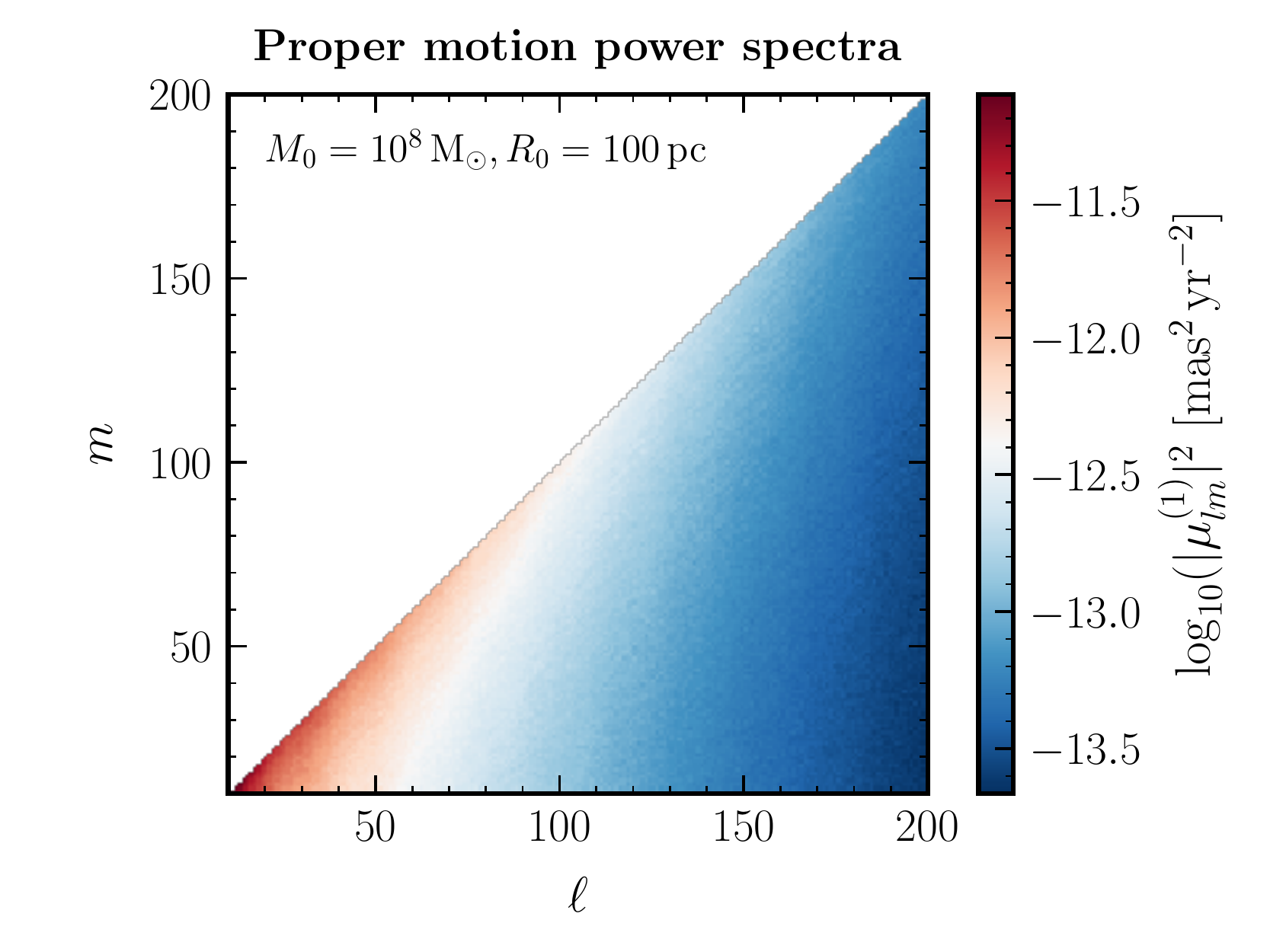}
\caption{\emph{(Left)} The magnitude of poloidal proper motion power spectrum coefficients $\mu_{\ell m}^{(1)}$ for the CDM setup. \emph{(Right)} The magnitude of poloidal proper motion power spectrum coefficients $\mu_{\ell m}^{(1)}$ for a population of compact objects of mass $M_0 = 10^8\,\mathrm{M}_\odot$ and size $R_0 = 100\,\mathrm{pc}$ making up all of the Galactic dark matter density and distributed following the smooth Milky Way DM halo. An azimuthal asymmetry in each case can be seen---with larger coefficients at higher $m$ for a fixed $\ell$---with greater asymmetry in the compact objects scenario. \nblink{01_simulations}}
\label{fig:m_abs}
\end{figure*}

\section{Signal discriminants}
\label{sec:handles}

\subsection{Toroidal modes as a control region}

As described in Sec.~\ref{sec:formalism}, the astrometric lensing signal is sourced from the gradient of the projected scalar lensing potential $\psi$, so it is expected to exclusively populate the poloidal (curl-free) component of the power spectrum decomposition. The noise on the other hand is expect to contribute to both the poloidal as well as toroidal modes. The toroidal (divergence-free) power spectrum can thus be used as a control channel to calibrate the noise spectrum and deal with unmodeled sources of noise of instrumental and/or astrophysical origin.

\subsection{Directional asymmetry}\label{sec:directionalasymm}

\begin{figure*}[!htbp]
\centering
\includegraphics[width=0.49\textwidth]{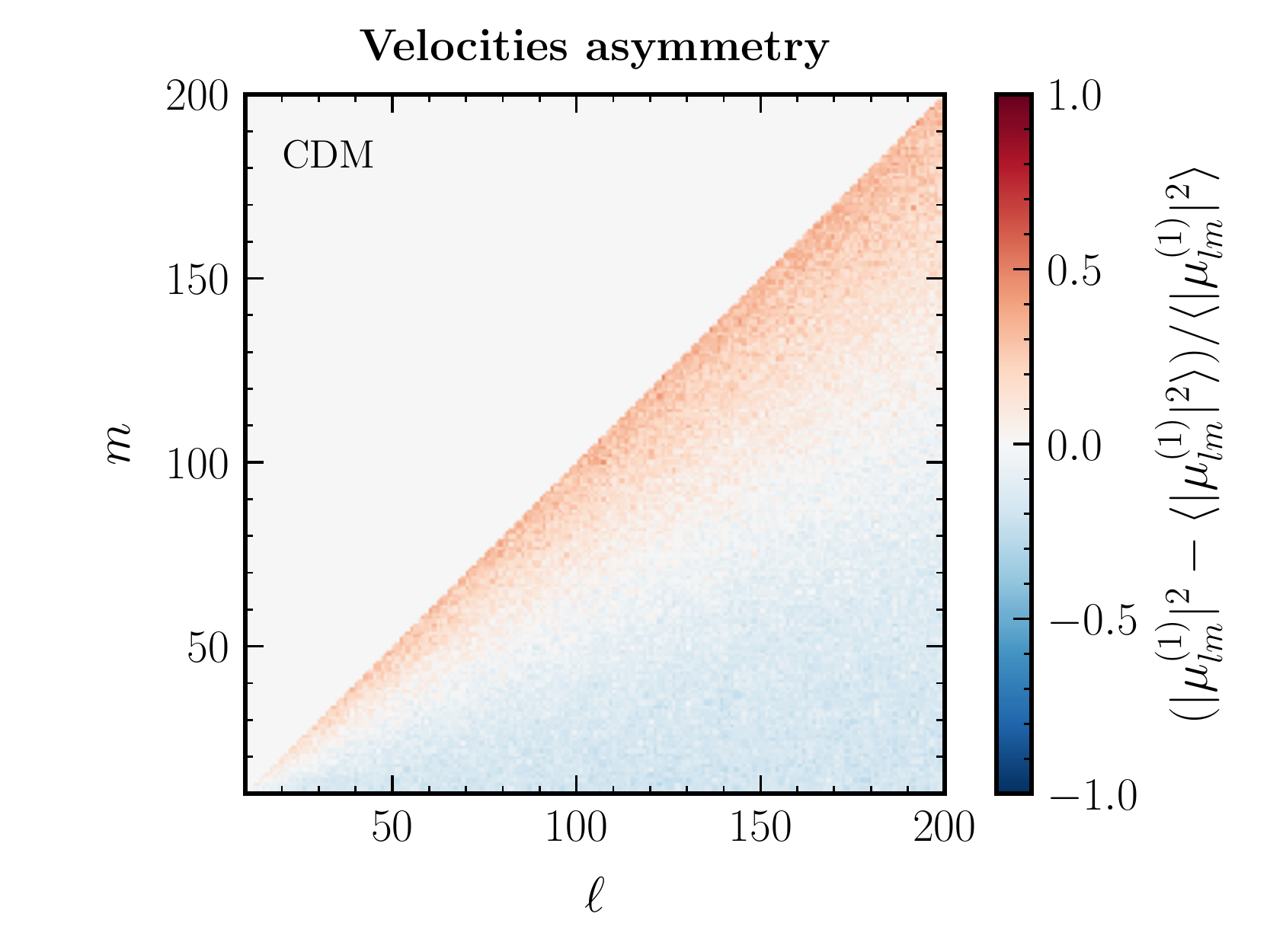}
\includegraphics[width=0.49\textwidth]{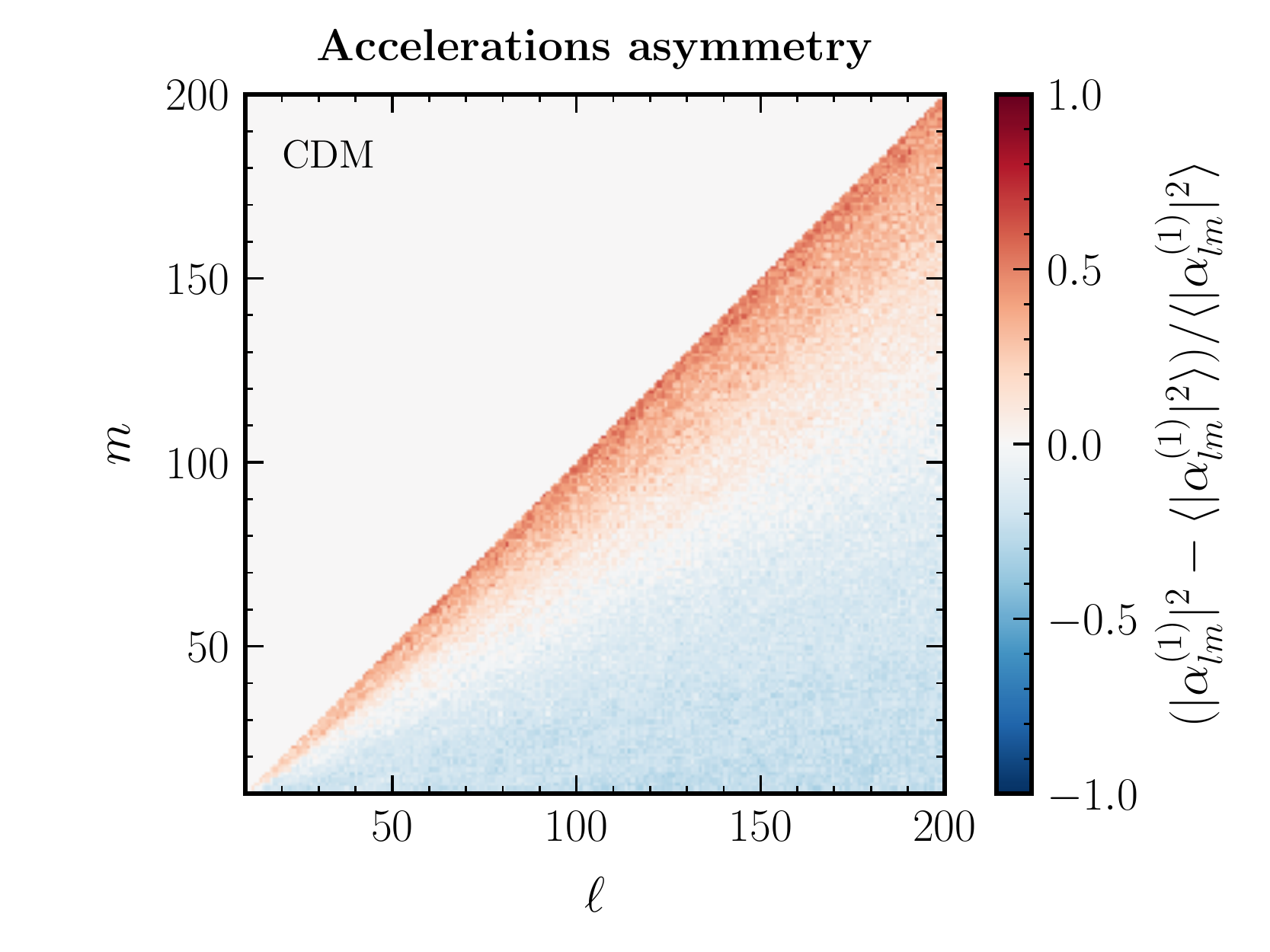}
\includegraphics[width=0.49\textwidth]{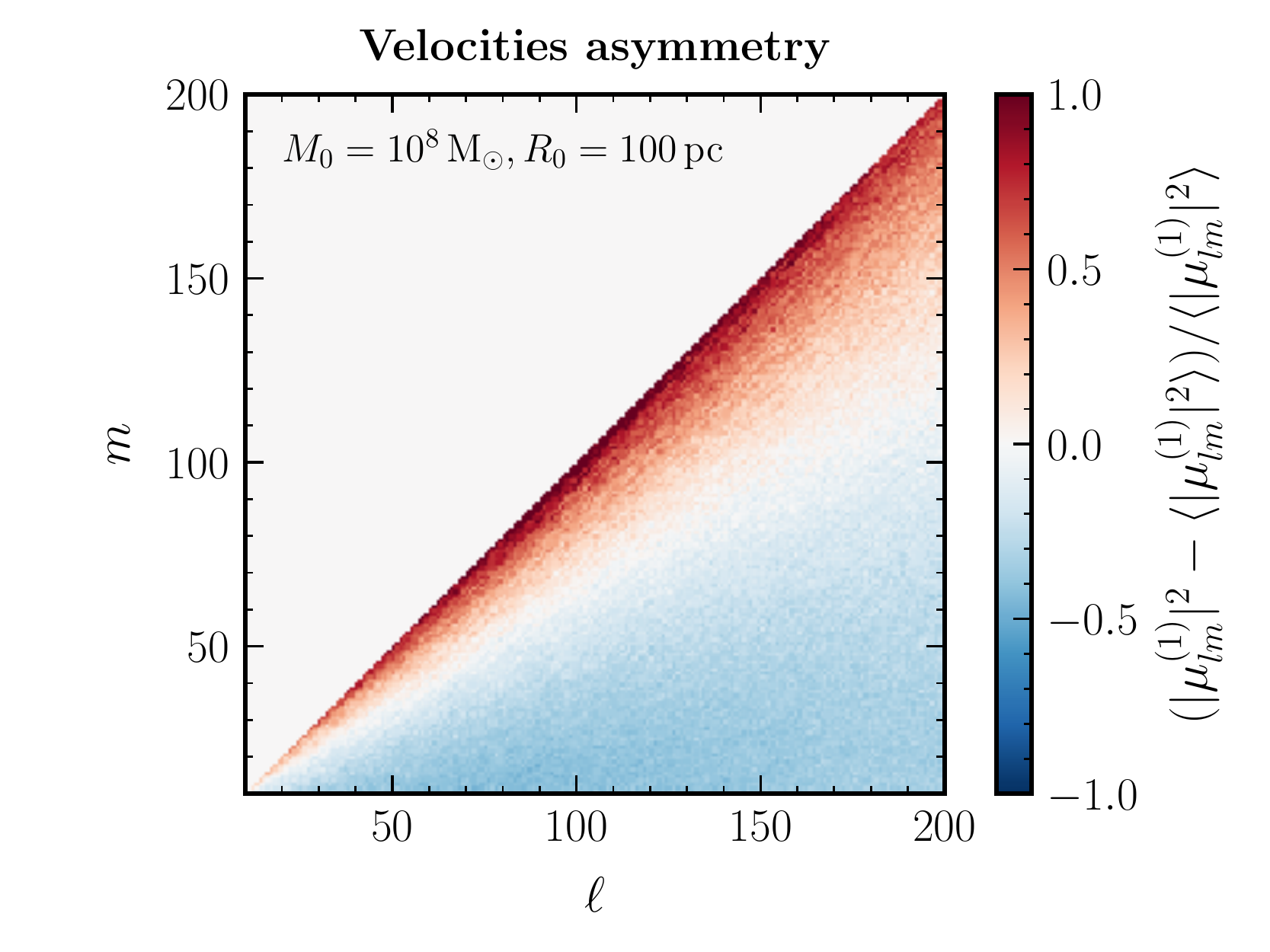}
\includegraphics[width=0.49\textwidth]{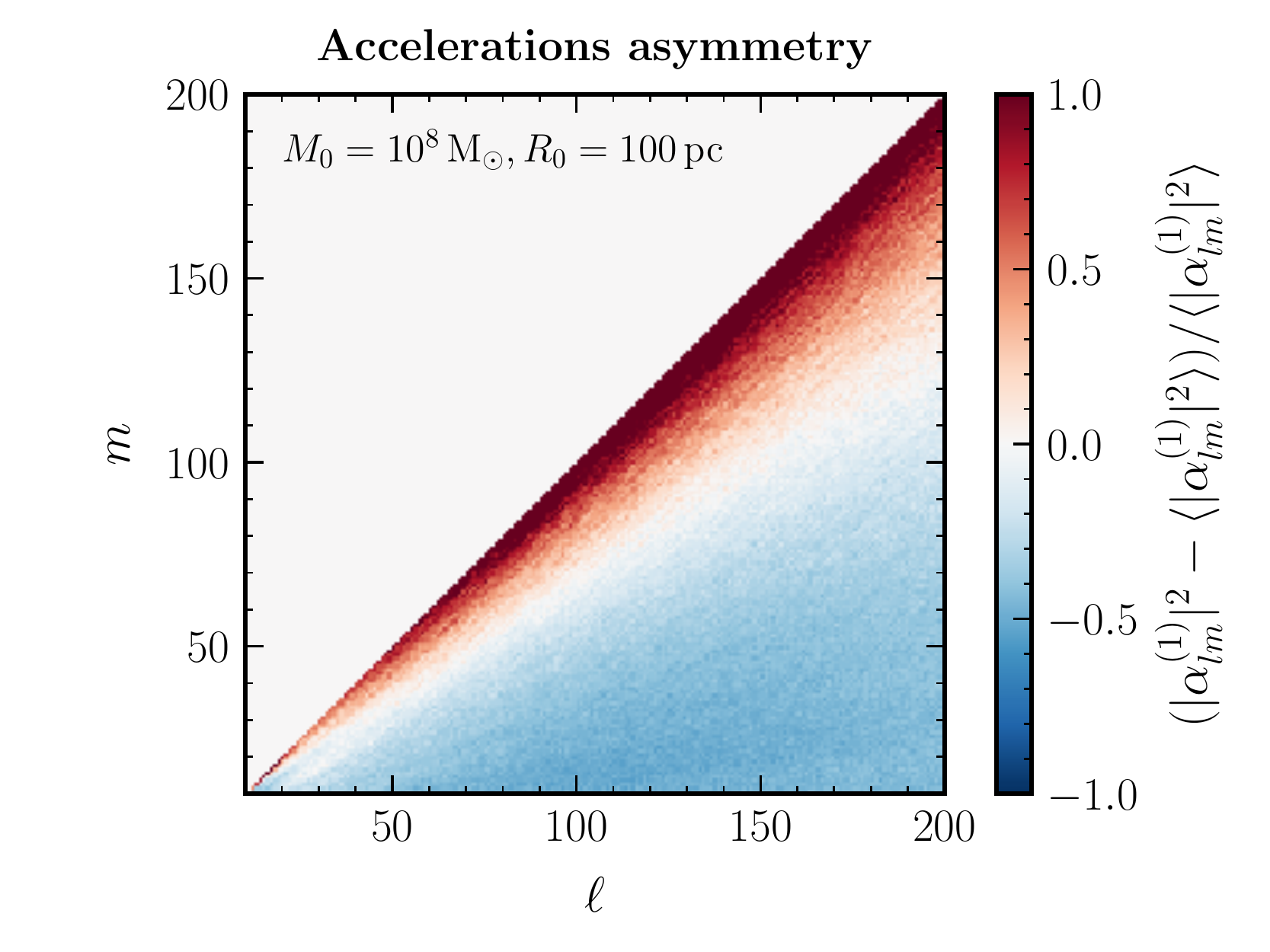}

\caption{The \emph{fractional} azimuthal asymmetry in the poloidal power spectrum, compared to the mean power spectrum coefficient at a given $\ell$, estimated from 500 simulated realizations. Shown for the fiducial CDM setup (top row) and for a population of compact objects of mass $M_0 = 10^8\,\mathrm{M}_\odot$ and extent $R_0 = 100\,\mathrm{pc}$ making up all of the Galactic dark matter and following the Milky Way DM halo spatially (bottom row). Shown for the proper motion power spectrum coefficients $|\mu_\ellm^{(1)}|^2$ (left column) and the proper acceleration power spectrum coefficients $|\alpha_\ellm^{(1)}|^2$ (right column). Greater asymmetry is expected for accelerations compared to velocities. \nblink{01_simulations}}
\label{fig:m_asymm}
\end{figure*}

After inserting Eq.~\eqref{eq:deflectionpotential} into Eq.~\eqref{eq:harmdecomposition}, integrating by parts, and using $\vect{\nabla}_{\vect{\theta}}^2 Y_{\ell m} = - \ell (\ell + 1) Y_{\ell m}$ and Eq.~\eqref{eq:PsiPhidef}, we find that for a single lens
\begin{align}
\mu_{\ell m}^{(1)} = - \frac{\ell (\ell + 1)}{D_l} \int \dd \Omega \, \psi(\vect{\beta}) \vect{v}_l \cdot \vect{\Psi}_{\ell m}^*(\vect{\theta}),
\end{align}
where $\psi(\vect{\beta}) = 4 \GN \int \dd \Omega' \, \Sigma(\vect{\theta}',\vect{\theta}_l) \ln \beta$ is the projected scalar lensing potential, $\Sigma(\vect{\theta}', \vect{\theta}_l)$ the projected surface mass density at $\vect{\theta}'$ from a lens at $\vect{\theta}_l$, $\vect{\beta} = \vect{\theta} - \vect{\theta}_l$ the angular impact parameter, $\int \dd \Omega$ an integral over $\vect{\theta}$ coordinates, and $\int \dd \Omega'$ an integral over $\vect{\theta}'$ coordinates.

Due to the Sun's motion around the Galactic center, the distribution $f_\oplus(\vect{v}_l)$ for the effective lens velocity is asymmetric---see Eqs.~\eqref{eq:fvinfty} and \eqref{eq:fvoplus}---with higher magnitudes expected for velocity components in the Galactic longitude direction than in the Galactic latitude direction. This will typically lead to an asymmetry in the expected power at different $m$ values at fixed $\ell$, because the high-$|m|$ (low-$|m|$) modes of $\vect{\Psi}_{\ell m}$ are preferentially oriented along the Galactic longitude (latitude) direction. 

Explicitly, we can see this directional asymmetry in the expected power by computing the expectation value of the square amplitudes $\mu_{\ell m}^{(1)}$ for a totality of lenses with number density distribution $n_l(\vect{\theta}_l, D_l)$:
\begin{align}
& \big \langle \big|\mu_{\ell m}^{(1)}\big|^2\big\rangle  = \ell^2 (\ell + 1)^2 \int \dd f_\mathrm{DM} \int \dd D_l \,
n_l(\vect{\theta}_l,D_l)\int \dd^2 \vect{v}_l  \nonumber \\
& \times  f_\oplus(\vect{v}_l,\vect{\theta}_l,D_l)  \bigg\lbrace v_{l,\theta}^2 \bigg| \int \dd\Omega\, \Psi_{\ell m,\theta}(\vect{\theta}) \psi(\vect{\beta})\bigg|^2 \label{eq:mlasymmetry}\\
& \phantom{\times f_\oplus(\vect{v}_l,\vect{\theta}_l,D_l)  \bigg\lbrace} +  v_{l,\varphi}^2  \bigg| \int \dd\Omega\, \Psi_{\ell m,\varphi}(\vect{\theta}) \psi(\vect{\beta}) \sin \theta  \bigg|^2 \bigg\rbrace. \nonumber
\end{align}
As stated above, the directional asymmetry in the rate of change in impact parameter stems from the asymmetric $f_\oplus(\vect{v}_l)$ in Eq.~\eqref{eq:fvoplus}, with higher expected proper components in the Galactic longitude direction: $\langle v_{l,\varphi}^2 \sin^2 \theta_l \rangle > \langle v_{l,\theta}^2 \rangle $.\footnote{Note that the $\int \dd \Omega$ integral in Eq.~\eqref{eq:mlasymmetry} at high $\ell$ is dominated by the region $\vect{\theta} \simeq \vect{\theta}_l$, due to the higher gradients in $\psi(\vect{\beta})$ there.} At high $|m|/\ell$, we also have that $\langle |\Psi_{\ell m,\varphi}|^2 \rangle > \langle |\Psi_{\ell m,\theta}|^2 \rangle$ over the celestial sphere, with the opposite inequality for low values of $|m|/\ell$. At low $|m|/\ell$, the dominant contributions arise from the second line in Eq.~\eqref{eq:mlasymmetry}, while at high $|m|/\ell$, they arise from the third line. We therefore deduce that for the astrometric lensing signal, $\big \langle \big|\mu_{\ell m}^{(1)}\big|^2\big\rangle$ is an increasing function of $|m|$ at fixed $\ell$, which is calculable from the 6D phase space distribution of the DM subhalos. This $|m|$ asymmetry is further exacerbated by the fact that the DM lens distribution $n(\vect{\theta}_l,D_l)$ is peaked towards the Galactic Center and thus latitudes $\theta_l \approx \pi / 2$, where the high-$|m|/\ell$ modes have more support than the low-$|m|/\ell$ modes.

Asymmetries in $|m|$ may also be caused by variations in exposure and noise across the celestial sphere. For optical astrometric surveys of quasars for example, there is a lower background source number density and higher astrometric noise per source near the Galactic equator, causing $\vect{\mu}$ to be more poorly measured there. Because of the support of the VSH functions, this would lead to somewhat similar asymmetry in $|m|/\ell$ for the noise, in both $\mu_{\ell,m}^{(1)}$ and $\mu_{\ell,m}^{(2)}$ components. Indeed, we observe this asymmetry in the quasar sample of the \textit{Gaia} DR2 data in Sec.~\ref{sec:gaia-quasars}. Nevertheless, because the lensing signal contributes only to $\mu_{\ell,m}^{(1)}$, one generally expects a \emph{difference} in $|m|$ asymmetry for the poloidal and toroidal mode power. Any excess power in the poloidal modes relative to the toroidal modes---as expected from a lensing signal---can then be tested to see if it conforms to the expected asymmetry implied by Eq.~\eqref{eq:mlasymmetry}. We expect a similar but quantitatively even higher $|m|/\ell$ asymmetry in the poloidal mode power of the lens-induced proper accelerations, due to the higher number of powers of $\vect{v}_l$---4 instead of 2---involved.

We illustrate the azimuthal asymmetry by plotting in Fig.~\ref{fig:m_abs} the square magnitude of the poloidal power spectrum amplitudes $|\mu_{\ell m}|^2$ for our fiducial CDM-like scenario (left panel) and for a population of $M_0 = 10^8\,\mathrm M_\odot$ subhalos of extent $R_0 = 100\,\mathrm{pc}$ making up all of the Galactic dark matter (right panel). In the latter case, the compact objects are distributed following the smooth dark matter profile of the Milky Way, without tidal evolution effects, which results in a larger concentration of subhalos towards the Galactic plane and an even larger azimuthal asymmetry compared to the CDM-like case, where subhalos appear largely isotropic in the sky. Figure~\ref{fig:m_asymm} further illustrates the fractional azimuthal asymmetry, plotting the fractional deviation of a given squared amplitude coefficient from the mean value at a given $\ell$. This is shown for the CDM-like model (top panels) and the compact objects population (bottom panels), in each case plotting the proper motion coefficients $|\mu_{\ell m}|^2$ (left panels) and proper acceleration coefficients $|\alpha_{\ell m}|^2$ (right panels). As anticipated above, a larger asymmetry for the acceleration spectra is seen, as well as a larger asymmetry for the compact objects configuration where there is more support near the Galactic plane. 

To summarize, the preferential motion of the Sun with respect to the stationary frame of Galactic subhalos would lead to a detectable asymmetry in the lens-induced correlation signal. Such a characteristic asymmetry is unlikely to be replicated by instrumental and non-lensing effects, and can be used as an additional handle to differentiate a putative signal from unmodeled noise.

\section{Power spectrum decomposition of \Gaia DR2 quasars}
\label{sec:gaia-quasars}

\begin{figure*}[tbp]
\centering
\includegraphics[trim = 40 0 0 0, clip, width=0.49\textwidth]{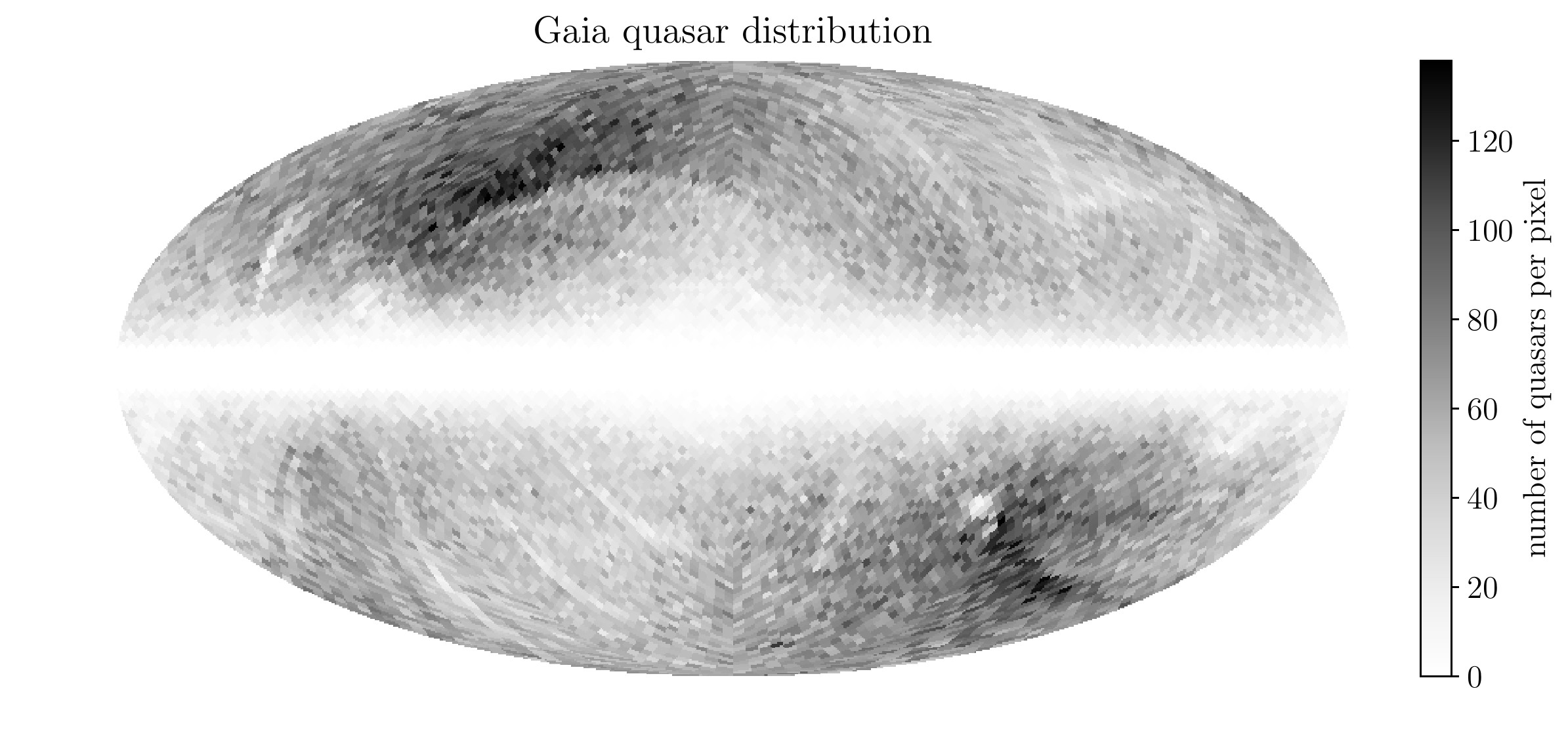} \\
\includegraphics[trim = 40 0 0 0, clip, width=0.49\textwidth]{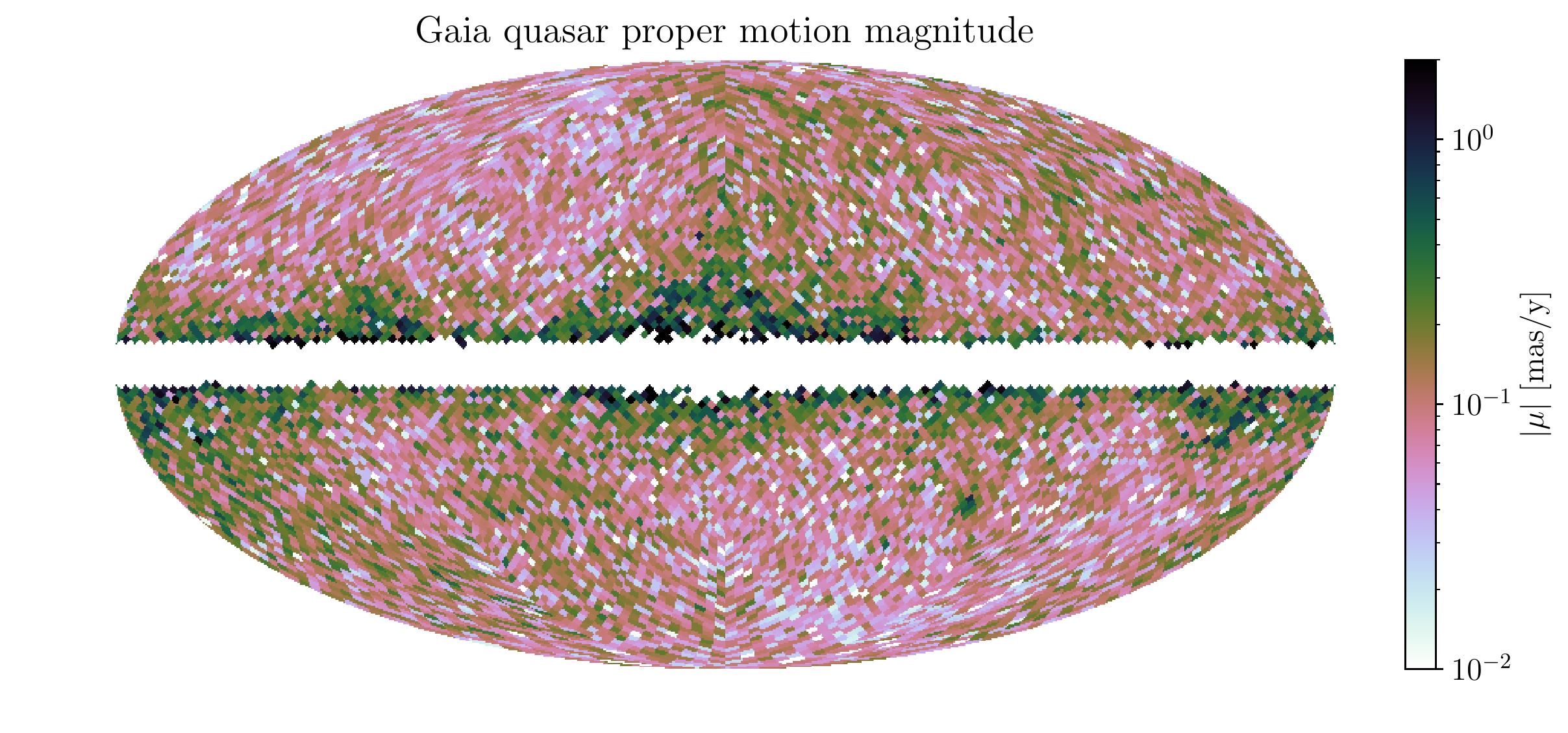}
\includegraphics[trim = 40 0 0 0, clip, width=0.49\textwidth]{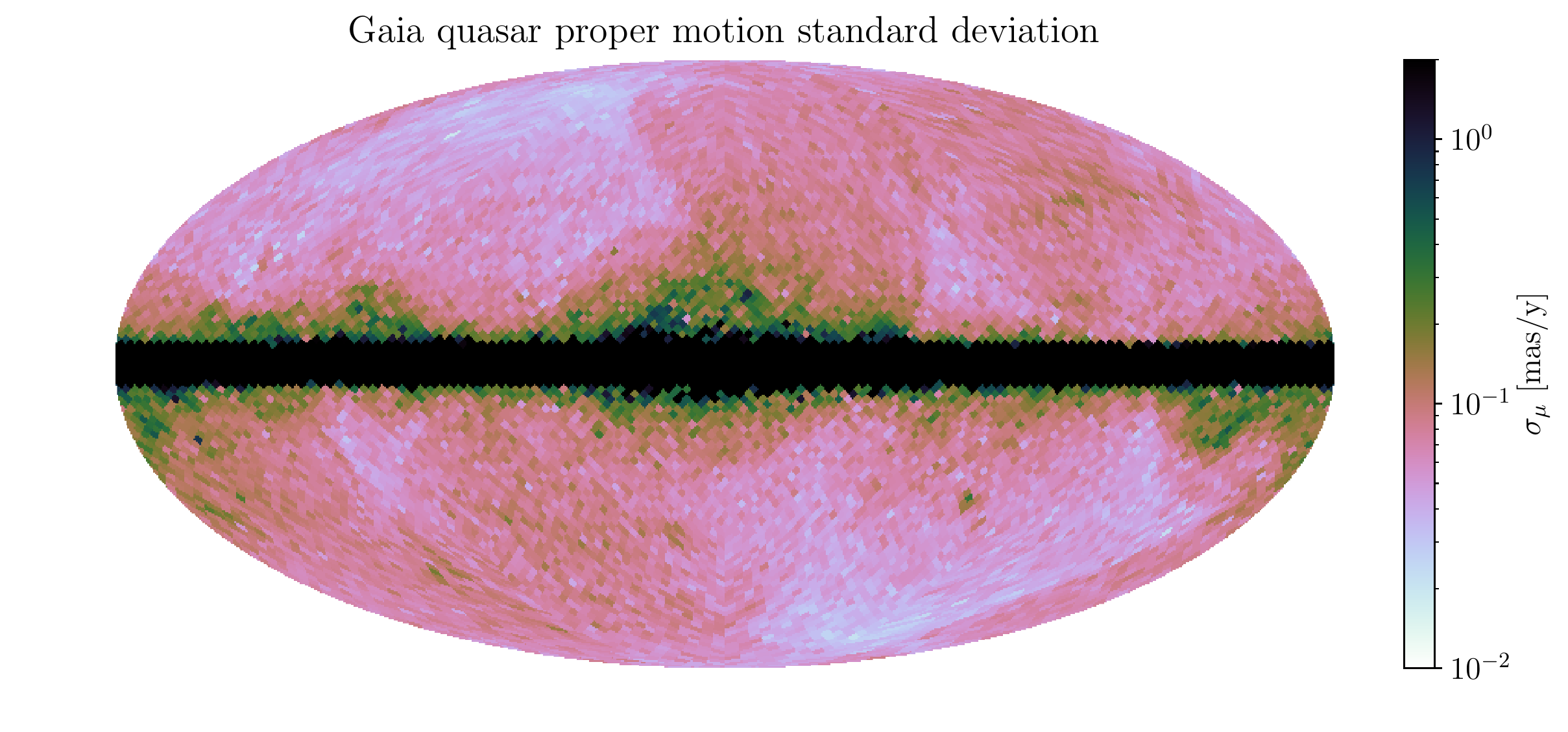}
\includegraphics[trim = 40 0 0 0, clip, width=0.49\textwidth]{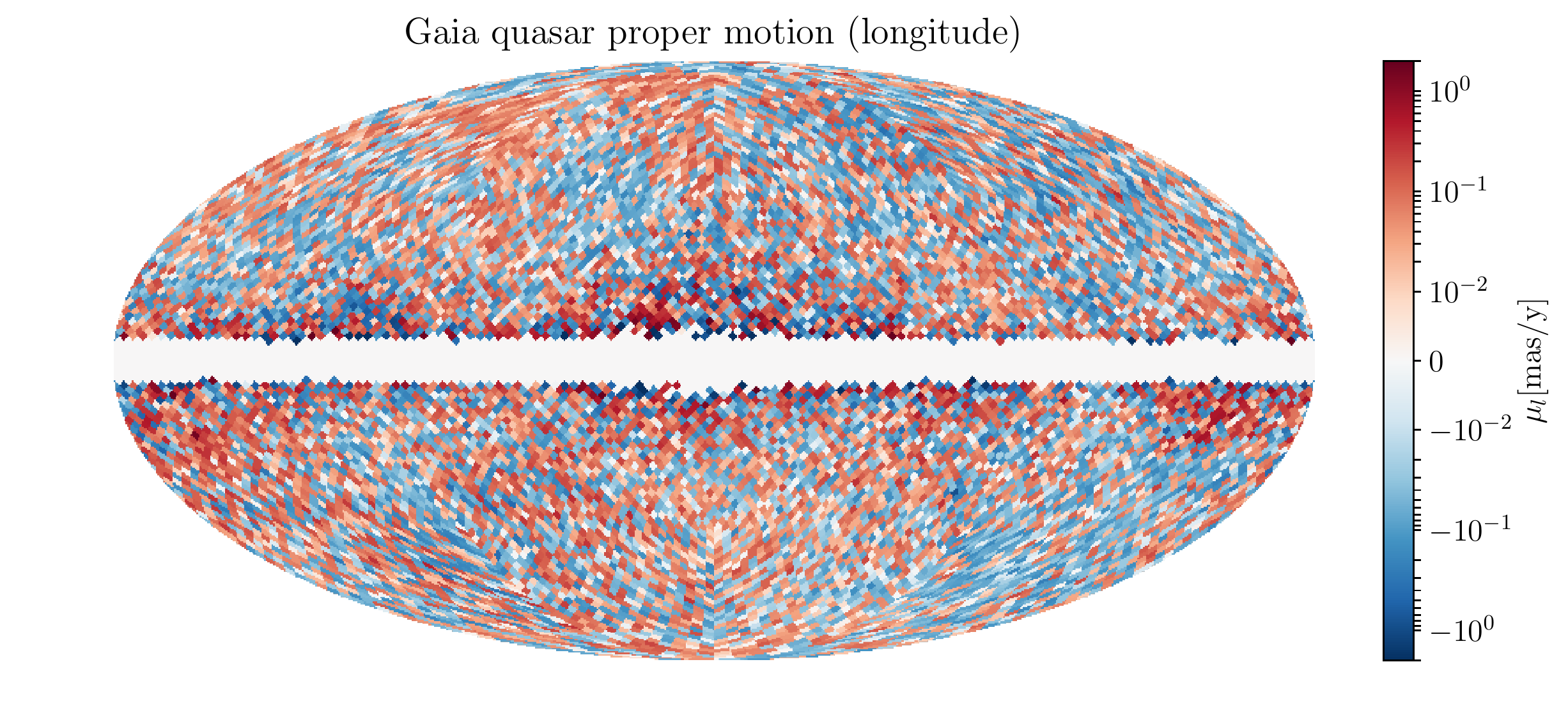}
\includegraphics[trim = 40 0 0 0, clip, width=0.49\textwidth]{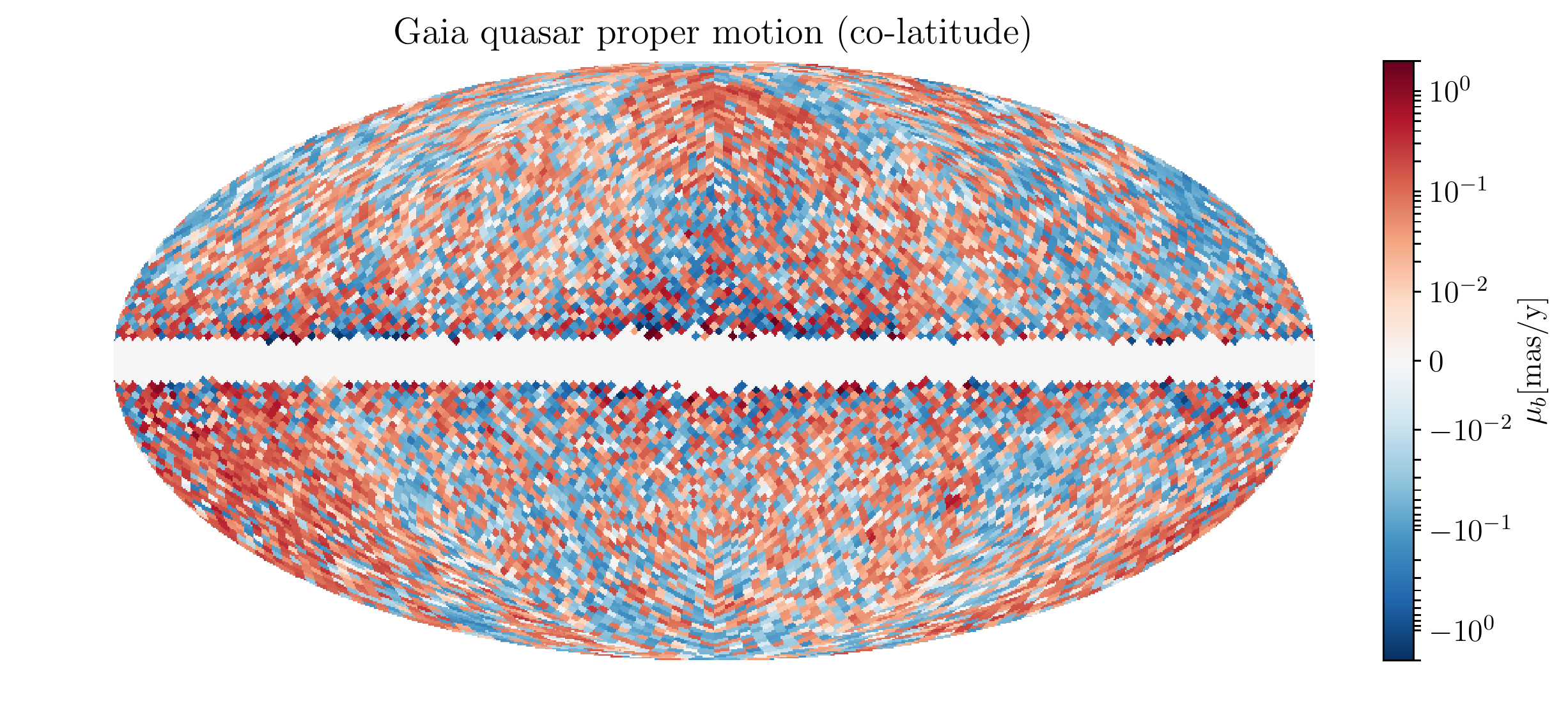}
\caption{Maps in Galactic coordinates of the number distribution of \Gaia DR2 quasars (top panel), their pixel-averaged proper motion magnitude $|\vect{\mu}|$ (left middle panel), their proper motion error $\sigma_\mu$ (right middle panel), and their proper motion components $\lbrace \mu_l, \mu_b \rbrace$ (bottom two panels). \nblink{07_vector_power_spectrum}}
\label{fig:gaia_maps}
\end{figure*}

In this section, we apply the technique proposed in the previous sections on an actual astrometric dataset. Appendix~\ref{app:estimator} derives in detail the optimal estimator for vector power spectra on the celestial sphere, accounting for nonuniform noise and incomplete coverage of the sky as well as ``spectral binning'', a coarse-graining technique that groups nearby $\ell m$ modes to obtain well-conditioned results out to high $\ell$ and/or with incomplete sky coverage. 

The dataset under consideration is the astrometric catalog of the 555,934 quasars in \textit{Gaia}'s second data release (DR2)~\cite{Prusti:2016bjo,Brown:2018dum,lindegren2018gaia}. Each quasar comes with a position, proper motion, and standard deviation $\lbrace \vect{\theta}_q, \vect{\mu}_q, \sigma_{\mu,q} \rbrace$ for $q = 1, \dots, \text{555,934}$.\footnote{For simplicity, we denote by $\sigma_{\mu,q}$ the average error in the ra and dec directions $[(\sigma_{\mu,q,\alpha}^2 +\sigma_{\mu,q,\delta}^2 )/2]^{1/2}$.} We bin the quasars into 12,288 equal-area pixels according to the \texttt{HEALPIX} scheme (corresponding to \texttt{nside}=32), as shown in the top panel of Fig.~\ref{fig:gaia_maps}. Each pixel $i$ is assigned the noise-weighted averages $\sigma_{\mu,i} = (\sum_{q \in i} 1/\sigma_{\mu,i}^2)^{-1/2}$ and $\vect{\mu}_i = \sigma_{\mu,i}^{-2} \sum_{q \in i} \vect{\mu}_q / \sigma_{\mu,q}^2$. Empty pixels are assigned infinite noise. The second and third panel of Fig.~\ref{fig:gaia_maps} illustrate maps of $|\vect{\mu}_i|$ and $\sigma_{\mu,i}$, respectively, showing smaller proper motion magnitudes in regions of high \textit{Gaia} exposure, which also show higher quasar counts. The fourth and fifth panel break down $\vect{\mu}_i$ into its two components; large-scale correlations (also seen in parallax) are easily visible by eye.

The VSH amplitudes of the data $\vect{\mu}_i = \vect{s}_i + \vect{n}_i$, namely $\mu_{\ell m}^{(1)} = s^{(1)}_{\ell m} + n^{(1)}_{\ell m}$ and $\mu_{\ell m}^{(2)} = s^{(2)}_{\ell m} + n^{(2)}_{\ell m}$, are presumed to be composed of both signal ($s$) and noise ($n$)  contributions. An astrometric lensing signal is expected to have $s_{\ell m}^{(2)} = 0$. We expect no cross-correlations between signal and noise:  $\langle s_{\ell m}^{(v)}  n_{\ell' m'}^{(v')} \rangle = 0 ~ \forall \, v,v',\ell,\ell',m,m'$. We assume the noise is independent between pixels and faithfully reported by \textit{Gaia}: $\langle \vect{n}_i \cdot \vect{n}_j \rangle = 2 \sigma_{\mu,i}^2 \delta_{ij}$. As we will see, the pixel-to-pixel independence assumption is incorrect on large angular scales. Futhermore, the uncertainties in \textit{Gaia}'s astrometric fit appear to not fully account for the noise budget. On the DR2 quasar dataset, $\langle \mu_q^2 / \sigma_{\mu,q}^2 \rangle \approx 2.43$, for example, in excess of the expectation of 2 for a bivariate normal random variable. 

We will apply the techniques outlined in App.~\ref{app:estimator-general} to evaluate the optimal estimator for the (poloidal) coarse-grained power spectrum:
\begin{align}
\left|\hat{\mu}^{(1)}_{B}\right|^2 = \sum_{B' i \alpha j \beta} \frac{1}{2}\left(F^{-1}\right)_{B B'} \frac{\mu_{i\alpha} \mu_{j \beta}}{\sigma_{\mu,i}^2 \sigma_{\mu,j}^2} P^{(1) B'}_{i \alpha j \beta}. \label{eq:muBestimator}
\end{align}
Specifically, we use noise weights $N_i = \sigma_{\mu,i}^2$ and data $d_{i\alpha} = \mu_{i\alpha}$ but omitting the last term in square brackets in Eq.~\ref{eq:estimatorlowS}, as we expect the latter noise subtraction to be an underestimate. We have also applied the spectral binning technique of App.~\ref{app:specbin} to coarse-grain the power spectrum with band matrices $W_{B\ell m}$: $\big|\hat{\mu}^{(1)}_{B}\big|^2 = \sum_{\ell m} W_{B \ell m} \big|\hat{\mu}^{(1)}_{\ell m}\big|^2$. The precise form of the band matrices is specified below Eq.~\ref{eq:Sbinned}. The binned response matrices $P^{(1)B}_{i \alpha j \beta}$ are specified in Eqs.~\ref{eq:Presponse} and \ref{eq:Pbinned}. The Fisher matrix of Eq.~\ref{eq:fisherlowS} is likewise binned as $F_{B B'} = \sum_{\ell m \ell' m'} F_{\ell m \ell' m'} W^\dagger_{\ell m B} W^\dagger_{\ell' m' B'}$. Finally, we can repeat the same procedure for the toroidal power spectrum estimator $\big|\hat{\mu}^{(2)}_{B}\big|^2$ in complete analogy with Eq.~\ref{eq:muBestimator} but with the replacement $P^{(1)} \leftrightarrow P^{(2)}$.

\begin{figure*}[tbp]
\centering
\includegraphics[trim = 69 0 56 0, clip, height=0.30\textwidth]{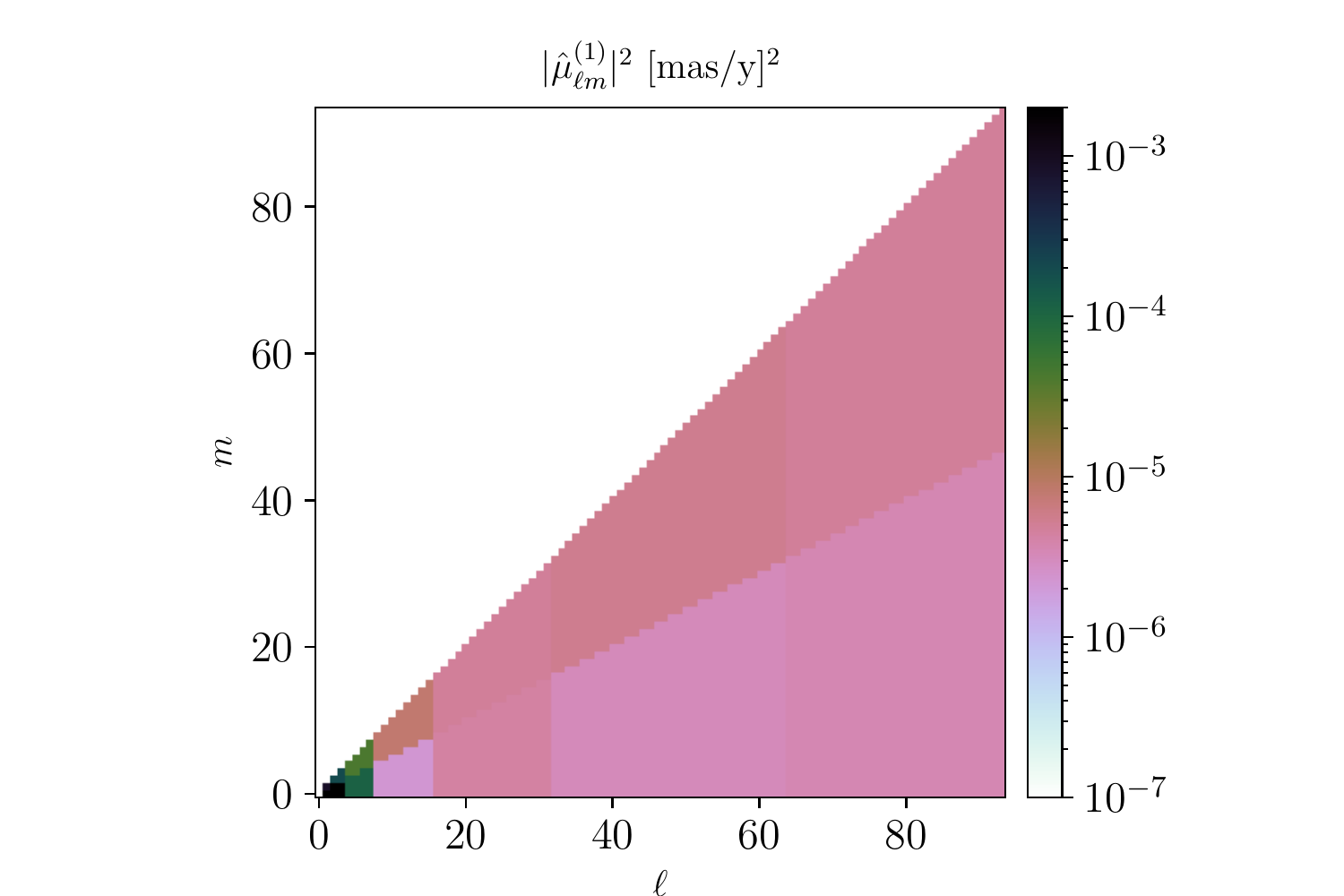} 
\includegraphics[trim = 69 0 56 0, clip, height=0.30\textwidth]{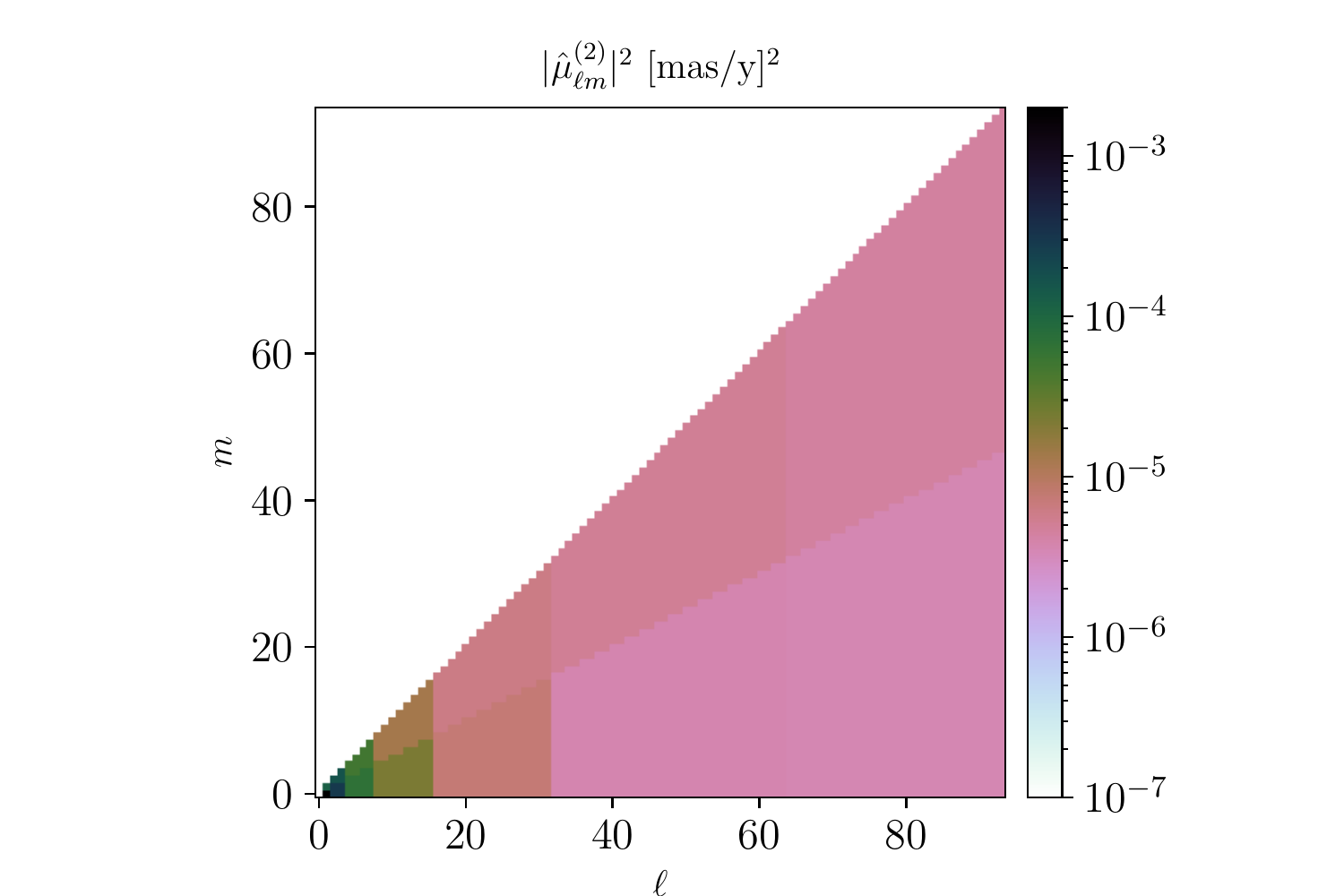}
\includegraphics[trim = 69 0 46 0, clip, height=0.30\textwidth]{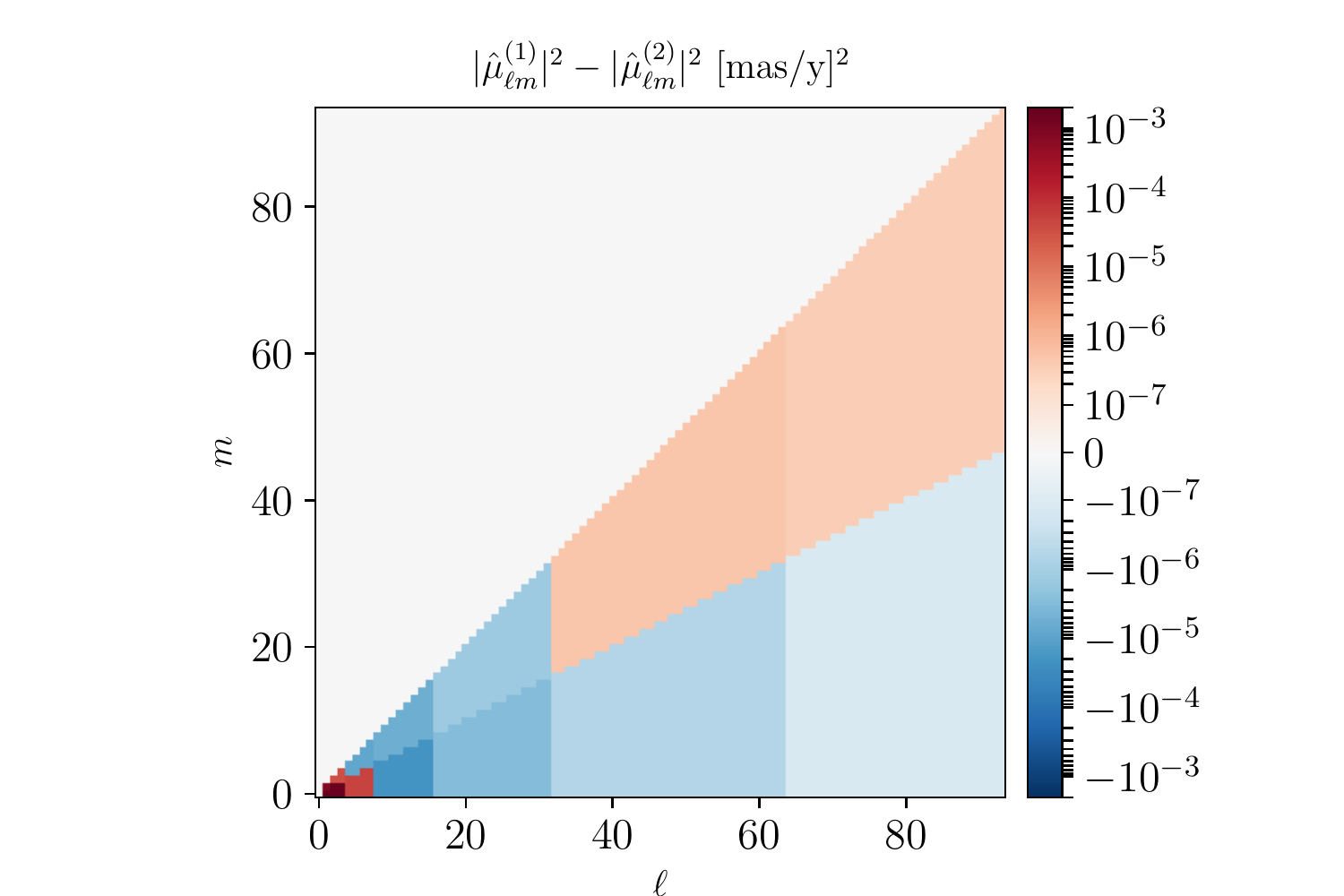}
\caption{Band-averaged estimates of the \Gaia DR2 quasar proper motion power spectra in poloidal modes (left panel) and toroidal modes (middle panel). A lensing signal would manifest itself as small additive contribution solely to the poloidal power spectrum (preferentially at high $|m|/\ell$, \emph{cf.}~Figs.~\ref{fig:m_abs} and~\ref{fig:m_asymm}), which can be revealed in the difference of the power spectra (right panel). This differences is currently consistent with noise-only statistical fluctuations at $\ell \gtrsim 16$. \nblink{07_vector_power_spectrum}}
\label{fig:pow_spec}
\end{figure*}

In the first two panels of Fig.~\ref{fig:pow_spec}, we plot the results of the spectrally binned estimators $\big|\hat{\mu}^{(1)}_{B}\big|^2$ and $\big|\hat{\mu}^{(2)}_{B}\big|^2$, respectively. Following the results of App.~\ref{app:estimator}, they are estimators of $S^{(v)}_B = \sum_{\ell m} W_{B \ell m} \big|s_{\ell m}^{(v)}\big|^2$ plus a noise contribution (from the second term in square brackets in Eq.~\ref{eq:estimatorlowS}) that can be shown to be the same for $v=1$ and $v=2$. This noise contribution is dominant in the individual power spectra, but the difference of the estimators shown in the third panel of Fig.~\ref{fig:pow_spec} is an unbiased estimator of the (coarse-grained) signal power spectrum under consideration $S^{(1)}_B - S^{(2)}_B$. We can see that this ``signal channel'' is substantially suppressed, and averages down (especially in the band averages $B$ with more $(\ell m)$ pairs) far below the mean power per mode. Finally, we see evidence for systematic excess power (above the expectation from Gaussian noise, \emph{i.e.,}~Eq.~\ref{eq:Ndiagonal}) for $\ell \lesssim 15$, systematic correlations previously reported by the \Gaia Collaboration and also present in both the parallax and proper motion power spectra~\cite{lindegren2018gaia,Undefined:2018amf}.

The signal channel features band-averaged power at the level of $10^{-7}\,\mathrm{mas^2\,\mathrm{yr}^{-2}}$ consistent with Gaussian noise in the two band averages between $64 \le \ell \le 93$, and only slightly more in the spectral bins of $32 \le \ell \le 63$ simply due to lower number of $(\ell m)$ pairs in those bins. This level of noise is not yet sufficient to tease out the small signals of those depicted in Fig.~\ref{fig:m_abs}. The baseline noise power scales as $\sigma_{\mu}^2/N_q$, the numerator of which is projected to improve with integration time as $\propto t_\mathrm{int}^{-3}$ (see V18 for details). Likewise, future \Gaia quasar catalogs will likely expand substantially (by at least a factor of four), as quasar identification methods mature.

\section{Conclusions}
\label{sec:conclusions}

Astrometry---the precise measurement of the positions and motions of celestial objects---offers a promising avenue to probe the nature of dark matter through induced lensing effects. In this paper, we have introduced a novel method to systematically leverage the measured correlated pattern of motions (transverse velocities and accelerations) induced by a population of Galactic subhalos on celestial bodies using the formalism of angular two-point correlation functions. We have shown how to calculate the lens-induced signal power spectrum for a population of lenses characterized by arbitrary population properties and internal characteristics through a vector spherical harmonic decomposition. This technique admits a number of checks and control channels: \emph{(i)} the signal should appear dominantly in the curl-free harmonic component, with the divergence-free harmonic component being populated exclusively by noise, and \emph{(ii)} the preferred motion of the Sun in the Milky Way should lead to a further a characteristic directional asymmetry in the signal channel.

Assuming putative noise properties based on current design specifications, we have shown that astrometric datasets deliverable by near-future surveys like SKA and WFIRST may harbor the imprint of substructure characteristic of a range of well-motivated new physics scenarios such as cold dark matter, the existence of compact dark objects, and scalar field dark matter. In particular, we showed that correlated astrometry has sensitivity to compact objects of much larger size and lower density than can be probed by conventional microlensing searches based on photometric measurements. Measurements by the ongoing \Gaia mission will already be able to access currently-unconstrained parameter space.

We have additionally demonstrated the feasibility of performing this measurement by constructing a vector spherical harmonics estimator and carrying out the harmonic decomposition of the proper motions of quasars in \Gaia's second data release (DR2). Although the current instrumental noise levels are not conducive to realistic searches for new physics, our proof-of-principle analysis can be carried over and applied to future astrometric datasets, including those in upcoming \Gaia data releases, in a straightforward manner.

Finally, we note that two-point correlations efficiently capture the statistical properties of a map only in the limit of the underlying signal being statistically Gaussian. While this is true to very good degree for the Cosmic Microwave Background for example~\cite{Akrami:2019izv}, our signal of interest is highly non-Gaussian (as apparent from Fig.~\ref{fig:population_maps}), and the spherical harmonic decomposition discards potentially large amounts of signal information. Methods accounting for statistics beyond the linear order---\emph{e.g.}, bispectra~\cite{Kamionkowski:2010me} and those based on convolutional filters~\cite{Krachmalnicoff:2019zjh,Perraudin:2018rbt}---may leverage this additional information in the substructure signal and significantly enhance sensitivity to dark matter substructure compared to that demonstrated in this paper. We leave the study and application of higher-order correlation statistics to astrometric lensing to future work.

The code used to obtain the results in this paper is available at \url{https://github.com/smsharma/
astrometry-lensing-correlations}.

\section*{Acknowledgements}
We thank Asimina~Arvanitaki, Masha~Baryakhtar, Laura~Chang, Doug~Finkbeiner, Junwu~Huang, Mariangela~Lisanti, Marilena~LoVerde, Cristina~Mondino, Robyn~Sanderson, Oren~Slone, and Anna-Maria~Taki for helpful conversations. SM and NW are supported by NSF grants PHY-1620727 and PHY-1915409, and the Simons Foundation. SM is additionally supported by the NSF CAREER grant PHY-1554858. KVT is supported by a Schmidt Fellowship funded by the generosity of Eric and Wendy Schmidt, by a grant-in-aid (No. de-sc0009988) from the U.S. Department of Energy, and funding by the Gordon and Betty Moore Foundation through Grant GBMF7392. This work made use of the NYU IT High Performance Computing resources, services, and staff expertise. This research has made use of NASA's Astrophysics Data System. This work has made use of data from the European Space Agency (ESA) mission {\it Gaia} (\url{https://www.cosmos.esa.int/gaia}), processed by the {\it Gaia} Data Processing and Analysis Consortium (DPAC, \url{https://www.cosmos.esa.int/web/gaia/dpac/consortium}). Funding for the DPAC has been provided by national institutions, in particular the institutions participating in the {\it Gaia} Multilateral Agreement. This project was developed in part at the 2019 Santa Barbara {\it Gaia} Sprint, hosted by the Kavli Institute for Theoretical Physics at the University of California, Santa Barbara. This research was supported in part at KITP by the Heising-Simons Foundation and the National Science Foundation under Grant No. NSF PHY-1748958. This research made use of the \texttt{astropy}~\cite{Robitaille:2013mpa,Price-Whelan:2018hus}, \texttt{CLASS}~\cite{Blas:2011rf}, \texttt{COLOSSUS}~\cite{Diemer:2017bwl}, \texttt{healpy}~\cite{Gorski:2004by,Zonca2019}, \texttt{IPython}~\cite{PER-GRA:2007}, \texttt{Jupyter}~\cite{Kluyver2016JupyterN}, \texttt{matplotlib}~\cite{Hunter:2007}, \texttt{mpmath}~\cite{mpmath}, \texttt{NumPy}~\cite{numpy:2011}, \texttt{pandas}~\cite{mckinney-proc-scipy-2010}, \texttt{seaborn}~\cite{seaborn}, \texttt{SciPy}~\cite{2020SciPy-NMeth}, and \texttt{tqdm}~\cite{da2019tqdm} software packages. 

\appendix 

\section{Derivation of equations}
\label{app:derivations}

\subsection{Velocity and acceleration power spectra}

In this appendix, we derive the formulae of Eqs.~\eqref{eq:pspec_mu} and \eqref{eq:pspec_alpha}. It is sufficient to calculate the induced $\mu_{\ell m}^{(1)}$ of a single lens at some distance $D_l$ and characterized by an enclosed mass function $M(b)$ at a conveniently chosen location and transverse velocity direction $\hat{v}_l$, as the contributions to $C^{\mu(1)}_\ell$ are additive among the lenses, and independent of location/direction by rotational invariance.

The first time derivative of the lensing deflection potential from a single lens is:
\begin{align}
\frac{\dd}{\dd t} \psi = \frac{4\GN M(\beta D_l)}{\beta D_l^2}(\hat{\vect{\beta}} \cdot {\vect{v}}_l ). \label{eq:psidot}
\end{align}
Using $\vect{\nabla}_{\vect{\theta}}^2 Y_{\ell m} = -\ell(\ell+1)Y_{\ell m}$ and integration by parts of Eq.~\eqref{eq:harmdecomposition}, we find:
\begin{align}
\mu_{\ell m}^{(1)} = \sqrt{\ell(\ell+1)}\int \dd\Omega\, \left(\frac{\dd}{\dd t} \psi(\vect{\theta})\right) Y_{\ellm}^*(\vect{\theta}) \label{eq:muperlens1}
\end{align}
We can evaluate this expression for a lens on the celestial north pole, which implies $\vect{\theta} = \vect{\beta}$, moving with a transverse velocity vector specified by $\vect{v}_l \cdot \hat{\vect{\theta}}  = v_l \cos \phi$.
That allows us to write Eq.~\eqref{eq:muperlens1} as:
\begin{align}
\mu_{\ell m}^{(1)} &= \frac{4 \GN v_l}{D_l^2} \sqrt{\ell(\ell+1)}\label{eq:muNP}\\
&\phantom{=}\times \int_0^\pi \dd \theta \, \int_0^{2\pi} \dd \phi \, \cos\phi  \frac{\sin \theta}{\theta} M(\theta D_l) Y^*_{\ell m}(\theta,\phi) \nonumber\\
&\simeq \frac{\GN v_l}{D_l^2} \ell \sqrt{8\pi \ell}\left[\delta_{m,-1}-\delta_{m,1}\right]\int_0^\infty \dd \theta \, M(\theta D_l) J_1(\ell \theta).\nonumber
\end{align}
To get to the final line, we made use of the approximation $P_\ell^m(\cos\theta) \simeq (-1)^m \ell^m J_m(\ell \theta)$ valid for $\theta \ll 1$ (but potentially large $\ell \theta$), and kept only the leading term in the  $1/\ell$ expansion. 

We can similarly find the acceleration power spectrum by computing the amplitudes
\begin{align}
\alpha_{\ell m}^{(1)} = \sqrt{\ell(\ell+1)}\int \dd\Omega\, \left(\frac{\dd^2}{\dd t^2} \psi(\vect{\theta})\right) Y_{\ellm}^*(\vect{\theta}), \label{eq:alphaperlens1}
\end{align}
using the fact that
\begin{align}
\frac{\dd^2}{\dd t^2} \psi &= \frac{4\GN}{D_l^3} \bigg\lbrace\frac{(\hat{\vect{\beta}} \cdot {\vect{v}}_l)^2}{\beta}  \partial_\beta M(\beta D_l) \\
&\phantom{=\frac{4\GN}{D_l^3} \bigg\lbrace} + \frac{v_l^2 - 2 (\hat{\vect{\beta}} \cdot {\vect{v}}_l)^2 }{\beta^2}M(\beta D_l)\bigg\rbrace. \nonumber
\end{align}
Evaluating at the north pole as in Eq.~\eqref{eq:muNP} yields
\begin{align}
\alpha_{\ell m}^{(1)}  = \frac{\GN v_l^2}{D_l^3} \ell^2 \sqrt{8\pi \ell}\left[\delta_{m,0}-\frac{\delta_{|m|,2}}{2} \right]\int_0^\infty \dd \theta \, M(\theta D_l) J_1(\ell \theta),\label{eq:alphaNP}
\end{align}
where we integrated by parts (and assumed $M(b)\to 0$ as $b\to 0$) to cast the acceleration amplitude in a similar form as the velocity amplitude. We note that the dependence on the enclosed mass function in Eqs.~\eqref{eq:muNP} and \eqref{eq:alphaNP} is identical, giving the simple scaling relation between the corresponding power spectra stated in Eq.~\eqref{eq:pspec_alpha}.

\subsection{Flat-sky velocity power spectrum}
For deep surveys over small patches of sky, it is more appropriate to construct flat-sky power spectra. We can express the proper motion field
\begin{align}
\vect{\mu}(\vect{\theta}) = \int \ddbar^2 k \, e^{i \vect{k} \cdot \vect{\theta}} \tilde{\vect{\mu}}(\vect{k})
\end{align}
in terms of its flat Fourier modes:
\begin{align}
\tilde{\vect{\mu}}(\vect{k}) &=  \int \dd^2 \theta \, e^{-i \vect{k} \cdot \vect{\theta}} \vect{\mu}(\vect{\theta})\\
&= -\frac{4\GN}{D_l^2} 2\pi (\vect{k} \cdot \vect{v}_l) \hat{\vect{k}} \int_0^\infty \dd \theta \, M(\theta D_l) J_1(\theta k). \nonumber
\end{align}
In the second line, we computed the Fourier amplitude for a lens at the origin. One can compute this via direct computation with Eq.~\eqref{eq:mureal} or from Eqs.~\eqref{eq:deflectionpotential} and \eqref{eq:psidot} using integration by parts. Just like for the vector spherical harmonic amplitude of Eq.~\eqref{eq:muNP}, the Fourier amplitude is proportional to the integral of the enclosed mass function times a Bessel function. The analogous Fourier amplitudes for the acceleration field $\vect{\alpha}(\vect{\theta})$ are easily found to be expressible in terms of the proper motion Fourier amplitudes:
\begin{align}
\tilde{\vect{\alpha}}(\vect{k}) = - \frac{i \vect{k} \cdot \vect{v}_l}{D_l} \tilde{\vect{\mu}}(\vect{k}), \label{eq:alphak}
\end{align}
valid for any one lens with a sufficiently smooth density profile. Equation~\eqref{eq:alphak} is the flat-sky analog of the identity in Eq.~\eqref{eq:pspec_alpha}. Here we have phrased the effects in terms of continuous Fourier transforms for simplicity, but in a data analysis one would of course compute the appropriate discrete Fourier transforms.

We further note that only the longitudinal components $\tilde{\vect{\mu}} \cdot \hat{\vect{k}}$ of the Fourier modes are populated, and that the transverse components are zero: $\tilde{\vect{\mu}} \times \hat{\vect{k}} = 0$. The same is true for accelerations. This fact is the flat-sky equivalent of the lensing signal not contributing to the toroidal amplitudes $\mu_{\ell m}^{(2)} = \alpha_{\ell m}^{(2)} = 0$. 

The directional asymmetry discussed in Sec.~\ref{sec:directionalasymm} is more easily seen than in the case of vector spherical harmonics. Each lens makes a contribution to the power in proper motion as $\propto (\vect{k}\cdot \vect{v}_l)^2$ and in acceleration as $\propto (\vect{k}\cdot \vect{v}_l)^4$. The expectation values $\langle v_{l,\varphi}^2 \rangle$ and $\langle v_{l,\varphi}^4 \rangle$ of the galactic longitude velocity components are larger than their galactic latitude equivalents of $v_{l,\theta}$, so more power is expected in modes with $\hat{\vect{k}}$ pointed parallel to the Galactic equator.

\subsection{Power spectrum for scalar dark matter perturbations}\label{app:scalar}

In this appendix, we compute the power spectrum of the velocity and acceleration distortion in a dark matter halo made up of a real scalar field alluded to in Sec.~\ref{sec:scalar}. We will assume that the scalar field ensemble is in a mixed state (such as a thermal state with ``temperature'' equal to the virial temperature) in which:
\begin{align} 
\left\langle a_{\vect{k}}^\dagger a_{\vect{q}} \right\rangle = n_0 f(\vect{k}) \deltabarthree(\vect{k}-\vect{q}). \label{eq:scalarensemble}
\end{align}
The total particle number density is $n_0$, and $f(\vect{k})$ is defined as the momentum distribution with $\int \ddbar^3 k \, f(\vect{k}) = 1$.
All other contractions with the external state, e.g.~those of the form $\langle a a \rangle$ and $\langle a^\dagger a^\dagger\rangle$ are assumed to be zero, as we expect virialization to scramble all phase information. We will also effectively take all commutators $[a_{\vect{k}}^\dagger, a_{\vect{q}}] = 0$, since it can be shown that terms involving commutators are suppressed by inverse powers of the occupation number $n_0 / \sigma_k^3\gg 1$ where $\sigma_k$ is a typical momentum, \emph{i.e.}~we are doing a classical expansion. Expressing the field as a superposition of momentum modes,
\begin{align}
\phi(x) = \int \frac{\ddbar^3 k}{\sqrt{2k^0}} \left(a_{\vect{k}} e^{-i k \cdot x} + a_{\vect{k}}^\dagger e^{+i k\cdot x}\right),
\end{align}
with $k^0 \equiv \sqrt{m^2 + \vect{k}^2}$, we can compute expectation values such as $\langle \phi \rangle = 0$ and that of the density $\rho = [\dot{\phi}^2 + (\nabla \phi)^2 + m^2 \phi^2] /2$:
\begin{align}
\langle \rho \rangle =  n_0 \int \ddbar^3 k\, f(\vect{k}) k^0 \simeq n_0 m \equiv \rho_0.
\end{align}
The approximate equality holds for a nonrelativistic momentum distribution, which is the case of interest. 

The aforementioned density fluctuations give rise to a nontrivial density correlation function
\begin{align}
&\hspace{-1em}\langle \rho(x) \rho(x') \rangle = \langle \rho \rangle^2 + \frac{n_0^2}{4} \int \frac{\ddbar^3 k_1}{k^0_1} \, \frac{\ddbar^3 k_2}{k^0_2} \,  f(\vect{k}_1) f(\vect{k}_2) \label{eq:rhocorr}
\\
&\times\bigg\lbrace \left[m^2 -  k^0_1 k^0_2 - \vect{k}_1 \cdot \vect{k}_2 \right]^2\cos\left[(k_1 + k_2)\cdot (x-x')\right] \nonumber\\
&\phantom{\times\bigg \lbrace} + \left[m^2 +  k^0_1 k^0_2 + \vect{k}_1 \cdot \vect{k}_2 \right]^2 \cos\left[(k_1 - k_2)\cdot (x-x')\right]  \Bigg\rbrace \nonumber
\end{align}
from which it can be read off that the fractional variance of $\rho$, namely ($\langle \rho^2 \rangle - \langle \rho \rangle^2)/\langle \rho\rangle^2$, is order unity. There are thus irreducible, \emph{unbound} $\mathcal{O}(1)$ density fluctuations in a virialized scalar dark matter halo, regardless of cosmological history and the \emph{bound} substructure of the halo, of which there is generally less with ultralight scalar dark matter than with CDM.\footnote{Note that large-misalignment scalar dark matter models have \emph{more} bound substructure~\cite{Arvanitaki:2019rax}.}
In what follows, we will take
\begin{align}
f(\vect{k}) = \frac{(2\pi)^{3/2}}{\sigma_k^3} \exp\left\lbrace-\frac{(\vect{k}-\vect{k}_\odot)^2}{2\sigma_k^2} \right\rbrace \label{eq:fmom}
\end{align}
where the momentum dispersion is $\sigma_k \equiv m \sigma_v$ with $\sigma_v \approx 166\,\mathrm{km \, s^{-1}}$ and the average momentum in the Sun's rest frame is $\vect{k}_\odot \equiv m \vect{v}_\odot$. From Eq.~\eqref{eq:rhocorr}, we can estimate the typical mass $M_0$ and size $R_0$ of these fluctuations to be those of Eqs.~\eqref{eq:scalarM0} and \eqref{eq:scalarR0}.
Since $\phi$ is a Gaussian random field, not all overdensities contain the same mass, but as we will see below, the constant $C$ can be made more precise in the context of the lens-induced power spectra.

We will now compute to what degree these density fluctuations cause distortions in proper motions and accelerations of luminous sources. Defining the Fourier transform of the observable $\mathcal{O}$ to be $\tilde{\mathcal{O}}(\vect{k}) = \int \ddbar^3k \, e^{-i \vect{k} \cdot \vect{x} } \mathcal{O}(\vect{x}) $ and its power spectrum $\langle \tilde{\mathcal{O}}(\vect{k}) \tilde{\mathcal{O}}(\vect{k}')^*  \rangle \equiv P_{\mathcal{O}}(\vect{k}) \deltabarthree(\vect{k}-\vect{k}')$, we can compute from Eq.~\eqref{eq:rhocorr} the power spectrum of $\dot{\rho}$:
\begin{align}
\hspace{-0.8em}P_{\dot{\rho}}(\vect{k}) &\simeq \frac{\rho_0^2}{8m^2} \int \ddbar^3 q\, f(\vect{q}) \big[(\vect{k}^2 - 2\vect{k}\cdot \vect{q})^2 f(\vect{k}-\vect{q}) \label{eq:Prhodot} \\
&\hspace{9em}+(\vect{k}^2 + 2\vect{k}\cdot \vect{q})^2 f(-\vect{k}-\vect{q}) \big] \nonumber\\
&= \frac{\pi^{3/2}}{4} \frac{\rho_0^2 (\vect{k}^2 + 8 \vect{k}_{\odot}^2)}{m^2 \sigma_k} \left[e^{-\frac{(\vect{k}+2\vect{k}_\odot)^2}{4\sigma_k^2}}+e^{-\frac{(\vect{k}-2\vect{k}_\odot)^2}{4\sigma_k^2}} \right]\nonumber.
\end{align}
In the second line, we evaluated the integral with the momentum distribution of Eq.~\eqref{eq:fmom}. The above power spectrum is computed from the equal-time correlation function $\langle \dot{\rho}(t,\vect{x}) \dot{\rho}(t,\vect{x}') \rangle$ keeping only the ``slow'' term in the third line of Eq.~\eqref{eq:rhocorr}, and dropping the ``fast'' term of the second line.\footnote{The fast term is not only suppressed in magnitude but averages down out severely when integrated over two lines of sight. It is justified to compute the power on the equal-time correlation function of the slow term because the coherence time $t_\mathrm{coh} \sim m/\sigma_k^2$ is much longer than the light crossing time of a de Broglie fluctuation $t_\mathrm{cross} \sim 1/\sigma_k$.} Along the same lines, we have the $\ddot{\rho}$ spectrum:
\begin{align}
\hspace{-1.3em} P_{\ddot{\rho}}(\vect{k}) &\simeq \frac{\rho_0^2}{32m^4} \int \ddbar^3 q\, f(\vect{q}) \big[(\vect{k}^2 - 2\vect{k}\cdot \vect{q})^2 f(\vect{k}-\vect{q}) \label{eq:Prhoddot}\\
&\hspace{9em}+(\vect{k}^2 + 2\vect{k}\cdot \vect{q})^2 f(-\vect{k}-\vect{q}) \big] \nonumber\\
&= \frac{3\pi^{3/2}}{8} \frac{\rho_0^2\sigma_k (\vect{k}^2 + 8 \vect{k}_{\odot}^2)^2}{m^4} \left[e^{-\frac{(\vect{k}+2\vect{k}_\odot)^2}{4\sigma_k^2}}+e^{-\frac{(\vect{k}-2\vect{k}_\odot)^2}{4\sigma_k^2}} \right]\nonumber. 
\end{align}
These spatio-temporal density fluctuations will give rise to corresponding fluctuations in the gravitational potential $\Phi$ through the gravitational Poisson equation: $\nabla^2 \Phi = 4 \pi \GN \rho$. The power spectra of the gravitational potential fluctuations can thus be written as $P_{\dot{\Phi}}(\vect{k}) = (4\pi \GN)^2 P_{\dot{\rho}}(\vect{k})/\vect{k}^4$ and likewise for higher time derivatives. The reduced lensing deflection potential $\psi$ of Eq.~\eqref{eq:deflectionpotential} is a line-of-sight integral of the gravitational potential, $\psi(\vect{\theta}) \equiv 2 \int_0^{z_s} \dd z \, \Phi(\vect{x}) (z_s - z)/(z_s z)$, where it is now understood that $\vect{x} = \lbrace z \vect{\theta},z\rbrace$, and $z_s$ is the source distance. Finally, we can write the harmonic coefficients of Eq.~\eqref{eq:harmdecomposition} as $\mu_{\ell m}^{(1)} = - \sqrt{\ell(\ell+1)} \int \dd^2\theta\, \dot{\psi}(\vect{\theta}) Y_{\ell m}^*(\vect{\theta})$ after integrating by parts, and analogously for $\alpha_{\ell m}^{(1)}$ and $\ddot{\psi}$.

We are now in a position to calculate the angular power spectra of the proper motion and acceleration, after having collected all the necessary ingredients above:
\begin{align}
\hspace{-0.8em} \big \langle \big|\mu_{\ell m}^{(1)}\big|^2\big\rangle &= 4 \ell(\ell+1)\int_0^{z_s} \dd z  \int_0^{z_s} \dd z'\, \frac{z_s -z}{z_s z} \frac{z_s - z'}{z_s z'} \\
&\int \dd^2 \theta \, \dd^2 \theta' \,  Y_{\ell m}^*(\vect{\theta}) Y_{\ell m}(\vect{\theta}') \int \ddbar^3 q\, P_{\dot{\Phi}}(\vect{q}) e^{i\vect{q}\cdot (\vect{x}-\vect{x}')}\nonumber,
\end{align}
and likewise for $ \langle |\alpha_{\ell m}^{(1)}|^2\rangle$ but with $P_{\ddot{\Phi}}(\vect{q})$. If we integrate the fluctuations over a sphere with constant density $\rho_0$, dispersion $\sigma_k$, and radius $D$, take $z_s \to \infty$ and $\vect{k}_\odot = 0$, and approximate $\int_0^D \dd z \, j_\ell (q z) / z \simeq \Theta(q - \ell / D) \sqrt{\pi/2\ell^3}$, we can evaluate all integrals and find:
\begin{align}
\big \langle \big|\mu_{\ell m}^{(1)}\big|^2\big\rangle &= \underbrace{32\pi^4 \frac{\GN^2 \rho_0^2}{m^2}}_{1.3 \times 10^{-8} \frac{\mu\mathrm{as}^2}{\mathrm{y}^{2}} m_{22}^{-2}} \frac{\mathrm{erfc}(\frac{\ell}{2\sigma_k D})}{\ell}  \label{eq:mulmscalar}
\\
\big \langle \big|\alpha_{\ell m}^{(1)}\big|^2\big\rangle &= \underbrace{96\pi^4 \GN^2 \rho_0^2 \sigma_v^4}_{8.4 \times 10^{-20} \frac{\mu\mathrm{as}^2}{\mathrm{y}^{4}}}\frac{\mathrm{erfc}(\frac{\ell}{2\sigma_k D})+\frac{\ell}{\sqrt{\pi}\sigma_k D} e^{-\frac{\ell^2}{4\sigma_k^2 D^2}}}{\ell} \nonumber
\end{align}
It can be shown that one gets exactly the same angular power spectrum (with the same assumptions on $\rho_0, \sigma_k$, $D$, $z_s$, and $\vect{v}_\odot$) for an ensemble of Gaussian lenses (Eq.~\eqref{eq:Gaussianrho}) with uniform number density $\rho_0/M_0$, and mass $M_0$ and radius $R_0$ from Eqs.~\eqref{eq:scalarM0} and \eqref{eq:scalarR0}, assuming $C = 4/3\pi^{3/2}$ for $\big \langle \big|\mu_{\ell m}^{(1)}\big|^2\big\rangle$ and $C = 32/15\pi^{3/2}$ for $\big \langle \big|\alpha_{\ell m}^{(1)}\big|^2\big\rangle$. This means we can plot the sensitivity to scalar dark matter in the parameter space for compact objects explored in Sec.~\ref{sec:compact}. This is shown in the right panel of Fig.~\ref{fig:compact_sens} as the dashed horizontal and vertical lines denoting the densities and masses respectively of scalar field dark matter particles, for two benchmark points $m = 10^{-21}\,\mathrm{eV}$ and $m = 10^{-22}\,\mathrm{eV}$.

\section{Signal significance and scalings} 
\label{app:significance}

\subsection{Fisher information formalism}

We appeal to the Fisher information formalism (see, \emph{e.g.}, Ref.~\cite{Edwards:2017mnf} for a review) to isolate and study the contribution of different multipoles in a power spectrum measurement. 
For a given all-sky equivalent signal $C_{\ell}$ and noise configuration $N_{\ell}$, the Fisher information contained in a mode $\ell$  simplifies to~\cite{Edwards:2017mnf}
\begin{equation}
F_\ell = f_\mathrm{sky}(\ell + 1/2) \left(\frac{C_{\ell}}{C_{\ell} + N_{\ell}}\right)^2
\label{eq:fisher_l}
\end{equation}
where $f_\mathrm{sky}\equiv \Omega_\mathrm{sky}/(4\pi)$ is the fraction of the sky over which the measurement is made for sky coverage solid angle $\Omega_\mathrm{sky}$. Formally, the Fisher information corresponds to the inverse of the minimum possible variance with which a measurement can be made, and quantifies the information extractable from each mode. Note that for a partial-sky measurement, both the signal and noise scale the same way $\propto f_\mathrm{sky}$ with sky coverage, and the information loss comes from the mode multiplicity pre-factor in Eq.~\eqref{eq:fisher_l}. Unless explicitly specified, the power spectra $C_{\ell}$ may refer to either the expected (poloidal) proper motion or proper acceleration signal, with $N_\ell$ referring to the corresponding noise spectrum.

For a power spectrum measurement of multipoles in the range $[\ell_\mathrm{min}, \ell_\mathrm{max}]$ the maximum significance of a given signal is given by the square root of the inverse covariance, and with each mode constituting an independent measurement can be computed from the Fisher information as
\begin{equation}
\sigma_\mathrm{sig}\equiv\mathrm{Cov}^{-1/2}=\sqrt{\sum_{\ell = \ell_\mathrm{min}}^{\ell_\mathrm{max}}F_\ell}.
\label{eq:signif}
\end{equation}
Thus, $F_\ell$ quantifies the contribution of each mode to the total signal significance. 

From Eqs.~\eqref{eq:fisher_l}--\eqref{eq:signif}, we can immediately understand how the signal significance scales with various measurement characteristics. In particular, we have the scalings $\sigma_\mathrm{sig}\propto \Omega_\mathrm{sky}^{1/2}\Sigma_q\sigma_{\mu/\alpha}^{-2}$ where  $\Sigma_q$ is the number density of observed sources and $\sigma_{\mu/\alpha}$ their effective astrometric measurement uncertainty. 

\subsection{Population of point lenses}

Next, we illustrate how the significance is affected by various \emph{signal} properties for a few toy examples to gain intuition for the various relevant scales in the problem.
For a population of point lenses, the Fisher information per $\ell$ mode of the velocity power spectrum approximately grows as $F_\ell^\mu \appropto \ell$ due to the scale invariance of the signal and of the noise. The total significance then also grows approximately linearly with maximum multipole $\ell$ probed, $\sigma_\mathrm{sig}^\mu \appropto \ell_\mathrm{max}$, in the noise-dominated ($N_\ell^\mu\gg C_{\ell}^{\mu (1)}$) regime. 
For accelerations, the significance with increasing maximum multipole $\ell_\mathrm{max}$ grows approximately as $\sigma_\mathrm{sig}^\alpha\propto\ell_\mathrm{max}^3$. Thus, the acceleration power spectrum is even more sensitive to smaller scales compared to the proper motion power spectrum. 
In practice, the maximum possible multipole $\ell_\mathrm{max}$ is limited by telescope resolution and finite source density.

\begin{figure}[htbp]
\centering
\includegraphics[width=0.45\textwidth]{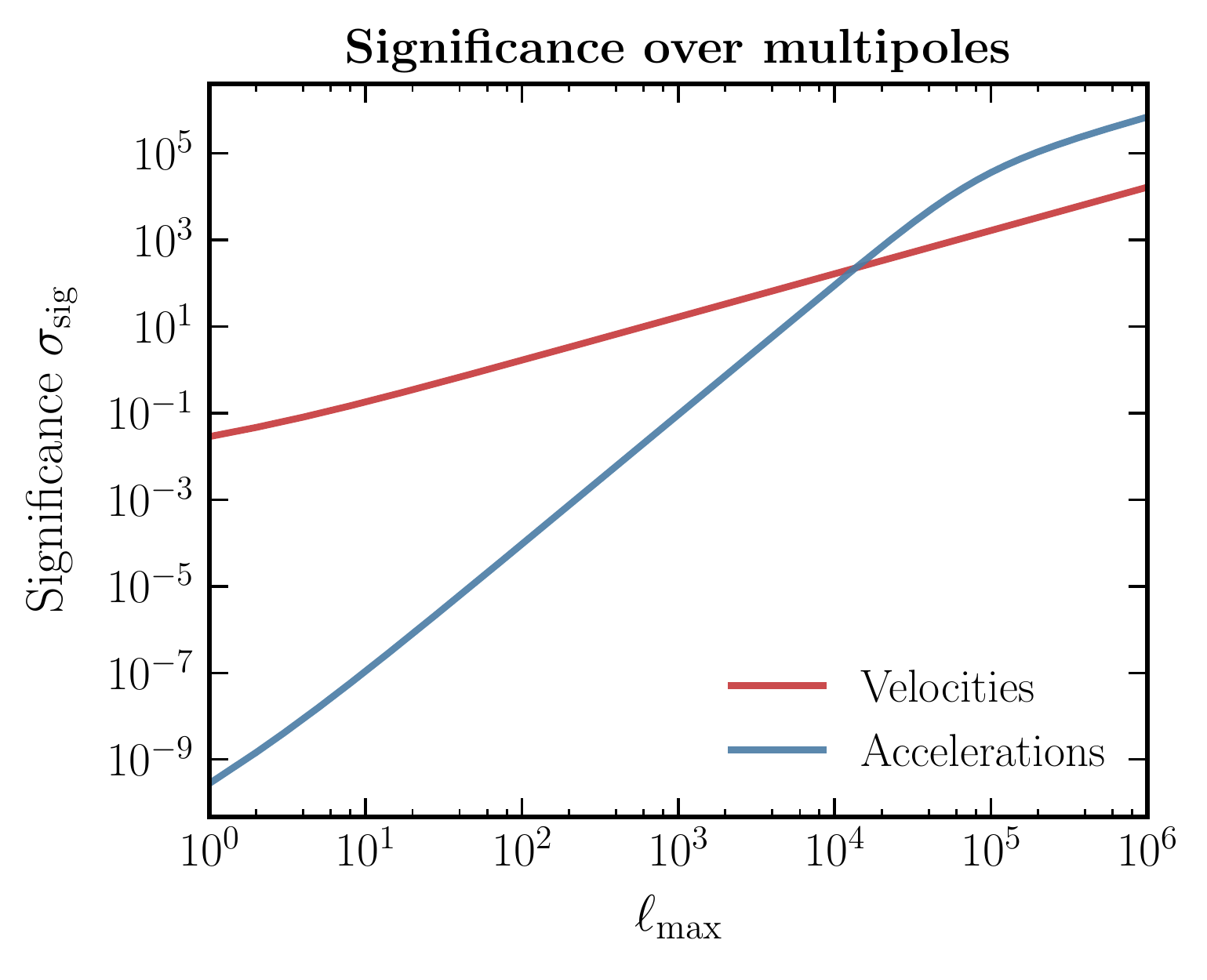}
\caption{Signal significance per lens as a function of maximum multipole $\ell_\mathrm{max}$ probed, shown for a population of $M_0=10^8$\,M$_\odot$ point source lenses uniformly distributed between $D_l^{\mathrm{min}}=0.1$\,kpc and $D_l^{\mathrm{max}}=10$\,kpc and normalized to the local dark matter density $\rho_\mathrm{DM}=0.4$\,GeV\,cm$^{-3}$ with $v_l=10^{-3}$. $N_q=10^8 (10^{11})$ background sources with measurement errors $\sigma_\mu=100\,\mu$as\,yr$^{-1}$ ($\sigma_\alpha=10\,\mu$as\,yr$^{-2}$) are assumed for proper motion and acceleration measurement. Shown are the contributions from the proper motion (red) and proper acceleration (blue) power spectra. The acceleration power spectrum probes comparatively smaller scales and smaller distances. \nblink{03_scalings}} 
\label{fig:mualpha_compact}
\end{figure}

Figure~\ref{fig:mualpha_compact} illustrates the detection significance as a function of maximum multipole $\ell_\mathrm{max}$, as defined in Eq.~\eqref{eq:signif}, for a population of $M_0=10^8$\,M$_\odot$ point source lenses uniformly distributed between $D_l^{\mathrm{min}}=0.1$\,kpc and $D_l^{\mathrm{max}}=10$\,kpc making up all of the local dark matter density $\rho_\mathrm{DM}=0.4$\,GeV\,cm$^{-3}$. $N_q=10^8 (10^{11})$ background sources with measurement errors $\sigma_\mu=100\,\mu$as\,yr$^{-1}$($\sigma_\alpha=10\,\mu$as\,yr$^{-2}$) assumed for proper motion (acceleration) measurements, for illustration. The proper acceleration power spectrum (blue line) is preferentially sensitive to smaller scales compared to the proper motion power spectrum (red line), and thus to smaller impact parameters (and therefore also more compact objects, as we will see below). 
For a general lens distribution, the smaller the typical angular scale that contributes to the total power, the greater the relative importance of accelerations. Its relative importance also grows with integration time $\tau$, since typically $\sigma_\alpha/\sigma_\mu\sim1/\tau^2$ (see V18 for details). These arguments naturally carry over to the case of extended lenses, which we consider next.

\begin{figure*}[htbp]
\centering
\includegraphics[width=0.9\textwidth]{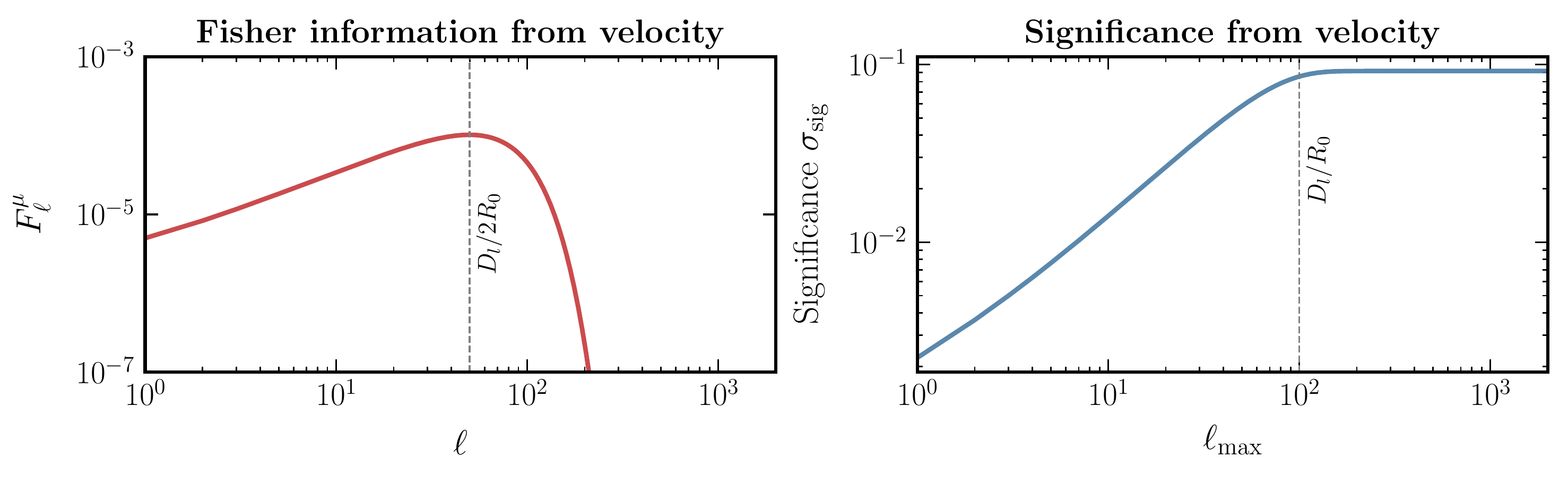}
\includegraphics[width=0.9\textwidth]{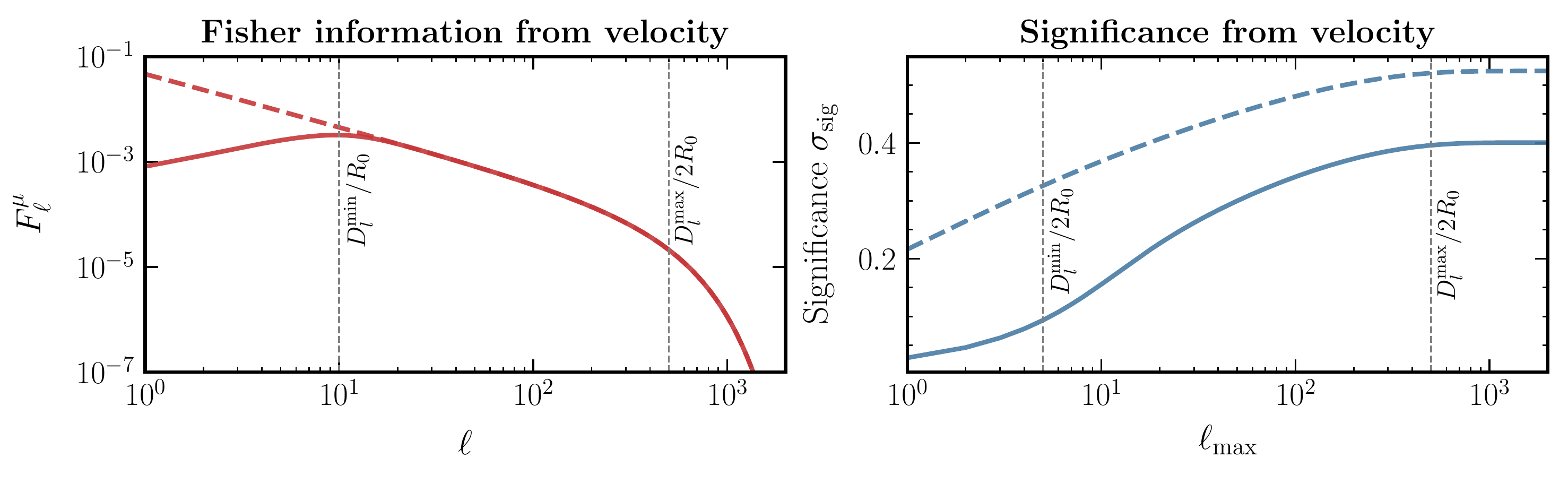}
\includegraphics[width=0.9\textwidth]{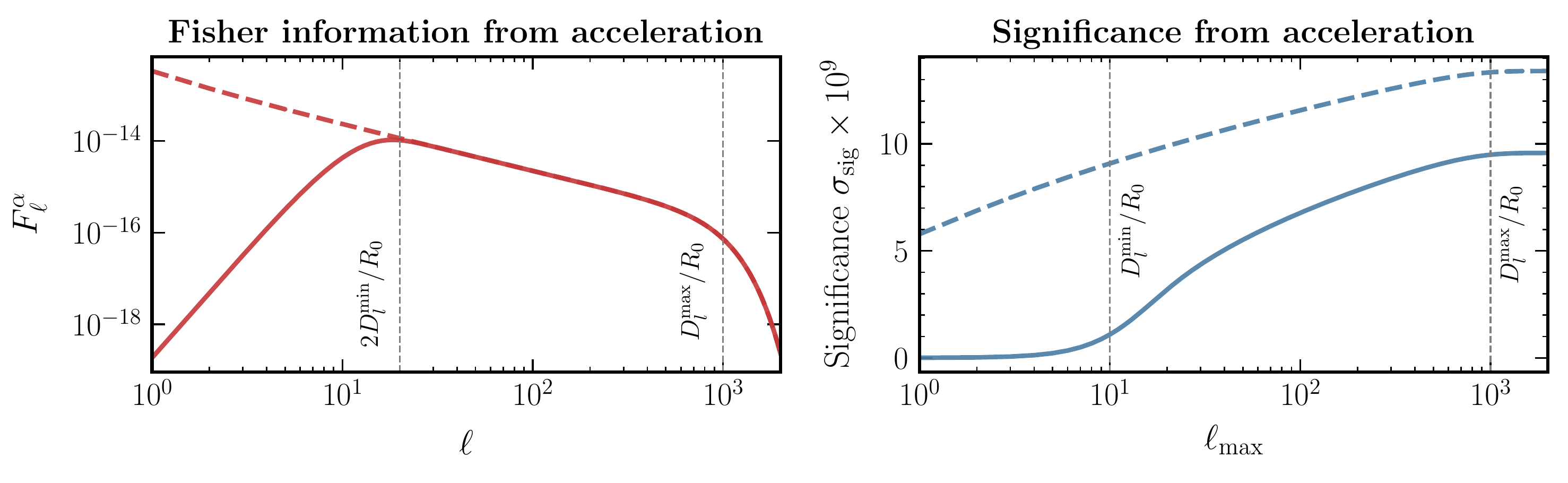}
\caption{Fisher information (left column) and significance (right column), as defined in Eq.~\eqref{eq:fisher_l} and Eq.~\eqref{eq:signif} respectively, for Gaussian lenses at equal-distance $D_l=1\,\mathrm{kpc}$ (top row, from velocity power spectra and shown per lens), a population between $D_{l}^{\mathrm{min}}=0.1\,\mathrm{kpc}$ and $D_{l}^{\mathrm{max}}=10\,\mathrm{kpc}$ (middle and bottom row, from velocity and acceleration power spectra respectively, normalized to the local dark matter density $\rho_\mathrm{DM}=0.4$\,GeV\,cm$^{-3}$). The same population parameters as in Fig.~\ref{fig:mualpha_compact} are assumed, with spatial extension $R_0=10$\,pc. Dotted lines correspond to taking $D_l^\mathrm{min}=0$. Measurement errors $\sigma_\mu=50\,\mu$as\,yr$^{-1}$ and $\sigma_\alpha=5\,\mu$as\,yr$^{-2}$ are assumed. \nblink{03_scalings}} 
\label{fig:fisher_sig}
\end{figure*}

\subsection{Population of extended lenses}\label{app:aristotle}

The Fisher information and significance for a population of extended subhalos with Gaussian internal density profiles (see Sec.~\ref{sec:extended_pop}) is illustrated on the left and right of the top row of Fig.~\ref{fig:fisher_sig}, respectively, shown as the per-lens contribution \emph{at fixed line of sight distance}. The same lens properties are assumed as for Fig.~\ref{fig:mualpha_compact}, with lens size $R_0=10$\,pc and lenses at $D_l=1$\,kpc, assuming noise properties $\sigma_\mu = 50\,\mu\mathrm{as}\,\mathrm{yr}^{-1}$ and $N_q = 10^8$ for illustration. Maximum Fisher information is contained at scales $\ell \approx {D_l}/{2R_0}$, with the significance growing linearly with $\ell_\mathrm{max}$ until $\ell \approx {D_l}/{R_0}$ when it plateaus and there is little information in higher modes. 

Next, the Fisher information and significance for a population of Gaussian lenses distributed in an Aristotelian ball between $D_l^{\mathrm{min}}$ and $D_l^{\mathrm{max}}$, otherwise with the same lens properties as in Fig.~\ref{fig:mualpha_compact}, are shown in the middle panel of Fig.~\ref{fig:fisher_sig}. These are normalized to the local dark matter density $\rho_\mathrm{DM}=0.4$\,GeV\,cm$^{-3}$.
In the noise-dominated regime, the Fisher information for this population peaks at $\ell \sim D_l^{\mathrm{min}}/R_0$, insensitive to other lens properties. The growth in significance until this point is again roughly linear. 
Multipoles $\ell > D_l^{\mathrm{min}}/R_0$ contribute logarithmically to the total significance, with a cutoff around $\ell \sim D_l^{\mathrm{max}}/R_0$ after which the significance plateaus. For accelerations, we show the Fisher information per $\ell$ mode and the cumulative significance in the bottom row of Fig.~\ref{fig:fisher_sig}, assuming $\sigma_\alpha = 5\,\mu\mathrm{as}\,\mathrm{yr}^{-2}$ and all other quantities the same as before. Note that for these parameters, the significance is very low for acceleration power spectra, which is more sensitive to lenses with smaller $R_0$.

Also unlike in the point lenses case, there is nothing preventing us from considering the limit $D_l^\mathrm{min}\rightarrow0$, since the singularity at the origin is regulated in the case of fluffier lenses. In this case, there is additional contribution from larger scales $\ell\lesssim D_l^\mathrm{min} / R_0$ compared to the case just considered. This is illustrated with the dotted lines for velocities (middle row) and accelerations (bottom row) in Fig.~\ref{fig:fisher_sig}.

It is instructive also to consider how the ``peak'' significance, $\sigma_\mathrm{sig}(\ell_\mathrm{max} = D_l^\mathrm{max}/R_0)$, scales with $D_l^\mathrm{max}$. This is clear from the previous section---significance receives equal contribution per decade in $\ell_\mathrm{max}$ until $\ell_\mathrm{max} \sim D_l^\mathrm{max}/R_0$. Hence, the peak significance also receives equal contribution per logarithmic distance interval probed. Because each decade in line-of-sight distance below $D_l^\mathrm{max}$ contributes equally to the significance, there is substantial fractional variation of the significance for different signal realizations.

\begin{figure}[htbp]
\centering
\includegraphics[width=0.45\textwidth]{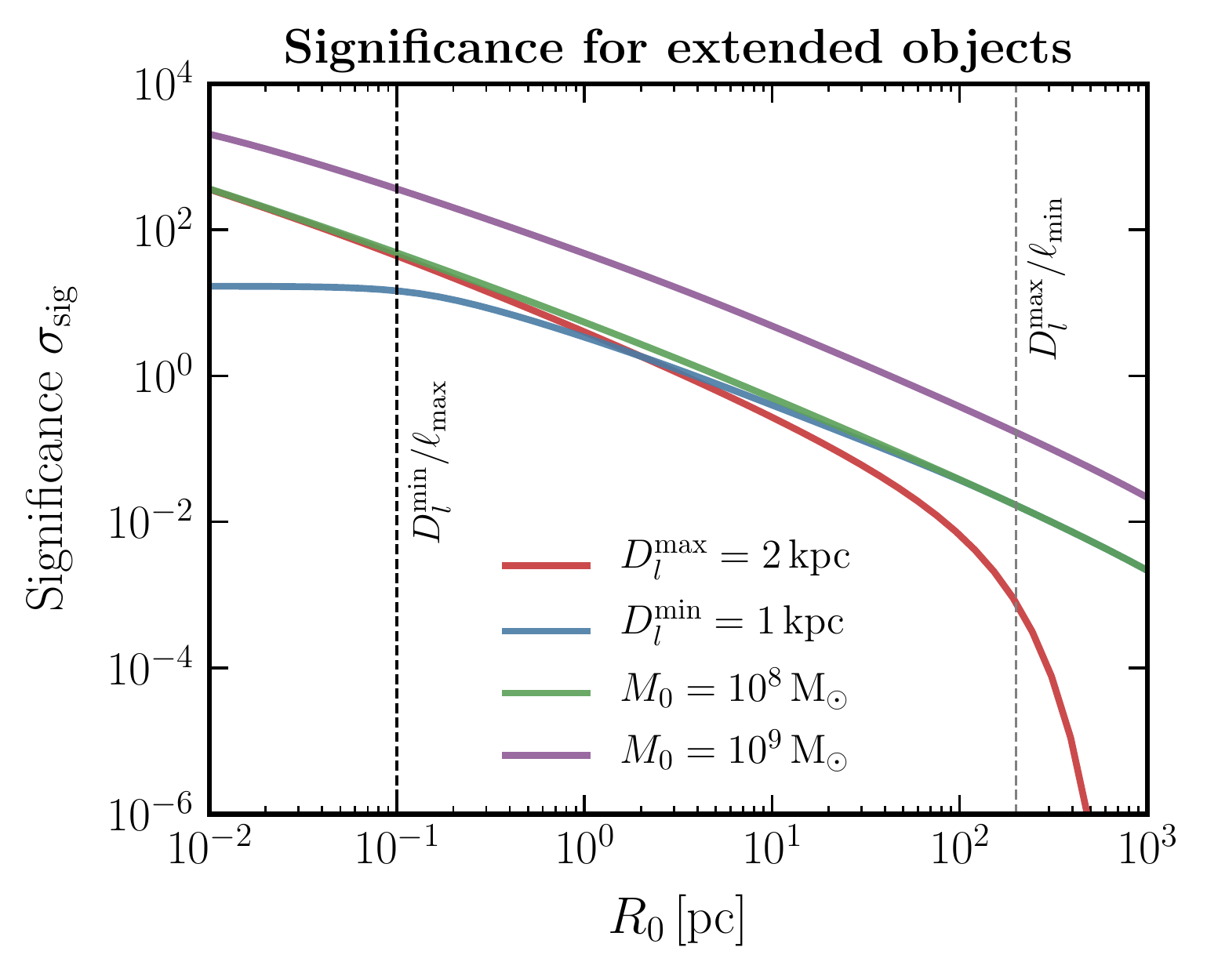}
\caption{Detection significance as a function of spatial extension $R_0$ illustrated for Gaussian lenses uniformly distributed with varying minimum and maximum lens distances, with proper motion power spectra measurements between $\ell \in [10, 10^4]$. Same lens properties as Fig.~\ref{fig:fisher_sig} (middle row). The red and blue lines show the effect of impossing a maximum or minimum distance to the lenses at 2 and 1\,kpc, respectively. Significance without a distance cutoff is shown for $M_0 = 10^{8(9)}\,\mathrm{M}_\odot$ lenses in green(purple). \nblink{03_scalings}} 
\label{fig:sig_R0}
\end{figure}

Finally, we consider the impact of lens extension on detection significance. The significance using velocity spectra in the range $\ell \in [10, 10^4]$ as a function of lens extension $R_0$ is shown in Fig.~\ref{fig:sig_R0} for uniformly distributed lenses with masses $M_0 = 10^{8}(10^{9})\,\mathrm{M}_\odot$ in green (purple). Two relevant scales can be seen. The significance plateaus for $R_0 \lesssim D_l^\mathrm{min}/\ell_\mathrm{max}$ (blue line) and falls off rapidly for $R_0 \gtrsim D_l^\mathrm{max}/\ell_\mathrm{min}$ (red line) when minimum and maximum distances are imposed. Note also from Fig.~\ref{fig:sig_R0} (and Eqs.~\eqref{eq:mu_ext} and~\eqref{eq:mu_extpop}) that the various scales of interest do not depend on the lens mass $M_0$, with the total significance scaling linearly with $M_0$ in the case of a uniformly distributed population of lenses.

{}

We can summarize the main takeaways of this appendix by approximate formulae of the signal significance for proper motions and accelerations induced by Gaussian lenses uniformly distributed in a sphere of radius $D_l^\mathrm{max}$. Using Eqs.~\ref{eq:mu_extpop} and \ref{eq:alpha_extpop} for the signal power spectra, Eq.~\ref{eq:noise_powspec} for the noise power spectra, and Eq.~\ref{eq:fisher_l} for the Fisher information at signal-to-noise ratio $C/N \ll 1$, we can approximate the discrete sum in the significance formula of Eq.~\ref{eq:signif} as an integral from $\ell_\mathrm{min}$ to $\ell_\mathrm{max}$. We then find the parametric significance for velocity and acceleration power spectra:
\begin{align}
\sigma_\mathrm{sig}^\mu &\simeq \frac{\Sigma_q f_\mathrm{sky}^{1/2}}{\sigma_\mu^2} 16 \pi^{5/2} G_N^2 \frac{M_0}{R_0} v_l^2 \rho_\mathrm{DM} f_\mathrm{DM} F \label{eq:signif_mu}\\
\sigma_\mathrm{sig}^\alpha &\simeq \frac{\Sigma_q f_\mathrm{sky}^{1/2}}{\sigma_\alpha^2} 6 \pi^{5/2} G_N^2 \frac{M_0}{R_0^3} v_l^4 \rho_\mathrm{DM} f_\mathrm{DM} F  \label{eq:signif_alpha}
\end{align}
where the factor $F$ is approximately given by:
\begin{align}
F \simeq \sqrt{\ln\left[\frac{\min\left( \ell_\mathrm{max}, D_l^\mathrm{max}/R_0\right)}{\ell_\mathrm{min}} \right]}.
\end{align}
These formulae are in concordance with the findings of this appendix as well as the signal-to-noise formulae of the correlation observables proposed in V18: \emph{cf.}, their Eqs.~(6.8) and (6.11) (which only had different form factors $I_1$ and $I_2$ due to the assumed NFW profile of the lenses).

\begin{figure}[htbp]
\centering
\includegraphics[width=0.45\textwidth]{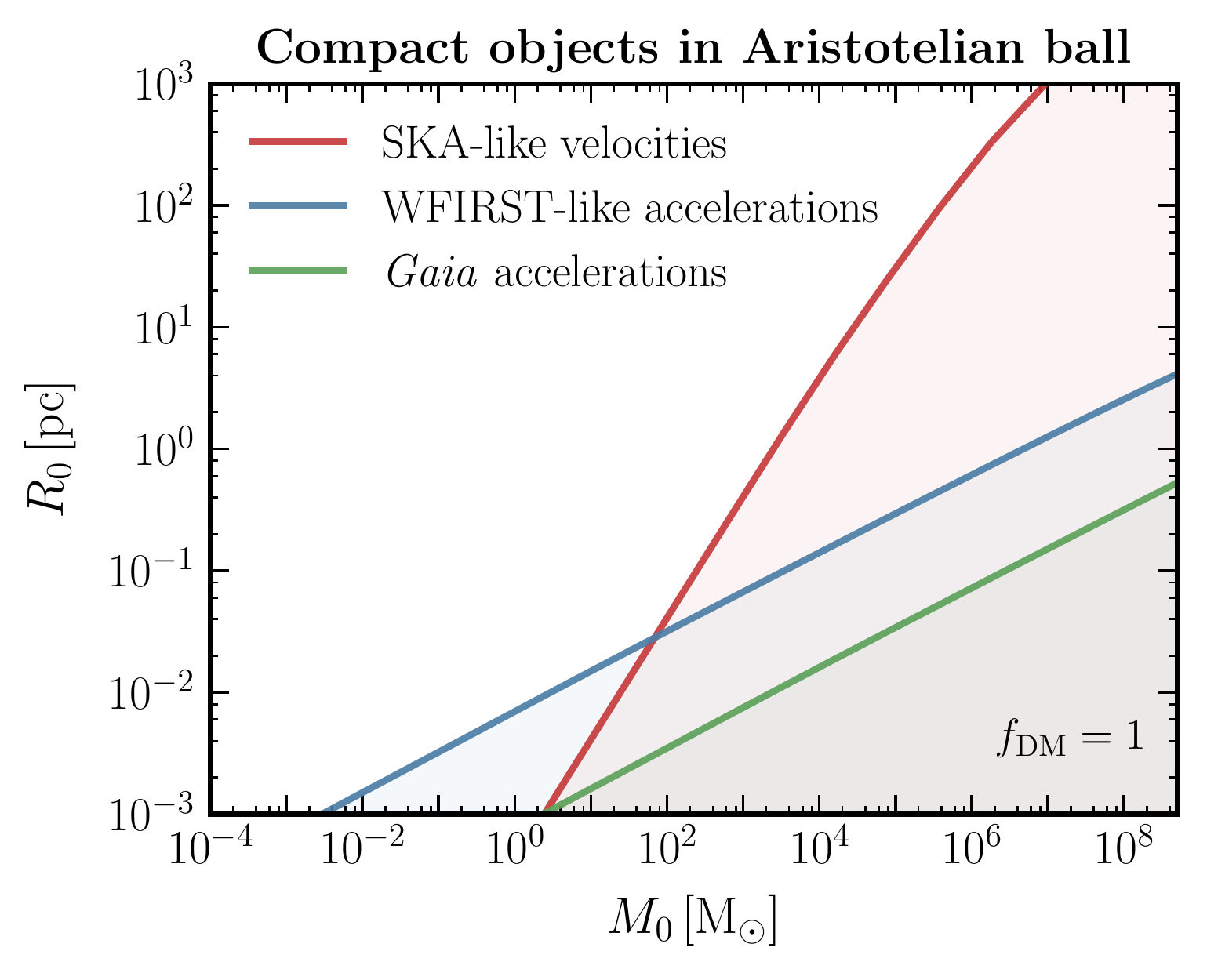}
\caption{Projected limits at the 95\% confidence level obtained for a uniform population of Gaussian lenses distributed in an Aristotelian ball, with signal power spectra given by Eqs.~\eqref{eq:mu_extpop} and~\eqref{eq:alpha_extpop}. Shown for extragalactic proper motion measurements from an SKA-like survey (red line), and Galactic proper acceleration measurements from a WFIRST-like survey and end-of-mission \Gaia (blue and green lines, respectively), assuming survey characteristics in Tab.~\ref{tab:noise_specs}. Lenses up to $D_l^{\mathrm{max}} = 50\,\mathrm{kpc}$ and 1\,kpc are considered for the velocity and acceleration measurements, respectively. \nblink{10_aristotle}} 
\label{fig:compact_sens_arist}
\end{figure}

In Sec.~\ref{sec:compact}, we derived projected sensitivities for a population of compact objects of different masses and sizes making up the dark matter in the Milky Way, integrating over the assumed Galactic spatial and velocity phase space distributions of the lenses. In Fig.~\ref{fig:compact_sens_arist} we show 95\% confidence level sensitivity projections for a simplified scenario of lenses uniformly distributed in an Aristotelian ball up to $D_l^{\mathrm{max}} = 1(50)$\,kpc for acceleration(velocity) measurements and assuming transverse velocity $v_l=10^{-3}$. Survey characteristics from Tab.~\ref{tab:noise_specs} are used for SKA-like extragalactic proper motions (shown as red), and WFIRST-like and end-of-mission \Gaia Galactic proper accelerations (shown in blue and green, respectively). Eqs.~\eqref{eq:mu_extpop} and~\eqref{eq:alpha_extpop} are directly used for these simplified estimates. Excellent agreement with the sensitivities derived using the full phase space in Fig.~\ref{fig:compact_sens} is seen, as well as with the approximate estimates of Eqs.~\ref{eq:signif_mu} and \ref{eq:signif_alpha}.

\section{Power spectrum estimator}\label{app:estimator}

In this appendix, we describe a fast, practical way to estimate the power spectrum of vector data on (a potential subset of) the celestial sphere. This is done through the construction of a quadratic maximum-liklihood VSH estimator. As an illustration, we apply this method to the proper motions of the quasar sample in \Gaia's second data release (DR2)~\cite{Prusti:2016bjo,Brown:2018dum} in Sec.~\ref{sec:gaia-quasars}.

\subsection{Setup}\label{app:estimator-setup}
Suppose we are given vector data on the sphere $d_{j\alpha} = s_{j\alpha} + n_{j\alpha}$ composed of a signal $s$ and noise $n$. The Greek index $\alpha = 1,2$ runs over the two vector components in the colatitude $\theta$ and longitude $\varphi$ directions.
The Roman index $j = 1, \dots, J$ runs over a list of equal-area pixels tessellating the celestial sphere.  We assume the number of pixels $J$ is taken large enough to resolve the smallest angular scales over which the signal power is to be estimated. If the data come as a list of objects (as in the \Gaia~quasar sample of Sec.~\ref{sec:gaia-quasars}), we first bin the objects in their corresponding pixels; pixels with multiple constituents receive a value $d_{j\alpha}$ corresponding to their noise-weighted average.

Our goal is to estimate the power spectrum of the signal contribution to the covariance matrix of the data:
\begin{align}
C_{i\alpha j \beta} = \langle d_{i \alpha} d_{j \beta} \rangle = \langle s_{i \alpha} s_{j \beta} \rangle + \langle n_{i \alpha} n_{j \beta} \rangle \equiv S_{i \alpha j \beta} + N_{i \alpha j \beta},
\end{align}
where in the second equality we have assumed the signal to be uncorrelated from the noise.
Using shorthand notation $\Psi^{\ell m}_{j \alpha} \equiv \Psi_{\ell m, \alpha}(\theta_j, \varphi_j)$ and $\Phi^{\ell m}_{j \alpha} \equiv \Phi_{\ell m, \alpha}(\theta_j, \varphi_j)$, we can write any field on the tessellated sphere, \emph{e.g.}~the signal, as
\begin{align}
s_{j\alpha} &=  s_{\ell m}^{(1)} \Psi^{\ell m}_{j \alpha} +  s_{\ell m}^{(2)} \Phi^{\ell m}_{j \alpha}; \\
s_{\ell m}^{(1)} &= \frac{4\pi}{J}  s_{j\alpha} \Psi^{\ell m *}_{j \alpha}; \quad s_{\ell m}^{(2)} = \frac{4\pi}{J} s_{j\alpha} \Phi^{\ell m *}_{j \alpha}.
\end{align}
Above and in the rest of this appendix, repeated indices are summed unless otherwise noted.
We can express the signal covariance matrix $S_{i\alpha j\beta}$ in terms of its power spectra $S^{(1,2)}_{\ell m}$ and response matrices $P^{(1,2)\ell m}_{i\alpha j \beta}$:
\begin{align}
S_{i \alpha j \beta} &=  S^{(1)}_{\ell m} P_{i \alpha j \beta}^{(1) \ell m } + S^{(2)}_{\ell m} P_{i \alpha j \beta}^{(2) \ell m }; \\
\langle s^{(1)}_{\ell m} s^{(1)*}_{\ell m} \rangle &\equiv S^{(1)}_{\ell m} \delta_{\ell \ell'} \delta_{m m'}; \quad P_{i \alpha j \beta}^{(1) \ell m } \equiv \Psi_{i \alpha}^{\ell m} \Psi_{j \beta}^{\ell m *}; \label{eq:Presponse}
\end{align}
and analogous formulas for $S^{(2)}_{\ell m}$ and $P^{(2)\ell m}_{i \alpha j \beta}$. The noise covariance matrix can be similarly expressed in terms of its power spectra $N^{(1,2)}_{\ell m}$. We shall assume here that the noise covariance matrix is known (or otherwise inferred). In what follows, we take the noise to be uncorrelated between pixels and isotropic in direction (but \emph{not} isotropic in location on the celestial sphere):
\begin{align}
N_{i \alpha j \beta} = N_i \delta_{ij} \delta_{\alpha \beta} \quad \text{(no sum over $i$)} \label{eq:Ndiagonal}.
\end{align}
For brevity in the analysis below, we will take 
\begin{align}
S^{(2)}_{\ell m} = 0 \label{eq:S2zero}
\end{align} 
(as in the case of a lensing signal) so that only $S^{(1)}_{\ell m}$ is to be estimated.
All formulas can be generalized with straightforward modifications if the assumptions of Eqs.~\eqref{eq:Ndiagonal} and \eqref{eq:S2zero} are relaxed.

\subsection{General likelihood estimator}\label{app:estimator-general}
The Gaussian likelihood of the data is given by:
\begin{align}
\mathcal{L} = \frac{\exp\left\lbrace -\frac{1}{2} d_{i \alpha} \Cinv_{i \alpha j \beta} d_{j \beta} \right\rbrace}{(2\pi)^J \sqrt{\det C}},
\end{align}
where $C$ is the covariance matrix, and the log likelihood is defined as $L \equiv - 2 \ln \mathcal{L}$. We are interested in finding the signal power spectra $S^{(1)}_{\ell m}$ and thus the correct covariance matrix $C$ that minimizes the log likelihood. To that end we compute the first derivative:
\begin{align}
&\hspace{-1em} \frac{\partial L}{\partial S^{(1)}_{\ell m}} =\\
& \Cinv_{i \alpha j \beta} P^{(1) \ell m}_{j \beta i \alpha} - d_{a \alpha} \Cinv_{a \alpha b \beta} P^{(1) \ell m}_{b \beta c \gamma} \Cinv_{c \gamma e \epsilon} d_{e \epsilon}.\nonumber
\end{align}
It can be shown that $\langle {\partial L}/{\partial S^{(1)}_{\ell m}} \rangle = 0$ for $C = \langle d d \rangle$. After calculating the Hessian $ {\partial^2 L}/{\partial S^{(1)}_{\ell m} \partial S^{(1)}_{\ell' m'}}$, one can devise an iterative procedure based on the Newton-Raphson method that finds a local minimum of $L$, as in Ref.~\cite{Dahlen:2007sv}. If the global minimum $\bar{S}^{(1)}_{\ell m}$ is found, then the maximum-likelihood estimator $\hat{S}^{(1)}_{\ell m} = \bar{S}^{(1)}_{\ell m}$ is guaranteed to be the optimal unbiased estimator:
\begin{widetext}
\begin{align}
\hat{S}^{(1)}_{\ell m} &= d_{i\alpha} Z^{(1) \ell m}_{i\alpha j \beta} d_{j \beta} - N_{i \alpha j \beta} Z^{(1) \ell m}_{j \beta i\alpha}; \\
Z^{(1) \ell m}_{i \alpha j \beta} &= \frac{1}{2}  (F_{(1)}^{-1})_{\ell m \ell' m'} \Cinv_{i \alpha i' \alpha'} P^{(1) \ell' m'}_{i' \alpha' j' \beta'} \Cinv_{j' \beta' j \beta}; \\
F^{(1)}_{\ell m \ell' m'} &= \frac{1}{2} \Cinv_{a \alpha b \beta} P^{(1) \ell m}_{b \beta c \gamma} \Cinv_{c \gamma e \epsilon} P^{(1) \ell' m'}_{e \epsilon a \alpha}; \\
\text{Cov}\left\lbrace \hat{S}^{(1)}_{\ell m} \hat{S}^{(1)}_{\ell' m'} \right\rbrace &= (F_{(1)}^{-1})_{\ell m \ell' m'}.
\end{align}
\end{widetext}
Note that the last equation is the covariance matrix of the estimator, thus providing error estimates on the measurement.

\subsection{Simplifications at low signal to noise}
In the above, $Z$ and the Fisher matrix $F$ are to be evaluated at $\bar{S}$ on the RHS of the $\hat{S}$ formula, which is why an iterative procedure is needed. However, in the limit where the signal power is small compared to the noise power $S \ll N$ in any one (spatial or spectral) bin, which is the regime of interest, we can make the approximation $C \simeq N$ at the cost of slightly biasing the estimator by a fractional amount of $S/N$. In this case, no iteration is needed. If we further assume that $N$ is diagonal as in Eq.~\eqref{eq:Ndiagonal}, we find:
\begin{align}
\hspace{-1em}&\hat{S}^{(1)}_{\ell m} = \frac{1}{2} (F_{(1)}^{-1})_{\ell m \ell' m'} \left[ \frac{ d_{i \alpha} d_{j \beta}}{N_i N_j} P^{(1)\ell' m'}_{i \alpha j \beta} -   \frac{P^{(1)\ell' m'}_{i \alpha i \alpha}}{N_i} \right] \label{eq:estimatorlowS}\\
\hspace{-1em}&F^{(1)}_{\ell m \ell' m'} = \frac{1}{2} \sum_{i j \alpha \beta} \frac{P^{(1)\ell m}_{i \alpha j \beta} P^{(1)\ell' m'}_{j \beta i \alpha}}{N_i N_j} \label{eq:fisherlowS}
\end{align}
For the toroidal power spectrum estimator $\hat{S}^{(2)}_{\ell m}$, the formulas are the same except for the replacements $P^{(1)} \leftrightarrow P^{(2)}$. One can actually show that $F^{(1)} = F^{(2)} \equiv F$ in this case.

If the noise were the same in each pixel, $N_i = \sigma^2~ \forall i$, then we would have a diagonal Fisher matrix $F_{\ell m \ell' m'} = (J/2\sigma^4)\delta_{\ell \ell'} \delta_{m m'}$ and an even simpler estimator: $\hat{S}^{(1,2)}_{\ell m} = d_{i \alpha} d_{j \beta} P^{(1,2)\ell m}_{i \alpha j \beta} - \sigma^2$. Empty pixels $d_i$ carry no information; it can be seen from the likelihood or its estimator that not summing over empty pixels $d_i$ is equivalent to taking their noise to be infinite $N_i = \infty$. This allows us to analyze vector data over a subset of the celestial sphere.

In principle, with Eqs.~\eqref{eq:estimatorlowS} and \eqref{eq:fisherlowS} we have assembled all the necessary ingredients to estimate the signal power spectrum, but there are two practical hurdles: computational complexity and invertibility of the Fisher matrix. Suppose we want to estimate the power spectrum up to $\ell = \ell_\mathrm{max}$, \emph{i.e.}~estimate the power for $(\ell_\mathrm{max}+1)^2$ values of $(\ell, m)$. That means we require at least as many pixels $J \gtrsim (\ell_\mathrm{max}+1)^2$. At first glance, it would appear from Eq.~\eqref{eq:fisherlowS} that the computational complexity of the Fisher matrix calculation scales as $\mathcal{O}(\ell_\mathrm{max}^8)$, which would be preclude computations up to even medium-high $\ell_\mathrm{max}$ values. In addition, there is no guarantee that the inverse of the Fisher matrix in Eq.~\eqref{eq:fisherlowS} exists for maps with nonuniform exposure, especially when $\ell_\mathrm{max}^2$ approaches the number of nonempty pixels. Even if the inverse exists, its computation may be numerically unstable. App.~\ref{app:complexity} details a parametrically faster method to compute the Fisher matrix, and App.~\ref{app:specbin} a spectral binning method that makes it invertible in most practically relevant cases; combined, they circumvent the issues of complexity and invertibility.

\subsection{Fast Fisher matrix computation}\label{app:complexity}
We outline a method for computing Eq.~\eqref{eq:fisherlowS} in $\mathcal{O}(\ell_\mathrm{max}^4 \log^2 \ell_\mathrm{max})$ time. First, define the $(\ell m)$ Fourier amplitude of a scalar map $f^{\ell' m'}_i$ to be $f_{\ell m}^{\ell' m'} \equiv f^{\ell' m'}_i Y_i^{\ell m*}$. Also observe that we can write the VSHs as:
\begin{align}
\Psi^{\ell m}_{i \alpha} &= \frac{\delta_{\alpha 1} \big[ c^{\ell m}_1 Y^{\ell-1, m}_i + c^{\ell m}_2 Y^{\ell+1, m}_i \big]  + \delta_{\alpha 2} c^{\ell m}_3 Y^{\ell m}_i}{-\sin \theta_i}; \nonumber \\
c^{\ell m}_1 &=  \sqrt{\frac{(\ell+1)(\ell - m)(\ell + m)}{\ell(2\ell -1)(2\ell +1)}}; \\
c^{\ell m}_2 &= - \sqrt{\frac{\ell(\ell - m+1)(\ell + m+1)}{(\ell+1)(2\ell +3)(2\ell +1)}}; \quad
c^{\ell m}_3 = \frac{i m}{\sqrt{\ell(\ell+1)}}.  \nonumber
\end{align}
The formula for $\Phi^{\ell m}_{i \alpha}$ is similar, via Eq.~\eqref{eq:PsiPhidef}.

The key observation is that one can write the Fisher matrix as the element-wise product:
\begin{align}
F_{\ell m \ell' m'} &= Q_{\ell m \ell' m'} Q_{\ell' m' \ell m} \quad \text{(no sum over $\ell, m, \ell', m'$)}; \nonumber\\
Q_{\ell m \ell' m'} &\equiv c^{\ell' m'}_1 A^{\ell m}_{\ell'-1,m'} + c^{\ell' m'}_2 A^{\ell m}_{\ell'+1,m'} - c^{\ell' m'}_3 B^{\ell m}_{\ell' m'}; \nonumber\\ 
A^{\ell m}_i &= \frac{1}{N_i \sin \theta_i} \left( c^{\ell m}_1 Y^{\ell-1,m}_i + c^{\ell m}_2 Y^{\ell+1,m}_i \right); \nonumber\\ 
B^{\ell m}_i &= \frac{1}{N_i \sin \theta_i} c^{\ell m}_3 Y^{\ell m}_i.
\end{align}
Brute-force computation of each $Q_{\ell m \ell' m'}$ amplitude individually takes $\mathcal{O}(\ell_\mathrm{max}^2)$, reducing the total complexity of $F_{\ell m \ell' m'}$ to $\mathcal{O}(\ell_\mathrm{max}^6)$. However, fast spherical Fourier transform algorithms~\cite{2013GGG....14..751S,2013A&A...554A.112R,doi:10.1029/2018GC007529} exist to compute $Q_{\ell m \ell' m'}$ for all $\lbrace \ell' m'\rbrace$ at once in $\mathcal{O}(\ell_\mathrm{max}^2 \log^2 \ell_\mathrm{max})$ steps, yielding a total complexity of $\mathcal{O}(\ell_\mathrm{max}^4 \log^2 \ell_\mathrm{max})$ for the $(\ell_\mathrm{max} + 1)^4$ components of $F$, making computations up to $\ell_\mathrm{max} \sim 10^3$ feasible. 

\subsection{Spectral binning}\label{app:specbin}
The maximum likelihood estimator requires the inverse of the Fisher matrix, which may not exist if the data are too sparse or not sufficiently uniform, and/or if $\ell_\text{max}$ is too large. Since we are not interested in the power in any one $(\ell,m)$ mode, but rather in the gross behavior, we can collect the power spectrum into ``spectral bins'': \emph{e.g.}~logarithmic bins in $\ell$, and two bins in $|m|$: a high-$|m|/\ell$ bin, and a low-$|m|/\ell$ bin. More precisely, we can compute ``band averages'':
\begin{align}
S^{(1)}_B \equiv  W_{B \ell m} S^{(1)}_{\ell m}, \label{eq:Sbinned}
\end{align}
with band matrices $W_{B \ell m}$ that have many fewer rows than columns and which satisfy $\sum_{\ell m} W_{B \ell m} = 1$. The binning suggested above could be spectral bins $B = 2n$ containing all $(\ell,m)$ satisfying $\lbrace 2^n \le \ell < 2^{n+1}, |m|\le\text{floor}(\ell/2)\rbrace$ as the low-$|m|/\ell$ bins, and $B = 2n+1$ with $\lbrace 2^n \le \ell < 2^{n+1}, |m|>\text{floor}(\ell/2)\rbrace$ as the high-$|m|/\ell$ ones. The band matrix $W_{B \ell m}$ would then be nonzero only for those $\{\ell m\} \in B$, and equal to the inverse number of elements in $B$.

This spectral binning does not lose essential information insofar that the power spectrum $S^{(1)}_{\ell m}$ can be adequately approximated by the coarse-grained power spectrum $S^{(1)\dagger}_{\ell m}$:
\begin{align}
S^{(1)\dagger}_{\ell m} =  W_{\ell m B}^\dagger S_B^{(1)}
\end{align}
where $W_{\ell m B}^\dagger$ is the pseudo-inverse
\begin{align}
W_{\ell m B}^\dagger \equiv W_{B \ell m} \left( W_{B' \ell' m'} W_{B \ell' m'} \right)^{-1}.
\end{align}
All formulas for $S_B$, $Z_{B B'}$, and $F_{BB'}$ are the same as in the previous section with the replacements $\ell m \leftrightarrow B$ and using the spectrally-binned response matrices:
\begin{align}
P^{(1)B}_{i\alpha j \beta} = P^{(1)\ell m}_{i\alpha j \beta} W^\dagger_{\ell m B}. \label{eq:Pbinned}
\end{align}
The exemplary logarithmic binning in $\ell$ and high/low-binning in $m$ compresses a power spectrum of $\mathcal{O}(\ell_\mathrm{max}^2)$ modes into a coarse-grained one with $\mathcal{O}(2 \log_2 \ell_\mathrm{max})$ band averages. The more compact band-averaged Fisher matrix $F_{BB'}$ is more likely to be well-conditioned such that its inverse exists and can be calculated quickly and reliably.

\bibliographystyle{apsrev4-1}
\bibliography{lenspower}

\end{document}